\documentclass{aa}
\usepackage{graphicx}
\usepackage{rotating}
\usepackage{longtable}
\def\lesssim{\mathrel{\hbox{\rlap{\hbox{\lower4pt\hbox{$\sim$}}}\hbox{$<$}}}}
\def\gtrsim{\mathrel{\hbox{\rlap{\hbox{\lower4pt\hbox{$\sim$}}}\hbox{$>$}}}}
\begin{document}
\onecolumn

  \title{The {\em Chandra} Deep Field South/GOODS survey
      \thanks{Based in part on observations obtained at the European Southern Observatory using the ESO Very Large Telescope on Cerro Paranal (ESO program 168.A-0485)}}

   \subtitle{Optically faint X-ray sources}

   \author{V. Mainieri\inst{1,2,3}
           \and P. Rosati\inst{2}
           \and P. Tozzi\inst{4}
            \and J. Bergeron\inst{5}
            \and R. Gilli\inst{6}
            \and G. Hasinger\inst{1}
            \and M. Nonino\inst{4}
            \and I. Lehmann\inst{1}     
            \and D.M. Alexander\inst{7}
            \and R. Idzi\inst{8}
            \and A.M. Koekemoer\inst{8}
            \and C. Norman\inst{9}
           \and G. Szokoly\inst{1}
           \and W. Zheng\inst{9}
          }

   \offprints{V. Mainieri, \email{vmainieri@mpe.mpg.de}}

   \institute{Max-Planck-Institut f\"ur extraterrestrische Physik, Giessenbachstrasse PF 1312, 
                85748 Garching bei Muenchen, Germany
        \and 
                European Southern Observatory,
                Karl-Schwarzschild-Strasse 2, D-85748 Garching, Germany
       \and
                Dip. di Fisica, Universit\`a degli Studi Roma Tre, 
                Via della Vasca Navale 84, I-00146 Roma, Italy
        \and 
                INAF, Osservatorio Astronomico di Trieste, via G.B. Tiepolo 11, I-34131, Trieste, Italy
        \and
            Institut d'Astrophysique de Paris, 98bis Boulevard, F-75014 Paris, France
        \and    
                INAF, Osservatorio Astrofisico di Arcetri, Largo E. Fermi 5, I-50125, Firenze, Italy
        \and
                Institute of Astronomy, Madingley Road, Cambridge CB3 0HA, UK
        \and
                Space Telescope Science Institute, 3700 San Martin Drive, Baltimore, MD 21218, USA
        \and
                Center for Astrophysical Sciences, Department of Physics and Astronomy, The Johns Hopkins University, Baltimore, MD 21218, USA
             }

   \date{Received 2 May 2004; Accepted 25 May 2005}

   \abstract{We provide important new constraints on the nature and
     redshift distribution of optically faint ( R$\ge 25$) X-ray
     sources in the Chandra Deep Field South Survey. We use a large
     multi-wavelength data set, including the GOODS/ACS survey, the
     recently released Hubble Ultra Deep Field (UDF) data, and the new
     public VLT/ISAAC imaging. We show that we can derive accurate
     photometric redshifts for the spectroscopically unidentified
     sources thus maximizing the redshift completeness for the whole
     X-ray sample.  Our new redshift distribution for the X-ray source
     population is in better agreement with that predicted by X-ray
     background synthesis models; however, we still find an
     overdensity of low redshift (z$< 1$) sources. The optically faint
     sources are mainly X-ray absorbed AGN, as determined from direct
     X-ray spectral analysis and other diagnostics.\\ Many of these
     optically faint sources have high ($>10$) X-ray-to-optical flux
     ratios. We also find that $\sim 71 \%$ of them are well fitted
     with the SED of an early-type galaxy with $<$z$_{\rm phot}> \sim
     1.9$ and the remaining $29 \%$ with irregular or starburst
     galaxies mainly at z$_{\rm phot}>3$. We estimate that $23 \%$ of
     the optically faint sources are X-ray absorbed QSOs.  The overall
     population of X-ray absorbed QSOs contributes a $\sim 15 \%$
     fraction of the [2-10] keV X-ray Background (XRB) whereas current
     XRB synthesis models predict a $\sim 38 \%$ contribution.
      }

   \maketitle
%

\begin{figure}  
 \begin{center}
\resizebox{18cm}{!}{\includegraphics{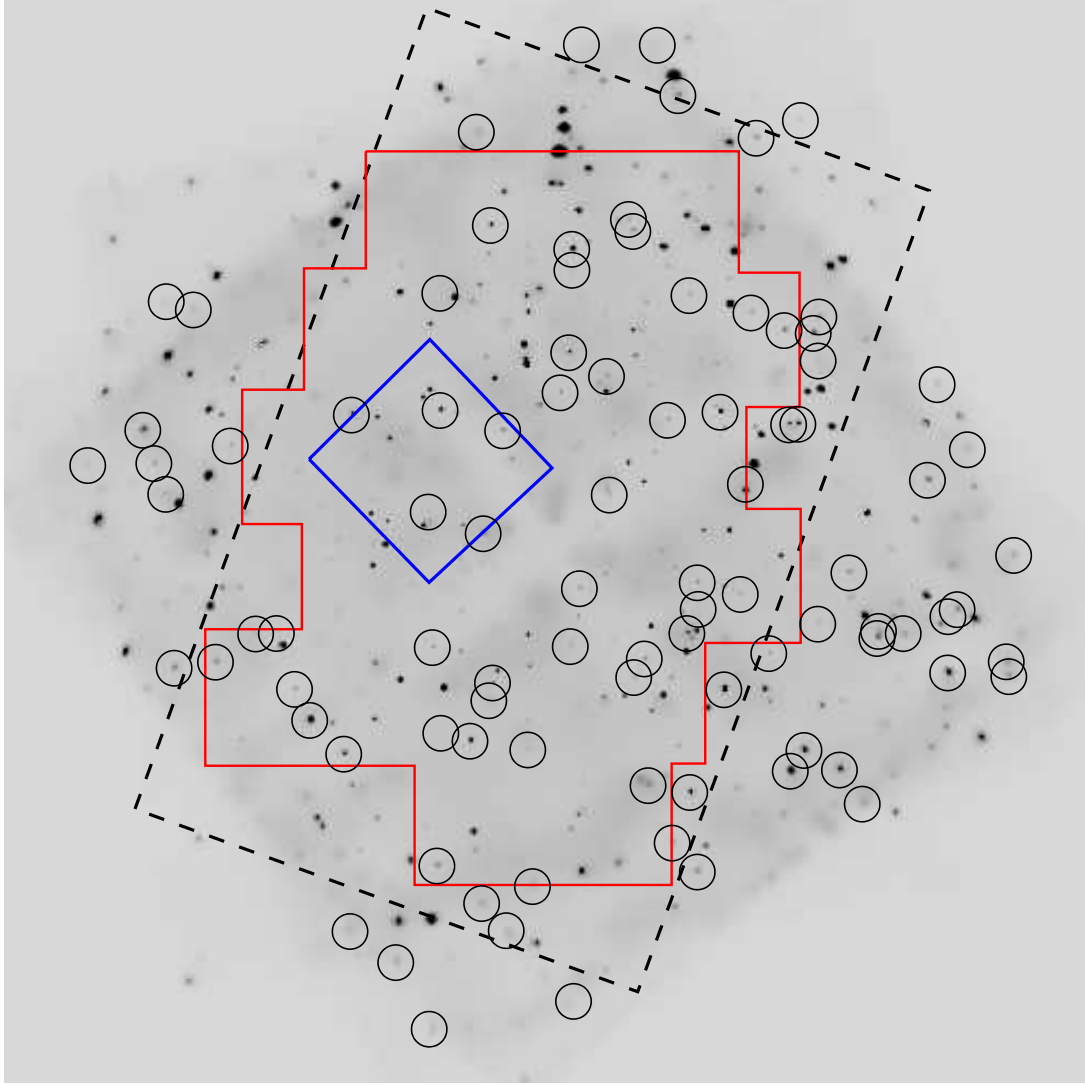}}
\end{center}
   \caption{Adaptively smoothed Chandra image of the CDF-S in the
     [0.5-7] keV band. The circles show the position of OFS. The big
     dashed rectangle is the area of the GOODS survey ({\it ``GOODS
       area''}); the polygon indicates the region of the deep public
     VLT/ISAAC observations currently available and the small square
     indicates the area of the Hubble Ultra Deep Field (UDF).}
\label{figure:fig_2} 
\end{figure}

\begin{figure}  
 \begin{center}
\resizebox{8cm}{!}{\includegraphics{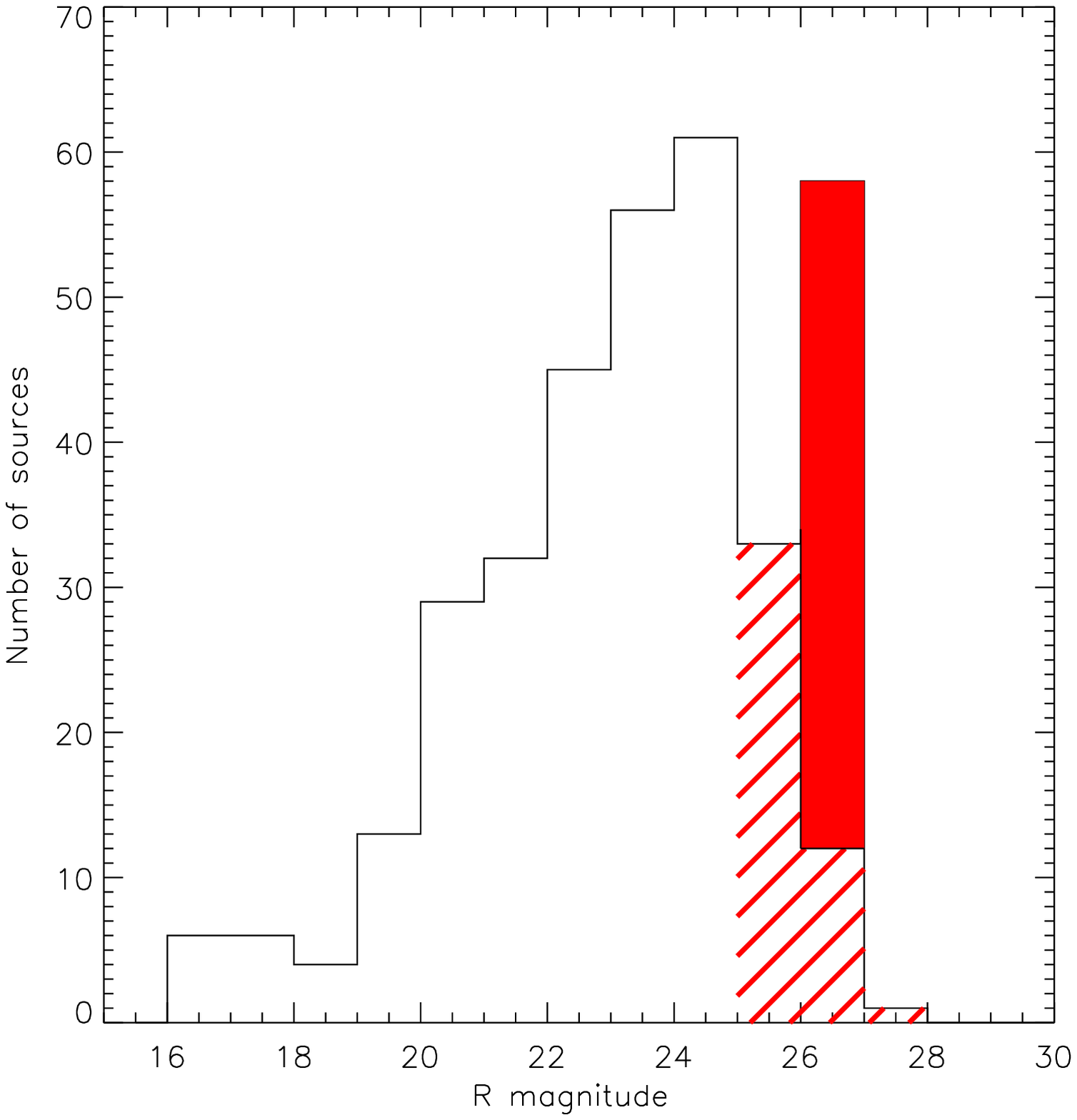}}
\end{center}
   \caption{The R-band magnitude distribution for the optical
     counterparts of the X-ray sample. The hatched box are the OFS,
     while the shaded part refers to the sources without a R band
     counterparts (down to R$=26.7$).}
\label{figure:fig_1} 
\end{figure}

\section{Introduction}

Recent {\it Chandra} and XMM-{\it Newton} (Brandt et al.
\cite{brandt01}, Rosati et al. \cite{rosati02}, Hasinger et al.
\cite{hasinger01}) deep surveys have almost resolved the entire $2-10$
keV X-ray background (XRB), forty years after its discovery (Giacconi
et al. \cite{giacconi62}). A large optical follow-up program of
hundreds of X-ray sources in these fields has lead to the
identification of a mixture of obscured and unobscured AGN, with an
increasing fraction of obscured AGN at faint X-ray fluxes (Barger et
al. \cite{barger03}, Szokoly et al. \cite{szokoly04}). The optical
counterparts of a significant fraction of the X-ray sources are too
faint (R$\geq$25) for optical spectroscopy, even for $\approx 5-6$
hours exposures with 8-10 m class telescopes. In this work, we focus
on this subsample of X-ray sources with R$\ge25$, which will be
referred to as {\it Optically Faint Sources} (OFS). This criterium
simply reflects an observational limit beyond which we need to rely on
accurate photometric measurements and X-ray spectral information to
estimate their redshift and to establish their physical nature. We
note that OFS represent a quarter of the entire X-ray sample,
therefore they have a significant impact on statistical studies, such
as the X-ray luminosity function, the evolution of the TypeI/TypeII
ratio and the overall N$_{\rm H}$ distribution. Moreover, many OFS are
relatively bright in the X-ray band, due to powerful AGN activity
[Alexander et al.  (\cite{alexander01}); Mignoli et al.
(\cite{mignoli04})]. It is possible that a sizeble fraction of these
sources belong to the long-sought class of high-redshift, high
luminosity, heavily obscured active galactic nuclei (Type II QSOs).\\
The first detailed study of OFS has been done by Alexander et al.
(\cite{alexander01}) in the Chandra Deep Field Nord (CDF-N). Here we
extend that work by estimating photometric redshifts for a large
fraction of the OFS population, providing direct constraints on their
intrinsic absorption and on their spectral energy distribution.\\ The
outline of the paper is as follows: in \S2 we describe the X-ray
sample. In \S3 we compare the X-ray properties of the optically bright
(R$<$25) and faint (R$\ge$25) X-ray sources. In \S4 we derive
photometric redshifts. In \S5 we discuss the few OFS with known
spectroscopic redshifts. In \S6 we use the X-ray to optical flux
ratios as a diagnostic tool. In \S7 we perform an X-ray spectral
analysis and in \S8 we derive the fraction of OFS that are likely to
be X-ray detected Type-2 QSOs. We study in detail the OFS inside the
Hubble Ultra Deep Field in \S9. Finally in \S10 we summarize and
discuss the results.  Throughout this paper we use Vega magnitudes (if
not otherwise stated) and assume $\Omega_{\rm M}=0.3$,
$\Omega_\Lambda=0.7$ and H$_0=70$ km/s/Mpc.

\section{The sample} 

The X-ray sample is obtained from the 1 Ms {\it Chandra} observation
of the Chandra Deep Field South (CDF-S; Giacconi et al.
\cite{giacconi02}; Rosati et al. \cite{rosati02}). We refer to Table 2
in Giacconi et al. (\cite{giacconi02}) for the X-ray quantities and
relative errors (e.g., X-ray fluxes, counts). In Figure
\ref{figure:fig_2} we show a smoothed 0.5-7 keV image of the CDF-S
field on which we have highlighted the OFS. The distribution of R band
magnitudes for the X-ray sources is reported in Figure
\ref{figure:fig_1}. From the entire X-ray sample of 346 sources, 92
are OFS ($\sim 27\%$), of which 46 ($\sim 13\%$ of the total sample)
do not have a R band counterpart (down to R$=$26.1-26.7). In the
CDF-N, the fraction of optically faint sources (I$\ge 24$) is slightly
larger ($\approx 33 \%$; Alexander et al.  \cite{alexander01});
however, this study was performed in the most sensitive region of the
CDF-N field, while here we are considering the whole of the CDF-S
field.\\ At present only six ($\sim 6\%$) of the OFS have been
spectroscopically identified, as compared to 151 ($\sim 60\%$) of the
optically bright (R$<25$) X-ray sources (Szokoly et al.
\cite{szokoly04}). We will discuss in detail the properties of these
six OFS in Section \ref{section:optf_spec}. For the remaining OFS it
will be challenging to obtain a spectroscopic redshift with 8-10 meter
telescopes, with the exception of those sources with strong emission
lines.  In order to obtain information on their redshift distribution
we have determined photometric redshifts (see Section
\ref{section:photo_z}).

\begin{figure}  
 \begin{center}
\resizebox{12cm}{!}{\includegraphics{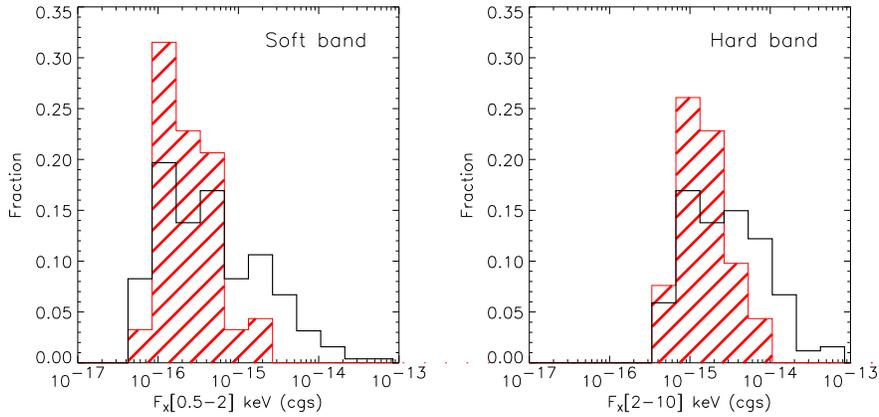}}
\end{center}
   \caption{The distributions of X-ray fluxes in the 0.5-2 keV (left
     panel) and 2-10 keV band (right panel) for the OFS (hatched
     histogram) and optically bright sources (open histogram).}
\label{figure:fig_3} 
\end{figure}

\section{Comparison of the optically bright and optically faint X-ray source samples}

Before we investigate in detail the nature of the optically faint
X-ray source population, we will compare their general X-ray
properties with those of the well-studied optically bright X-ray
source population (see also Section 4 of Alexander et
al. \cite{alexander01}).

\subsection{X-ray fluxes}

In Figure \ref{figure:fig_3} we compare the X-ray fluxes distributions
of the optically bright and faint sources in the 0.5$-$2 keV (left
panel) and 2$-$10 keV bands (right panel). According to a
Kolmogorov-Smirnov test the probability that the distributions are
drawn from the same population is extremely small ($\sim0.02 \%$ in
the 0.5-2 keV band and $\sim 0.001 \%$ in the 2-10 keV band).
According to Figure \ref{figure:fig_3}, almost all the bright X-ray
sources are part of the optically bright sample.

The total flux in the 2-10 keV band of the OFS is $\sim 1.6 \times
10^{-12}$ erg cm$^{-2}$ s$^{-1}$ deg$^{-2}$ after correcting for the
sky coverage in the CDF-S (see Fig. 5 of Giacconi et al.
\cite{giacconi02}). This accounts for a $\sim 7-10 \%$ fraction of the
[2-10] keV XRB; the estimated error range corresponds to the
uncertainty in the measurement of the XRB flux (HEAO$-1$, Marshall et
al.  \cite{marshall80} and Revnivtsev et al.  \cite{revnivtsev04};
SAX, Vecchi et al.  \cite{vecchi99}; XMM-{\it Newton}, De Luca \&
Molendi \cite{deluca04}).

\subsection{Hardness ratios distibution}

The hardness ratio (HR) is a useful tool to characterize the spectral
shape of the AGN X-ray continuum. We adopt the definition
HR$\equiv$(H$-$S)/(H$+$S), where H and S are the net count rates in
the 2-7 keV and 0.5-2 keV band, respectively. We have recomputed the
HR values over those given in Giacconi et al.  (\cite{giacconi02}) by
performing aperture photometry for each source in both the 0.5-2 keV
and 2-10 keV bands irregardless of whether the source is detected in
either of these bands. Therefore, several values previously set to
$+1$ or $-1$ (Giacconi et al.  \cite{giacconi02}) have a different
value, still consistent with the old one within 1$\sigma$ error. When
HR$=\pm1$, we plot the 1$\sigma$ upper/lower limits. Other minor
differences come from the new reduction of the CDF-S data performed
after the release of CALDB 2.26 and CIAO3.1. In particular, we applied
the correction for the degraded effective area of ACIS--I chips due to
material accumulated on the ACIS optical blocking filter at the epoch
of the observation using the recently released time--dependent gain
correction (see http://asc.harvard.edu/ciao/threads/acistimegain/).\\ 
We have corrected the HR values for the off-axis angle of the source,
normalizing the soft and hard counts to refer to an on-axis source
falling on the aimpoint (chip3). To do that, we have used the soft and
hard exposure maps computed for the monochromatic energies of 1.5 and
4.5 keV respectively.

\begin{figure}  
 \begin{center}
\resizebox{8cm}{!}{\includegraphics{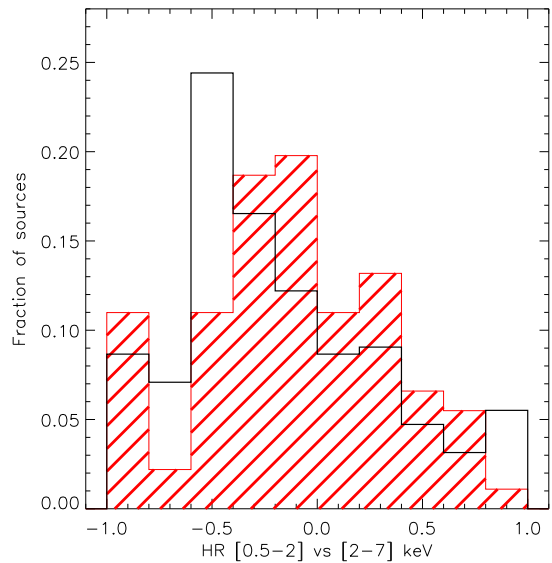}}
\end{center}
   \caption{The distribution of hardness ratios (HR) for the optically
   faint sample (hatched histogram) and the optically bright sample
   (open histogram).}
\label{figure:fig_4} 
\end{figure}

In Figure \ref{figure:fig_4} we show the hardness ratio distributions.
A larger fraction of OFS show a high HR, typical of intrinsically
absorbed AGN emission, than found in the optically bright
sample.\footnote{The HR is a reasonable indicator of the intrinsic
  column density of an AGN only for $z \lesssim 1.5$. At higher
  redshifts an absorbed AGN can have a lower HR value due to the
  higher rest-frame energies probed in the observed X-ray band.} The
two HR distributions are distinguishable according to the
Kolmogorov-Smirnov test at the $98 \%$ significance level; a similar
result has been obtained by Alexander et al.  (\cite{alexander01}) in
the CDF-N. These findings are also in agreement with the general trend
toward flatter X-ray spectral slopes at fainter X-ray fluxes (i.e.
Tozzi et al.  \cite{tozzi01}). It has been shown that this flattening
is due to the fainter X-ray sources having more absorbed X-ray spectra
(see e.g. Mainieri et al.  \cite{mainieri02}; Kim et al.
\cite{kim04}).\\ We conclude that there is a larger fraction of OFS
with flat X-ray spectral slopes. To confirm that this flattening is
due to high intrinsic absorption we fit the X-ray spectra of each
source. We discuss this in more detail in \S\ref{section:xspec}.

\section{Photometric redshift technique applied to an X-ray selected sample}
\label{section:photo_z}

Given the optically faintness of the OFS, the only viable way to
determine their redshifts is to use photometric redshift techniques.
In recent years, these techniques have achieved good accuracy,
particularly when high quality multiwavelength imaging data is
available; e.g. $\sigma _{z} = 0.06(1+\rm{z})$ for the {\it Hubble}
Deep Field North (Fern\'andez-Soto et al. \cite{fernandez99}; Benitez
\cite{benitez00}; Furusawa et al. \cite{furusawa00}). These procedures
rely on detecting the passage of continuum features within the
spectral energy distribution (SED) of sources across a series of
photometric passbands (e.g., the ${\rm 4000 \AA}$ break).\\
Recently these techniques have also been applied to the optical
counterparts of X-ray sources. For these sources the contribution to
the optical/near-IR emission from the AGN nucleus can be significant.
Gonzalez \& Maccarone (\cite{gonzalez02}) studied a sample of 65
sources detected by {\it Chandra} in the {\it Hubble} Deep Field North
and flanking fields. By using a set of galaxy templates, and excluding
objects dominated by the emission from the QSO, they were able to
obtain photometric redshifts to an accuracy similar to that achieved
for non-active galaxies. Mobasher et al. (\cite{mobasher04}) used the
wide multiwavelength photometry from the Great Observatories Origins
Deep Survey (GOODS)\footnote{see: http://www.stsci.edu/science/goods/}
to derive photometric redshifts for a sample of 19 AGN with
spectroscopic identification. They found an {\it rms} scatter of
$\approx 0.13$, good enough to be useful for many science
applications. Finally, Zheng et al. (\cite{zheng04}), using the GOODS
photometry, estimated photometric redshifts for the full sample of 346
X-ray sources detected in the CDF-S (Giacconi et al.
\cite{giacconi02}). By comparison with known spectroscopic redshifts
(137 sources from Szokoly et al. \cite{szokoly04}) we derived an
average dispersion $\Delta z/(1+z)$ $\approx 0.08$.
  
  In Zheng et al. (\cite{zheng04}), we presented photometric redshifts
  for the full sample of X-ray sources in the CDF-S. In this work, we
  fully describe our methodology developed to produce the best results
  for the OFS. By determining likelihood contours in the {\it
    redshift} vs {\it template} plane, our approach clearly elucidate
  degeneracies/dependencies between redshift and SED/reddening which
  often affect the photometric redshift determination.

\subsection{Multicolour catalogue}
\label{section:catalogue}

A key ingredient for deriving reliable photometric redshifts is broad,
multi-wavelength coverage and accurate photometry. An area of $\approx
160$ arcmin$^2$ of GOODS/CDF-S has been imaged with the Advance Camera
for Surveys (ACS) on board of the Hubble Space Telescope (HST) in the
F435W, F606W, F775W and F850LP bands (Giavalisco et al.
\cite{giavalisco03}). A large program with the VLT is under way to
image the GOODS area in the $J,H,K_s$ bands, using some 32 ISAAC
fields to mosaic a 150 arcmin$^2$ region (Vandame et al. in
preparation)\footnote{see:
  http://www.eso.org/science/goods/releases/20040430/}; in addition,
an extensive spectroscopic campaign has been completed with FORS2 at
ESO (Vanzella et al. \cite{vanzella04}) and additional spectroscopy
with VIMOS is planned in 2005.\footnote{see:
  http://www.eso.org/science/goods/} The CDF-S has also been selected
as one of the target fields in the Spitzer legacy program GOODS
(Dickinson et al. \cite{dickinson02}). For an overview of the
available data in this field we refer the reader to Giavalisco et al.
(\cite{giavalisco03}).\\ We will restrict our photometric redshifts
estimation only to the region of the CDF-S covered by deep HST/ACS
imaging (see Figure \ref{figure:fig_2}) and we will refer to it as the
{\it ``GOODS area''}. In this region there are 192 X-ray sources, 112
of which have a spectroscopic redshift (Szokoly et al.
\cite{szokoly04}). For this sample we have built a multicolour
catalogue in B$_{435}$, V$_{606}$, i$_{775}$, z$_{850}$, J, H, K$_{s}$
bands.  For the four optical bands we refer to the publicly available
GOODS catalogue\footnote{http://www.stsci.edu/science/goods/}, while
in the near-IR we use the deep VLT/ISAAC observations (Vandame et al.
in preparation), when available, or the shallower NTT/SOFI imaging
(Vandame et al. \cite{vandame01}).

 \begin{figure*}

   \parbox{12.3cm}{\resizebox{\hsize}{!}{\includegraphics[angle=0]{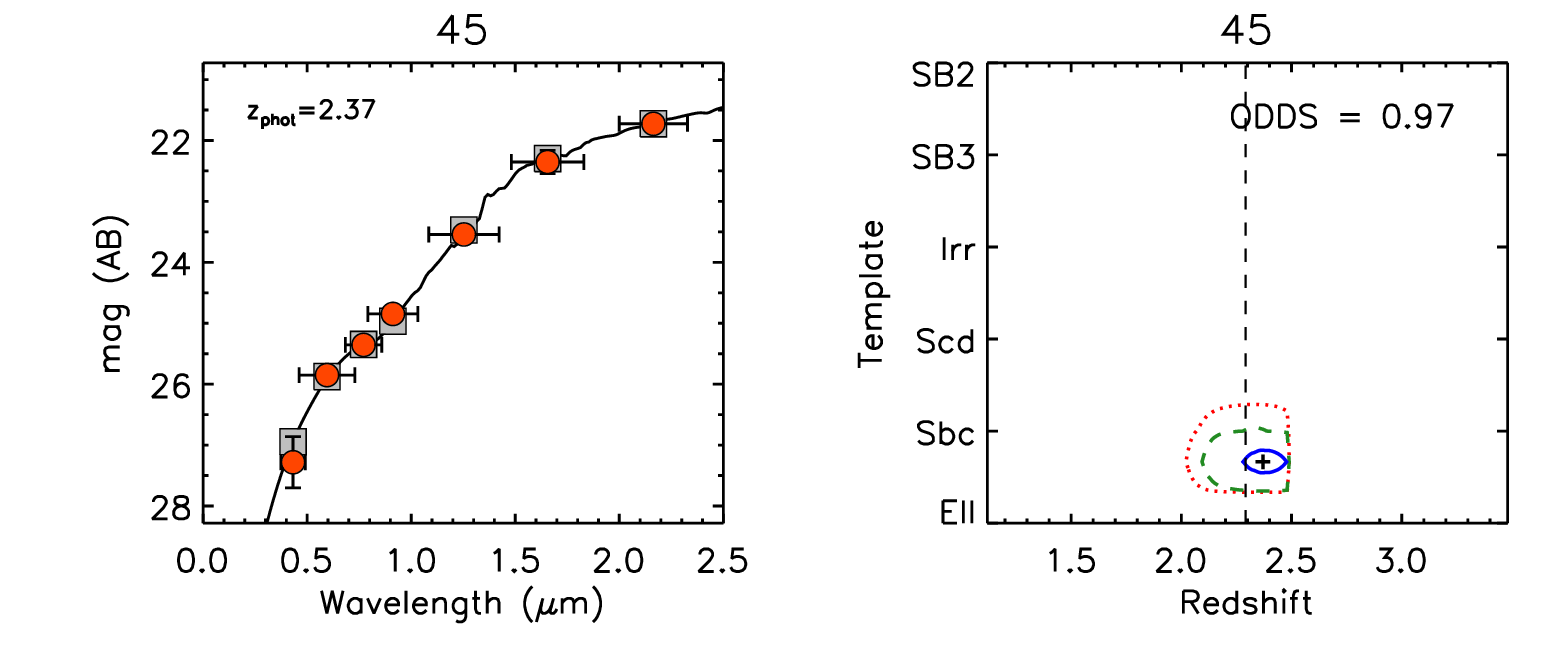}}}
   \\
   \parbox{12.3cm}{\resizebox{\hsize}{!}{\includegraphics[angle=0]{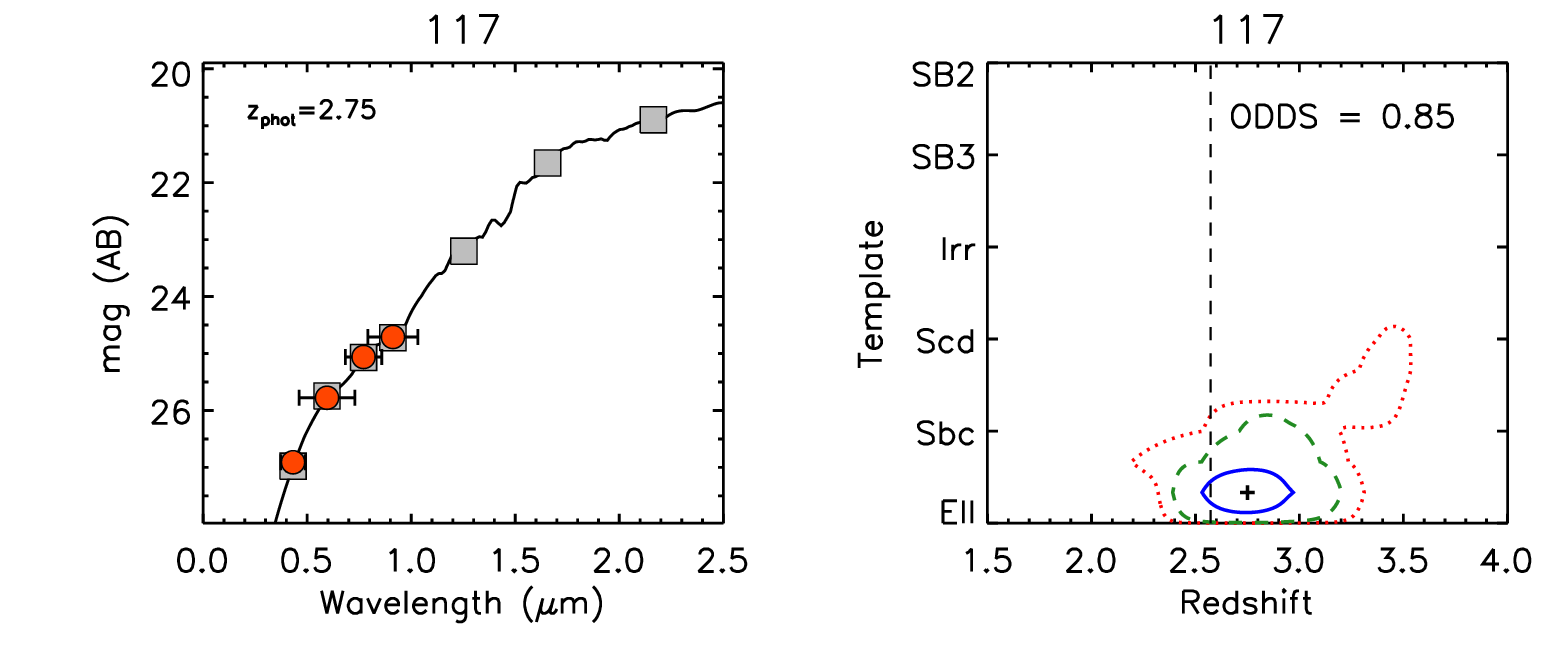}}}
   \\
   \parbox{12.3cm}{\resizebox{\hsize}{!}{\includegraphics[angle=0]{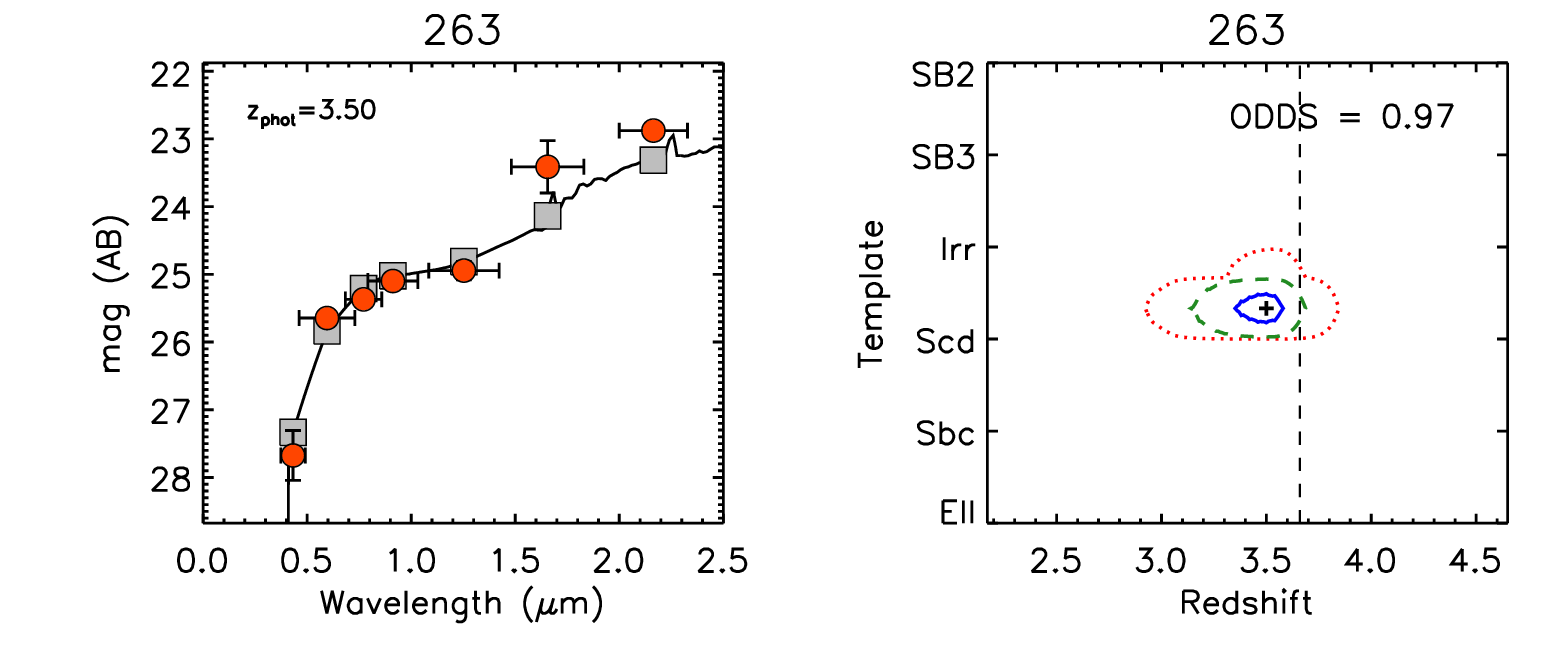}}}
   \\ \caption{ Comparison of the photometric and spectroscopic
     redshifts of the spectroscopically identified OFS in the ``GOODS
     area''. {\it Left side}: best fit template (continuum line) with
     the observed photometry (filled circles with error bars) and the
     best fit model photometry (filled squares) overplotted. {\it
       Right side}: $1 \sigma$ (dotted line), $2 \sigma$ (dashed line)
     and $3 \sigma$ (continuum line) confidence contours for the
     photometric redshift determination in the SED template {\it vs}
     redshift plane. The dashed line indicates the spectroscopic
     redshift and the cross shows the best fit solution.  The shaded
     area refers to a low-probability region in the solution space
     (see Sec. \ref{section:bpz}). XID are from Giacconi et al.
     (2002).  }
\label{figure:fig_cont_spec_1} 
\end{figure*}

 \begin{figure*} 
   \parbox{12.3cm}{\resizebox{\hsize}{!}{\includegraphics[angle=0]{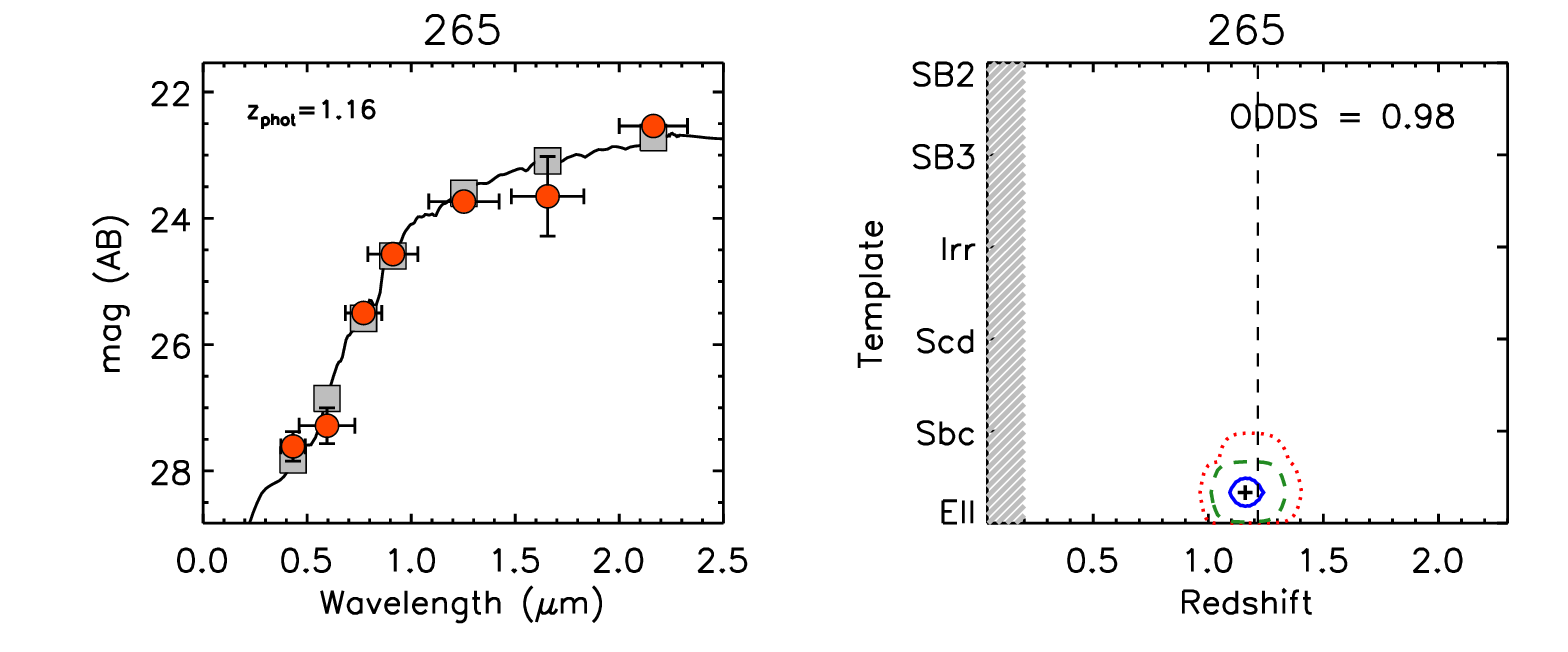}}}
\\

 \parbox{12.3cm}{\resizebox{\hsize}{!}{\includegraphics[angle=0]{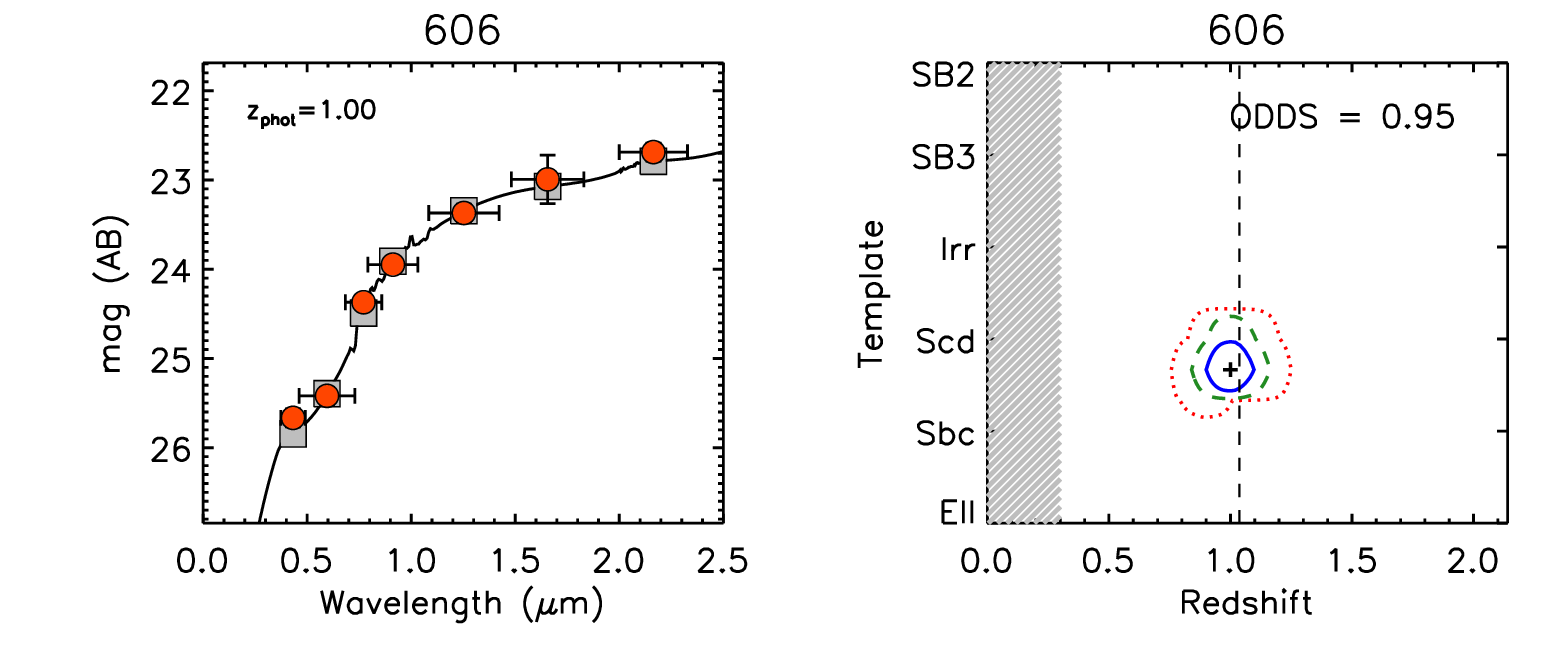}}}
 \\ \caption{As in Figure \ref{figure:fig_cont_spec_1}}
\label{figure:fig_cont_spec_2} 
\end{figure*}

We used SExtractor (Bertin \& Arnouts, \cite{bertin96}) for source
detection in each band. To obtain reliable source colours the spread
in seeing conditions for images in different wavebands has been taken
into account. We obtained PSF-matched magnitudes as follows: 1) we
computed aperture magnitudes in each waveband using the available
images ("original" images); 2) we degraded the point spread function
(PSF) of each image to match the worst condition ("degraded" images);
3) we recomputed the aperture magnitudes using the "degraded" images;
4) we derived corrections for the different seeing conditions by
comparing the magnitudes of ``bright'' stars in the ``original'' and
in the ``degraded'' images. These corrections ( $\lesssim 0.1$ mag)
are applied to the "original" magnitudes. We checked that those
corrections are constant over a large magnitude range. We also
corrected for Galactic extinction: the values in the different filters
have been obtained from the NASA/IPAC Extragalactic Database which are
taken from Schlegel et al.  (\cite{schlegel98}). The corrections in
the CDF-S region are small (from $\sim0.03$ mag in the B band to
$\sim0.003$ mag in the K$_{\rm s}$ band).

In the following, we adopt an X-ray based classification as suggested
by Szokoly et al.  (\cite{szokoly04}), but using unabsorbed
luminosities and intrinsic absorption as opposed to uncorrected
luminosities and hardness ratio (see Tozzi et al.  \cite{tozzi05}).
We introduce the following classes: {\it X-ray unabsorbed QSO}:
L$_{\rm X} [0.5-10] \ge 10^{44}$ erg/s and N$_{\rm H} < 10^{22}$
cm$^{-2}$; {\it X-ray unabsorbed AGN}: $10^{42} \le {\rm L_X [0.5-10]}
< 10^{44}$ erg/s and N$_{\rm H} < 10^{22}$ cm$^{-2}$; {\it X-ray
  absorbed QSO}: L$_{\rm X} [0.5-10] \ge 10^{44}$ erg/s and N$_{H} \ge
10^{22}$ cm$^{-2}$; {\it X-ray absorbed AGN}: $10^{42} \le {\rm L_X}
[0.5-10] < 10^{44}$ erg/s and N$_{\rm H} \ge 10^{22}$ cm$^{-2}$; {\it
  galaxy}: L$_{\rm X} [0.5-10] < 10^{42}$ erg/s and N$_{\rm H} <
10^{22}$ cm$^{-2}$.

\begin{table}
\caption{\footnotesize{Photometric redshifts for Optically Faint Sources in CDF-S/GOODS}}
\begin{center}
\begin{tabular}{ccccccccccl}
\hline
\hline
\small{XID$^{a}$} & \small{R mag} & \small{z$_{spec}$} & \small{z$_{phot}$} & \small{1 $\sigma$$^b$} & \small{template$^c$} & \small{odds$^d$} & \small{X/O$^e$} & \small{N$_H$$^f$} & \small{L$_{\rm X}$$^g$} & \small{Class X$^h$} \\
 & \scriptsize{(Vega)} & & & & & & & & \scriptsize{$([0.5-10]$ keV)} & \\
\hline
\noalign{\smallskip}
\small{45} & \small{25.4} & \small{2.291} & \small{2.37} & \small{2.23-2.43} & \small{1.67} & \small{0.97} & \small{23.8} & \small{22.91$^{23.05}_{22.74}$} & \small{44.18} & \small{QSO abs} \\
\small{58} & \small{26.0} & \small{...} & \small{0.92} & \small{0.75-1.07} & \small{2.00} & \small{0.60} & \small{19.2} & \small{22.40$^{22.49}_{22.31}$} & \small{43.16} & \small{AGN abs} \\
\small{79} & \small{26.5} & \small{...} & \small{2.00} & \small{1.91-2.07} & \small{1.00} & \small{0.98} & \small{36.4} & \small{$<$21.75} & \small{43.67} & \small{AGN unabs} \\
\small{79A$^{\ast}$} & \small{26.5} & \small{...} & \small{1.82} & \small{1.76-1.86} & \small{1.33} & \small{1.00} & \small{36.4} & \small{$<$21.75} & \small{43.67} & \small{AGN unabs} \\
\small{81} & \small{26.0} & \small{...} & \small{2.59} & \small{2.47-2.59} & \small{2.33} & \small{0.95} & \small{10.2} & \small{22.63$^{22.86}_{22.26}$} & \small{43.81} & \small{AGN abs} \\
\small{82} & \small{25.9} & \small{...} & \small{1.89} & \small{1.79-1.97} & \small{1.33} & \small{0.97} & \small{12.1} & \small{23.47$^{23.52}_{23.42}$} & \small{43.79} & \small{AGN abs} \\
\small{108} & \small{25.9} & \small{...} & \small{1.56} & \small{1.48-1.66} & \small{2.33} & \small{0.96} & \small{8.4} & \small{21.79$^{22.28}_{19.90}$} & \small{43.15} & \small{AGN unabs} \\
\small{117} & \small{25.5} & \small{2.573} & \small{2.75} & \small{2.50-3.00} & \small{1.33} & \small{0.85} & \small{8.5} & \small{22.49$^{22.70}_{22.16}$} & \small{43.99} & \small{AGN abs} \\
\small{124} & \small{25.3} & \small{...} & \small{0.61} & \small{0.45-0.77} & \small{3.00} & \small{0.53} & \small{0.3} & \small{$<$21.62} & \small{41.75} & \small{galaxy} \\
\small{133} & \small{99.0} & \small{...} & \small{1.21} & \small{0.78-4.08} & \small{2.00} & \small{0.11} & \small{21.9} & \small{22.75$^{22.85}_{22.66}$} & \small{43.32} & \small{AGN abs} \\
\small{146} & \small{99.0} & \small{...} & \small{2.67} & \small{2.55-2.66} & \small{1.67} & \small{0.97} & \small{42.3} & \small{$>$24.18} & \small{45.11} & \small{QSO abs} \\
\small{147} & \small{25.1} & \small{...} & \small{0.99} & \small{0.89-1.10} & \small{2.00} & \small{0.87} & \small{16.3} & \small{23.39$^{23.45}_{23.33}$} & \small{43.67} & \small{AGN abs} \\
\small{148} & \small{27.3} & \small{...} & \small{1.74} & \small{1.62-1.88} & \small{1.67} & \small{0.87} & \small{69.9} & \small{$>$24.18} & \small{45.09} & \small{QSO abs} \\
\small{159} & \small{99.0} & \small{...} & \small{3.30} & \small{3.17-3.46} & \small{3.67} & \small{0.98} & \small{142.8} & \small{23.00$^{23.11}_{22.88}$} & \small{44.81} & \small{QSO abs} \\
\small{178} & \small{26.0} & \small{...} & \small{0.29} & \small{0.23-3.54} & \small{3.67} & \small{0.32} & \small{0.7} & \small{21.68$^{22.02}_{20.52}$} & \small{41.00} & \small{galaxy} \\
\small{210} & \small{99.0} & \small{...} & \small{1.73} & \small{1.60-1.95} & \small{1.33} & \small{0.74} & \small{18.1} & \small{22.32$^{22.56}_{21.87}$} & \small{43.36} & \small{AGN abs} \\
\small{217} & \small{25.8} & \small{...} & \small{7.72} & \small{1.08-7.62} & \small{5.00} & \small{0.30} & \small{0.8} & \small{23.06$^{23.31}_{22.61}$} & \small{43.80} & \small{AGN abs} \\
\small{221} & \small{99.0} & \small{...} & \small{2.51} & \small{2.34-2.62} & \small{1.00} & \small{0.90} & \small{1.6} & \small{22.58$^{23.00}_{21.08}$} & \small{43.32} & \small{AGN abs} \\
\small{226} & \small{99.0} & \small{...} & \small{1.45} & \small{1.14-1.87} & \small{5.67} & \small{0.39} & \small{24.4} & \small{22.17$^{22.35}_{21.88}$} & \small{43.35} & \small{AGN abs} \\
\small{227} & \small{99.0} & \small{...} & \small{2.18} & \small{1.98-2.36} & \small{1.33} & \small{0.77} & \small{33.8} & \small{23.83$^{23.91}_{23.74}$} & \small{44.24} & \small{QSO abs} \\
\small{240} & \small{25.0} & \small{...} & \small{1.41} & \small{1.35-1.53} & \small{1.33} & \small{0.95} & \small{4.4} & \small{22.42$^{22.62}_{22.22}$} & \small{43.13} & \small{AGN abs} \\
\small{243} & \small{99.0} & \small{...} & \small{7.89} & \small{7.56-7.93} & \small{4.00} & \small{0.99} & \small{40.2} & \small{23.26$^{23.36}_{23.16}$} & \small{44.11} & \small{QSO abs} \\
\small{263} & \small{25.3} & \small{3.660} & \small{3.50} & \small{3.32-3.56} & \small{3.33} & \small{0.97} & \small{5.5} & \small{$>$24.18} & \small{44.54} & \small{QSO abs} \\
\small{265} & \small{99.0} & \small{1.215} & \small{1.16} & \small{1.09-1.24} & \small{1.33} & \small{0.98} & \small{51.2} & \small{23.20$^{23.27}_{23.13}$} & \small{43.70} & \small{AGN abs} \\
\small{508} & \small{99.0} & \small{...} & \small{2.50} & \small{1.90-2.85} & \small{1.00} & \small{0.58} & \small{18.8} & \small{23.88$^{23.97}_{23.79}$} & \small{44.15} & \small{QSO abs} \\
\noalign{\smallskip}
\hline
\end{tabular}
\end{center}
\scriptsize{$^a$ XID from Giacconi et al. (\cite{giacconi02}).} \\
\scriptsize{$^b$ 1 $\sigma$ confidence range.} \\
\scriptsize{$^c$ Best fit template. BPZ performs two interpolations between each one of the following templates:\\ \hspace*{1cm}1: Elliptical (CWW)\\ \hspace*{1cm}2: Sbc (CWW)\\ \hspace*{1cm}3: Scd (CWW)\\ \hspace*{1cm}4: Irregular (CWW)\\ \hspace*{1cm}5: SB3 (Kinney)\\ \hspace*{1cm}6: SB2 (Kinney).} \\
\scriptsize{$^d$ Odds parameter which gives a metric on the reliability of the photometric redshift determination (see Sec. 4.2).} \\
\scriptsize{$^e$ X-ray-to-optical flux ratios.} \\
\scriptsize{$^f$ Logaritm of the column density.} \\
\scriptsize{$^g$ Logaritm of the unabsorbed X-ray luminosity in the [0.5-10] keV band. The uncertainties are smaller than 0.01 in the adopted units.} \\
\scriptsize{$^h$ X-ray classification (see Sec. 4.1).} \\
\scriptsize{$^{\ast}$ Photometric redshift obtained using the UDF photometry (see Sec. \ref{section:udf} and Fig. \ref{figure:fig_UDF_cutouts},\ref{figure:fig_cont_UDF}).} \\
\label{table:zphot_optf_1}
\end{table}

\begin{table}
\caption{\footnotesize{Photometric redshifts for Optically Faint Sources in CDF-S/GOODS}}
\begin{center}
\begin{tabular}{ccccccccccl}
\hline
\hline
\small{XID$^{a}$} & \small{R mag} & \small{z$_{spec}$} & \small{z$_{phot}$} & \small{1 $\sigma$$^b$} & \small{template$^c$} & \small{odds$^d$} & \small{X/O$^e$} & \small{N$_H$$^f$} & \small{L$_{\rm X}$$^g$} & \small{Class X$^h$} \\
 & \scriptsize{(Vega)} & & & & & & & & \scriptsize{$([0.5-10]$ keV)} & \\
\hline
\noalign{\smallskip}
\small{510} & \small{25.4} & \small{...} & \small{2.51} & \small{2.34-2.59} & \small{2.33} & \small{0.93} & \small{5.1} & \small{23.44$^{23.62}_{23.24}$} & \small{43.79} & \small{AGN abs} \\
\small{513} & \small{26.1} & \small{...} & \small{3.56} & \small{3.41-3.68} & \small{1.00} & \small{0.96} & \small{7.6} & \small{24.16$^{24.31}_{24.06}$} & \small{44.19} & \small{QSO abs} \\
\small{515} & \small{99.0} & \small{...} & \small{2.26} & \small{2.13-2.46} & \small{1.67} & \small{0.84} & \small{19.7} & \small{23.50$^{23.67}_{23.32}$} & \small{43.75} & \small{AGN abs} \\
\small{515$^{\ast}$} & \small{99.0} & \small{...} & \small{2.30} & \small{2.24-2.38} & \small{2.00} & \small{1.00} & \small{19.7} & \small{23.50$^{23.67}_{23.32}$} & \small{43.75} & \small{AGN abs} \\
\small{518} & \small{99.0} & \small{...} & \small{0.84} & \small{0.58-1.37} & \small{2.00} & \small{0.33} & \small{9.5} & \small{21.91$^{22.31}_{20.68}$} & \small{42.23} & \small{AGN unabs} \\
\small{523} & \small{99.0} & \small{...} & \small{1.32} & \small{0.87-6.22} & \small{2.00} & \small{0.10} & \small{11.3} & \small{22.98$^{23.11}_{22.86}$} & \small{43.08} & \small{AGN abs} \\
\small{524} & \small{99.0} & \small{...} & \small{2.36} & \small{2.17-2.58} & \small{1.67} & \small{0.75} & \small{18.5} & \small{23.39$^{23.51}_{23.23}$} & \small{43.84} & \small{AGN abs} \\
\small{537} & \small{99.0} & \small{...} & \small{1.54} & \small{1.31-1.63} & \small{6.00} & \small{0.65} & \small{9.0} & \small{22.62$^{22.96}_{22.21}$} & \small{42.89} & \small{AGN abs} \\
\small{555} & \small{25.4} & \small{...} & \small{2.28} & \small{2.11-2.46} & \small{3.00} & \small{0.79} & \small{0.3} & \small{$<$22.68} & \small{42.83} & \small{AGN unabs} \\
\small{557} & \small{25.4} & \small{...} & \small{1.81} & \small{1.75-1.96} & \small{1.33} & \small{0.92} & \small{0.3} & \small{22.66$^{23.00}_{22.39}$} & \small{42.88} & \small{AGN abs} \\
\small{561} & \small{99.0} & \small{...} & \small{0.62} & \small{0.50-0.87} & \small{2.33} & \small{0.47} & \small{1.6} & \small{$<$21.38} & \small{41.82} & \small{galaxy} \\
\small{564} & \small{99.0} & \small{...} & \small{0.43} & \small{0.23-0.54} & \small{4.33} & \small{0.51} & \small{1.4} & \small{21.79$^{22.13}_{21.10}$} & \small{41.53} & \small{galaxy} \\
\small{572} & \small{27.0} & \small{...} & \small{2.73} & \small{2.21-2.68} & \small{2.00} & \small{0.53} & \small{1.8} & \small{23.43$^{23.68}_{23.17}$} & \small{43.62} & \small{AGN abs} \\
\small{583} & \small{99.0} & \small{...} & \small{2.77} & \small{2.66-2.88} & \small{1.00} & \small{1.00} & \small{2.2} & \small{23.32$^{23.49}_{23.11}$} & \small{43.77} & \small{AGN abs} \\
\small{589} & \small{99.0} & \small{...} & \small{1.33} & \small{0.89-6.51} & \small{2.00} & \small{0.10} & \small{1.2} & \small{22.90$^{23.11}_{22.57}$} & \small{42.81} & \small{AGN abs} \\
\small{593} & \small{25.9} & \small{...} & \small{2.07} & \small{1.83-2.15} & \small{2.67} & \small{0.77} & \small{0.6} & \small{23.32$^{23.61}_{22.95}$} & \small{43.31} & \small{AGN abs} \\
\small{599} & \small{25.2} & \small{...} & \small{2.84} & \small{2.48-2.82} & \small{1.33} & \small{0.76} & \small{1.9} & \small{24.05$^{24.17}_{23.92}$} & \small{43.99} & \small{AGN abs} \\
\small{605} & \small{99.0} & \small{...} & \small{4.71} & \small{4.39-4.83} & \small{5.00} & \small{0.85} & \small{8.2} & \small{24.70$^{24.93}_{24.49}$} & \small{44.52} & \small{QSO abs} \\
\small{605A$^{\ast}$} & \small{99.0} & \small{...} & \small{4.29} & \small{4.21-4.32} & \small{3.67} & \small{0.99} & \small{8.2} & \small{24.70$^{24.93}_{24.49}$} & \small{44.52} & \small{QSO abs} \\
\small{606} & \small{25.3} & \small{1.037} & \small{1.00} & \small{0.92-1.06} & \small{2.67} & \small{0.95} & \small{4.8} & \small{23.27$^{23.40}_{23.14}$} & \small{43.16} & \small{AGN abs} \\
\small{610} & \small{99.0} & \small{...} & \small{2.04} & \small{1.94-2.17} & \small{1.00} & \small{0.90} & \small{8.7} & \small{$>$24.18} & \small{43.87} & \small{AGN abs} \\
\small{614} & \small{99.0} & \small{...} & \small{1.13} & \small{0.60-1.57} & \small{4.33} & \small{0.27} & \small{19.8} & \small{21.30$^{21.93}_{20.62}$} & \small{42.67} & \small{AGN unabs} \\
\small{618} & \small{25.6} & \small{...} & \small{4.66} & \small{4.59-4.72} & \small{6.00} & \small{1.00} & \small{0.8} & \small{23.86$^{24.10}_{23.75}$} & \small{44.30} & \small{QSO abs} \\
\small{626} & \small{25.4} & \small{...} & \small{0.59} & \small{0.55-0.62} & \small{6.00} & \small{1.00} & \small{0.4} & \small{22.45$^{22.89}_{21.25}$} & \small{42.94} & \small{AGN abs} \\
\small{628} & \small{99.0} & \small{...} & \small{2.07} & \small{2.01-2.17} & \small{2.00} & \small{0.99} & \small{1.2} & \small{23.93$^{23.95}_{23.93}$} & \small{43.68} & \small{AGN abs} \\
\small{629} & \small{25.3} & \small{...} & \small{0.56} & \small{0.49-0.61} & \small{6.00} & \small{0.97} & \small{0.3} & \small{$<$22.37} & \small{41.39} & \small{galaxy} \\
\noalign{\smallskip}
\hline
\end{tabular}
\end{center}
\label{table:zphot_optf_2}
\end{table}

\subsection{Photometric redshifts}
\label{section:bpz}

Using the extraordinarily deep and wide photometric coverage in the
``GOODS area'', we have computed photometric redshifts for the OFS.
This allows us to estimate the redshift distribution of OFS to a
greater precision than has been performed before. For example,
Alexander et al. (\cite{alexander01}) assumed that most OFS reside in
$\sim$L$^{\star}$ host galaxies to infer that the majority of the
population is in the redshift range of z$\approx 1-3$.\\ We used the
publicly available code BPZ (Benitez \cite{benitez00})\footnote{BPZ is
  available from http://acs.pha.jhu.edu/$\sim$txitxo/}. This code
combines $\chi^2$ minimization and Bayesian marginalization, using
prior probabilities to include {\it a priori} knowledge of the
distribution of galaxy magnitudes and spectral types with redshift. We
used the default library of spectral templates in BPZ: four ( E, Sbc,
Scd, Irr) are spectral energy distributions from Coleman, Wu \&
Weedman (\cite{coleman80}) and two are derived from spectra of
starburst galaxies in Kinney at al. (\cite{kinney96}). We allowed the
code to calculate two interpolated SEDs between each pair of these
templates.\\ BPZ provides the ``odds'' parameter to characterize the
accuracy of the redshift estimation. This parameter is defined as the
integral of the redshift probability distribution within the interval
$|z-z_b|< 2 \times \sigma (z)$, where $z_b$ is the value which
maximizes the probability distribution and $\sigma (z)$ the observed
$rms$. A low value ($<0.6$) of the odds parameter is a warning that
the probability distribution is spread over a large redshift range or
is double-peaked.

\begin{figure}  
 \begin{center}
\resizebox{8cm}{!}{\includegraphics{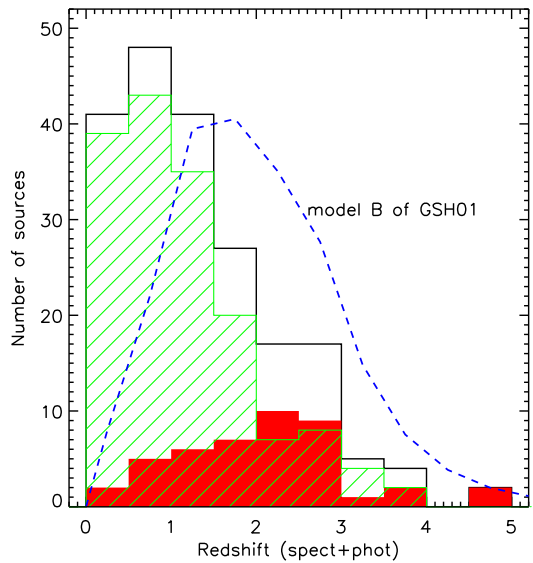}}
\end{center}
   \caption{Distribution of redshifts (spectroscopic and photometric)
     for the OFS (shaded histogram), optically bright sources (hatched
     histogram) and the total X-ray sample (open histogram). Sources
     belonging to the large scale structures in the CDF-S have been
     excluded (see text). The dashed line is the redshift distribution
     predicted by the model B of Gilli, Salvati and Hasinger
     (\cite{gilli01}) normalized to the total number of sources in the
     ``GOODS area'' for which we have either a spectroscopic or
     photometric redshift.}
\label{figure:fig_zdistr} 
\end{figure}

The reliability of the OFS photometric redshifts was estimated using
the five spectroscopically identified OFS in the ``GOODS area'':
CDF-S/XID 45, 117, 263, 265 and 606 (see Table \ref{table:optfaintS}
and Figure \ref{figure:fig_cont_spec_1}-\ref{figure:fig_cont_spec_2}).
In the left column the best fit template is plotted together with the
observed photometry. On the right column, likelihood contours
($1\sigma$, $2\sigma$, $3\sigma$) are reported in the {\it SED
  template vs redshift} plane. In all of the five cases the agreement
between the photometric and spectroscopic redshift is extremely good,
even for source CDF-S/XID 117 for which we do not have reliable
near-IR photometry. Therefore we are confident that we can apply this
procedure to the whole of OFS class. In Table \ref{table:zphot_optf_1}
and \ref{table:zphot_optf_2} we report the derived photometric
redshifts and the 1 $\sigma$ confidence range.  Best fits and
confidence contours for each source are shown in Figure
\ref{figure:fig_cont_1}. \\ We have also used X-ray information to set
a posteriori constraints in the redshift solution space. In a plot of
the Hardness Ratio {\it versus} the X-ray luminosity of the sources in
the CDF-S (see Figure 10 of Szokoly et al. \cite{szokoly04}) almost
all of the sources spectroscopically identified as AGN have: HR$>
-0.6$ and L$_{\rm X}[0.5-10 ~ keV]> 10^{41.5}$ erg s$^{-1}$.
Consequently for all the objects with HR$> -0.6$ we have imposed a
minimum X-ray luminosity limit (L$^{\rm min}_{\rm X}[0.5-10 ~ keV]=
10^{41.5}$ erg s$^{-1}$) that converts to a minimum redshift for each
source (z$_{\rm min}$). In the confidence contour plots the shaded
area corresponds to z$<{\rm z_{min}}$, where the probability to have
the correct solution is low ( see Figure \ref{figure:fig_cont_spec_1}
and Figure \ref{figure:fig_cont_1}).\footnote{$99 \%$ of the X-ray
  sources in the CDF-S with spectroscopic redshifts and HR$>
  -0.6$ have L$_{\rm X}[0.5-10 ~ keV]>{\rm L}^{\rm min}_{\rm X}$} \\
We compare in Figure \ref{figure:fig_zdistr} the redshift distribution
of optically bright sources and OFS. We use the spectroscopic
redshift, if known, or the derived photometric redshift, if the source
is still unidentified. The solid histogram shows the redshift
distribution of the OFS, the hatched histogram shows the distribution
of optically bright sources, and the open histogram shows the
distribution of the whole X-ray sample. We have excluded sources
belonging to the two large scale structures at z $= 0.67$ and 0.73
discovered by Gilli et al.  (\cite{gilli03a}) in the CDF-S. The
uncertainties in the photometric redshifts are too large to determine
whether a source belongs to these structures. Since $82 \%$ of the
spectroscopically identified sources with $0.6 \leq$ z$_{\rm spec}
\leq 0.8$ belong to the redshift spikes, we assumed that a similar
fraction of sources with $0.6 \leq$ z$_{\rm phot} \leq 0.8$ are part
of the same structures.

In Figure \ref{figure:fig_zdistr}, we have overplotted for comparison
the prediction of a synthesis model for the XRB [model B from Gilli,
Salvati \& Hasinger (\cite{gilli01})].

The majority of the OFS lie at z$=1-3$, with a small fraction at
z$>3$, as previously predicted by Alexander et al.
(\cite{alexander01}). Perhaps surprisingly, a small fraction of OFS
lie at z$<1$. We note that a larger fraction ($76\%$) of OFS are at
z$>1$ than found for the optically bright sources ($49\%$).  According
to a K-S test the probability that these two distributions are drawn
from the same population is extremely small ( $\sim 0.00002 \%$).
These have to be taken into account when comparing the redshift
distributions of recent deep {\it Chandra} and XMM-{\it Newton}
surveys with XRB synthesis models predictions. Almost all of the
spectroscopically identified sources in the two deepest {\it Chandra}
pointings are optically bright: $\sim 97 \%$ both in the CDF-S and
CDF-N. It has been noted (Hasinger \cite{hasinger02}; Gilli
\cite{gilli03b}) that the redshift distribution of these new surveys
is in disagreement with XRB models predictions based on the ROSAT
X-ray luminosity function. These models predict that the distribution
peaks at z$\sim 1.3 - 1.5$, whereas the observed N(z) of the sources
identified to date peaks at z$\lesssim 1$. Since OFS appear to have a
N(z) peaking at z$\sim 1.7$ ( Figure \ref{figure:fig_zdistr}) and they
make up the majority of still unidentified sources, the disagreement
with the models predictions is attenuated. Nevertheless, a significant
discrepancy with the models remains even with the addition of
photometric redshifts (Figure \ref{figure:fig_zdistr}). A solution for
this problem requires a new determination of the X-ray luminosity
function of AGN ( one of the main input parameters of XRB synthesis
models), particularly exploring the X-ray fainter regime not covered
by previous surveys (Gilli \cite{gilli03b}). A similar result has been
recently found by Fiore et al. (\cite{fiore03}) using a different
approach to derive the redshift information for unidentified sources
in the HELLAS2XMM 1dF sample.

Finally, a significant fraction (46 out of 346, $\sim 13 \%$) of the
X-ray sources in the 1 Msec {\it Chandra} exposure (Giacconi et al.
\cite{giacconi02}), are not detected in deep VLT optical images down
to R$\lesssim 26.1 - 26.7$. Yan et al. (\cite{yan03}) identified in
the near-IR six of these objects using the first release of deep
VLT/ISAAC JHK$_{\rm s}$ data, which covered an area 2.5 times smaller
than the new extended ISAAC imaging shown in Figure
\ref{figure:fig_1}. Using optical/near-IR colour-colour diagrams the
authors concluded that they were likely to be E/S0 galaxies at $0 \leq
z \leq 3.5$. Taking advantage of the deep optical ACS photometry we
can now set tighter constraints on the redshift of these six sources:
CDF-S/XID 201 has a spectroscopic redshift of z$=0.679$ (Szokoly et
al.  \cite{szokoly04})\footnote{This source is not reported in Table
  \ref{table:optfaintS} and \ref{table:zphot_optf_1} because, as shown
  in Szokoly et al.  (\cite{szokoly04}), an optical counterpart (201b)
  with R$=24.34$ has been identified inside the {\it Chandra} error
  circle.}; the best fit SED of CDF-S/XID 79 and CDF-S/XID 221 is an
unreddened early-type galaxy with redshift $1.82_{1.76}^{1.86}$ and
$2.51_{2.34}^{2.62}$ respectively and the remaining three sources
(XID/CDFS 515,561,593) are best fitted with the template of a spiral
galaxy and have photometric redshifts of $2.30_{2.24}^{2.38}$,
$0.62_{0.50}^{0.87}$ and $2.07_{1.83}^{2.15}$ respectively.

\begin{table}
 \caption{\footnotesize{OFS with a spectroscopic
 redshift in the CDF-S.}}
\begin{center} 
\begin{tabular}{rcccccclccc}
\hline 
\hline 
\scriptsize{XID$^{a}$} & \scriptsize{CXO CDFS$^{b}$} &
 \scriptsize{R$^{c}$} & \scriptsize{R$-$K$^{d}$} &
 \scriptsize{z$^{e}$} & \scriptsize{Q$^{f}$} & \scriptsize{Opt$^{g}$} &
 \scriptsize{X-ray$^{h}$} & \scriptsize{FS$^{i}$}
 &\scriptsize{FH$^{j}$} & \scriptsize{HR$^{k}$}  \\ 
\hline
\noalign{\smallskip}
\scriptsize{45} & \scriptsize{J033225.8$-$274306} & \scriptsize{25.3} & \scriptsize{5.1} & \scriptsize{2.291} & \scriptsize{1} & \scriptsize{LEX} & \scriptsize{QSO abs} &\scriptsize{$1.02\pm0.08$} & \scriptsize{$4.67\pm0.46$} & \scriptsize{$-0.12\pm0.06$}  \\
\scriptsize{54} & \scriptsize{J033214.7$-$275422} & \scriptsize{25.7} & \scriptsize{$<$5.4} & \scriptsize{2.561} & \scriptsize{3} & \scriptsize{HEX} & \scriptsize{QSO abs} & \scriptsize{$0.61\pm0.07$} & \scriptsize{$3.37\pm0.46$} & \scriptsize{$-0.01\pm0.09$}  \\ 
\scriptsize{117} & \scriptsize{J033203.1$-$274450} & \scriptsize{25.5} & \scriptsize{4.9} & \scriptsize{2.573} & \scriptsize{3} & \scriptsize{HEX} & \scriptsize{AGN abs} & \scriptsize{$0.61\pm0.07$} & \scriptsize{$1.19\pm0.34$} & \scriptsize{$-0.52\pm0.11$} \\
\scriptsize{263} & \scriptsize{J033218.9$-$275136} & \scriptsize{25.3} & \scriptsize{5.1} & \scriptsize{3.660} & \scriptsize{3} & \scriptsize{HEX} & \scriptsize{QSO abs} & \scriptsize{$0.10\pm0.04$} & \scriptsize{$1.23\pm0.31$} & \scriptsize{$+0.35\pm0.20$} \\
\scriptsize{265} & \scriptsize{J033233.4$-$274236} & \scriptsize{$>26$} & \scriptsize{$>5.2$} & \scriptsize{1.215$^m$} & \scriptsize{1} & \scriptsize{LEX} & \scriptsize{AGN abs} & \scriptsize{$0.19\pm0.05$} & \scriptsize{$3.33\pm0.45$} & \scriptsize{$+0.50\pm0.11$} \\
\scriptsize{606} & \scriptsize{J033225.0$-$275009} & \scriptsize{25.3} & \scriptsize{4.3} & \scriptsize{1.037} & \scriptsize{1} & \scriptsize{LEX} & \scriptsize{AGN abs} & \scriptsize{$<0.06$} & \scriptsize{$1.25\pm0.30$} & \scriptsize{$+0.68\pm0.19$} \\
\noalign{\smallskip} 
\hline 
\end{tabular} 
\end{center}
\scriptsize{$^a$ XID from Giacconi et al. (\cite{giacconi02}).} \\
\scriptsize{$^b$ IAU registered name, based on original X-ray coordinates.}\\ 
\scriptsize{$^c$ Vega R magnitude for the optical counterpart} \\ 
\scriptsize{$^d$ Vega R-K colour for the optical counterpart}\\ 
\scriptsize{$^e$ Optical spectroscopic redshift for the optical counterpart (Szokoly et al. \cite{szokoly04}).}\\
\scriptsize{$^f$ Quality flag of the optical spectrum (see Szokoly et al. \cite{szokoly04}).}\\
\scriptsize{$^g$ Optical classification (see Szokoly et al. \cite{szokoly04}).} \\ 
\scriptsize{$^h$ X-ray classification.} \\
\scriptsize{$^i$ Flux in soft (0.5-2 keV) band, in units of $10^{-15}$ erg cm$^{-2}$ s$^{-1}$.}\\ 
\scriptsize{$^j$ Flux in hard (2-10 keV) band, in units of $10^{-15}$ erg cm$^{-2}$ s$^{-1}$.}\\
\scriptsize{$^k$ Hardness ratio, defined as (H$-$S)/(H$+$S) where H and S are the net counts in the hard and soft bands, respectively.}\\
\label{table:optfaintS} 
\end{table}

\begin{table}
 \caption{\footnotesize{X-ray spectral parameters of OFS with a
 spectroscopic redshift in the CDF-S.}}
\begin{center} 
\begin{tabular}{rcccccc}
\hline 
\hline 
\small{XID$^{a}$} & \small{CXO CDFS$^{b}$} & \small{counts$^c$} & 
 \small{$\Gamma ^d$} & \small{N$_{\rm H}^{e}$} &
 \small{L$_{\rm X}^{f}$} & \small{cstat$^g$} \\ 
\hline
\noalign{\smallskip}
\small{45} & \small{J033225.8$-$274306} & \small{$302\pm18$}  & \small{1.46$^{1.68}_{1.26}$} & \small{22.91$^{23.05}_{22.74}$} & \small{$1.51 \times 10^{44}$} & \small{150.7}\\
\small{54} & \small{J033214.7$-$275422} & \small{$186\pm17$} & \small{1.38$^{1.72}_{1.10}$} & \small{23.03$^{23.21}_{22.79}$} & \small{$1.18 \times 10^{44}$} & \small{153.8}\\
\small{117} & \small{J033203.1$-$274450} & \small{$136\pm15$} & \small{1.8} & \small{22.49$^{22.70}_{22.16}$} & \small{$9.72 \times 10^{43}$} & \small{126.4}\\
\small{263} & \small{J033218.9$-$275136} & \small{$46\pm9$} & \small{1.75$^{2.10}_{1.40}$} & \small{$>24.18$} & \small{$3.48 \times 10^{44}$} & \small{60.1}\\
\small{265} & \small{J033233.4$-$274236} & \small{$114\pm14$} & \small{1.8} & \small{23.20$^{23.27}_{23.13}$} & \small{$5.00 \times 10^{43}$} & \small{107.9}\\
\small{606} & \small{J033225.0$-$275009} & \small{$33\pm8$} & \small{1.8} & \small{23.27$^{23.40}_{23.14}$} & \small{$1.45 \times 10^{43}$} & \small{59.9}\\
\noalign{\smallskip} 
\hline 
\end{tabular} 
\end{center}
\scriptsize{$^a$ XID from Giacconi et al. (\cite{giacconi02}).}\\ 
\scriptsize{$^b$ IAU registered name, based on original X-ray coordinates.}\\ 
\scriptsize{$^c$ Net counts in the [0.5-10] keV band.}\\
\scriptsize{$^d$ Spectral Index and 90\% confidence range (Tozzi et al. (2005)).}\\
\scriptsize{$^e$ Logaritm of the intrinsic absorption (and 90\% confidence range) [Tozzi et al. (2005)]. }\\
\scriptsize{$^f$ X-ray luminosity in [0.5-10] keV band, deabsorbed and in erg s$^{-1}$ [Tozzi et al. (2005)].}\\
\scriptsize{$^g$ C-statistic coefficient from the model fitting.}\\
\label{table:optfaintS_xray} 
\end{table}

\begin{figure}  
 \begin{center}
\resizebox{8cm}{!}{\includegraphics[angle=-90]{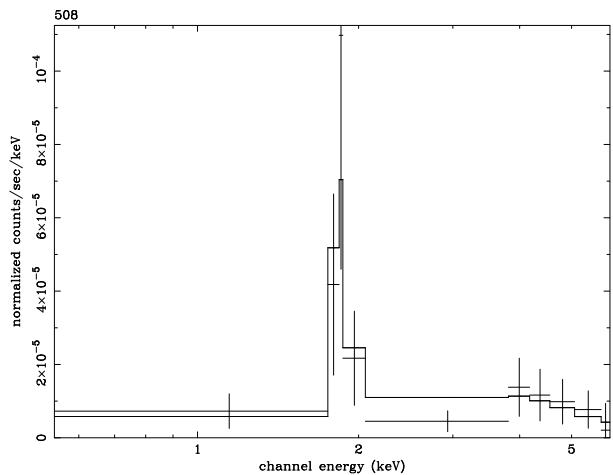}}
\end{center}
   \caption{X-ray spectrum of source CDF-S/508 with a significant feature that we identify with the Fe line at rest-frame 6.4 keV.}
\label{figure:fig_Feline} 
\end{figure}

\subsection{EXOs}

The objects with extreme X-ray-to-optical ratios (EXOs) studied by
Koekemoer et al. (\cite{koekemoer04}) are also included in our sample.
These sources were selected to be undetected in the ACS z band of the
GOODS survey. In this work, by supplementing the HST imaging with new
deep VLT/ISAAC data in J, H and K$_{\rm s}$ bands we can improve the
photometric redshift accuracy. Since in several bands we have only
upper limits, the confidence contours for these sources are generally
large and the low value of the odds parameter reflects the uncertainty
in the redshift determination. However, in three cases we have
odds$\gtrsim$0.6. The first, CDF-S/XID 243 has a single solution at
high redshift ( z$_{\rm phot}\sim 7.9$, which gives an absorbed
L$_{\rm X} [0.5-10]=2.0 \times 10^{45}$ erg s$^{-1}$, see Figure
\ref{figure:fig_cont_1}) and is best fitted with the SED of an
irregular galaxy. Instead CDF-S/XID 508 has a double solution, with
the peak of the redshift distribution corresponding at z$\sim 8.0$ (
absorbed L$_{\rm X} [0.5-10]=9.6 \times 10^{44}$ erg s$^{-1}$) but
with a secondary peak at lower redshift ($\sim 2.4$, absorbed L$_{\rm
  X} [0.5-10]=5.6 \times 10^{43}$ erg s$^{-1}$). We have inspected the
X-ray spectrum of this source, and have detected a clear feature
which, if identified with the Fe line at rest-frame 6.4 keV, would
indicate a redshift of z$\simeq 2.5$ (see Fig.
\ref{figure:fig_Feline}). We therefore suspect that the lower redshift
solution is correct.  Finally, CDF-S/XID 583 is detected in the {\it
  v, i} and {\it z} ACS bands\footnote{In Koekemoer et al.
  (\cite{koekemoer04}) only $3/5$ of the ACS observation time were
  considered while here we have used the entire ACS data set.}  and
this favours a low redshift solution ( z $\sim 2.7$, absorbed L$_{\rm
  X} [0.5-10]=9.4 \times 10^{42}$ erg s$^{-1}$).  These results do not
exclude the high-redshift ( z$=6-7$) AGN scenario proposed by
Koekemoer et al.  (\cite{koekemoer04}) for a small fraction of the
EXOs population. We conclude that using the best multiwavelength
imaging data set available to date, we have found a candidate at z$\ge
7$; the other two EXOs are most likely at lower redshifts
(z$\sim2.5$). Forthcoming Spitzer observations of this field shall
determine more accurate photometric redshifts for the EXO population.

\section{The properties of OFS with spectroscopic redshifts}
\label{section:optf_spec}

For the few OFS with known redshifts, we can study their X-ray
spectral properties. In the CDF-S, there are six of these objects and
we report in Table \ref{table:optfaintS} their main X-ray and optical
properties. The depth of the 1 Msec CDF-S data enables us to perform
an X-ray spectral analysis of these sources. We adopt a power-law
model plus an absorption component. The fit yields the power-law
photon index $\Gamma$, the intrinsic column density N$_{\rm H}$, and
the X-ray luminosity in the [0.5-10] keV rest-frame band corrected for
absorption. We give the results of these fits in Table
\ref{table:optfaintS_xray}. Adopting the X-ray classification
presented in Section \ref{section:catalogue}, the sample comprises
three X-ray absorbed QSOs, and three X-ray absorbed AGNs.

\begin{figure}  
 \begin{center}
\resizebox{8cm}{!}{\includegraphics{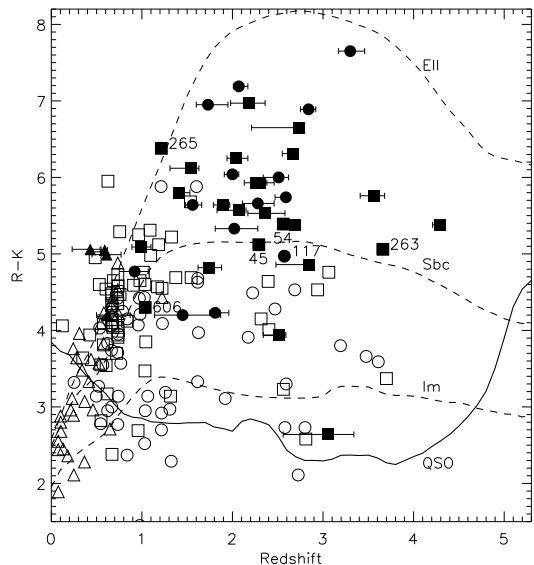}}
\end{center}
    \caption{R$-$K colour versus redshift. Following the X-ray
      classification (Szokoly et al. \cite{szokoly04}) described in
      \S\ref{section:catalogue}: circles are X-ray unabsorbed
      AGN/QSOs, squares X-ray absorbed AGN/QSOs, triangles galaxies.
      Filled symbols refer to OFS, empty symbols to optically bright
      sources.  The numbers are the XIDs of OFS spectroscopically
      identified in the CDF-S (see Table \ref{table:optfaintS}). The
      four evolutionary tracks correspond to an unreddened QSO [solid
      line; Vanden Berk et al. (2001)], and to unreddened elliptical,
      Sbc and irregular galaxies from the Coleman, Wu \& Weedman
      (1980) template library (dashed lines). The bars indicate 1
      $\sigma$ errors in the photometric redshift estimates.}
\label{figure:fig_8} 
\end{figure}

The identified OFS show a variety of optical classes in Table
\ref{table:optfaintS}. Following the optical classification introduced
by Szokoly et al.  (\cite{szokoly04})\footnote{The spectroscopic data
  of the CDF-S are publicly available at this URL:
  http://www.mpe.mpg.de/$\sim$mainieri/cdfs\_pub/}, there are: three
objects showing high excitation narrow lines (C\,{\sc
  iv}$\lambda$1549) together with narrow Ly$\alpha$ emission (HEX),
CDF-S/XID 54,117,263, and three objects showing only low excitation
lines (LEX) either Ly$\alpha$, CDF-S/XID 45, or [O\,{\sc
  ii}]$\lambda$3727, CDF-S/XID 265,606. For the sources in the LEX
class, the presence of an AGN is only revealed by their high X-ray
luminosities.  \\ In strongly absorbed X-ray sources, the host galaxy
dominates the optical and near-IR emission.  The R$-$K versus $z$
diagram can be used to contrain the nature of the X-ray sources. We
present this diagram in Figure \ref{figure:fig_8} for all of the CDF-S
sources with spectroscopic redshifts, highlighting the OFS.  For
comparison, we also plot the evolutionary tracks expected for
classical Type-1 QSOs and galaxies of various morphological types. The
optically faint X-ray population has on average redder colours than
the optically bright population. For the 92 OFS of our sample, we find
that 60 ($\sim 65\%$) of them are Extremely Red Objects (EROs),
R$-$K$\ge$5, as compared to only 28 ($\sim 11\%$) for the optically
bright population (see also Alexander et al.  \cite{alexander01} and
\cite{alexander02}).  The OFS with spectroscopic identification cover
a wide redshift range (z$=1-4$).

In Figure \ref{figure:fig_8}, we plot also the OFS for which we have
determined a photometric redshift. Almost all of them have colours
dominated by the host galaxy (elliptical or spiral) and, except for a
few candidates at high redshift, the bulk of them lie in the redshift
range of $1.5 \lesssim z \lesssim 2.5$, filling in a region that is
poorly sampled by spectroscopically identified X-ray sources.

\begin{figure}  
 \begin{center}
\resizebox{8cm}{!}{\includegraphics{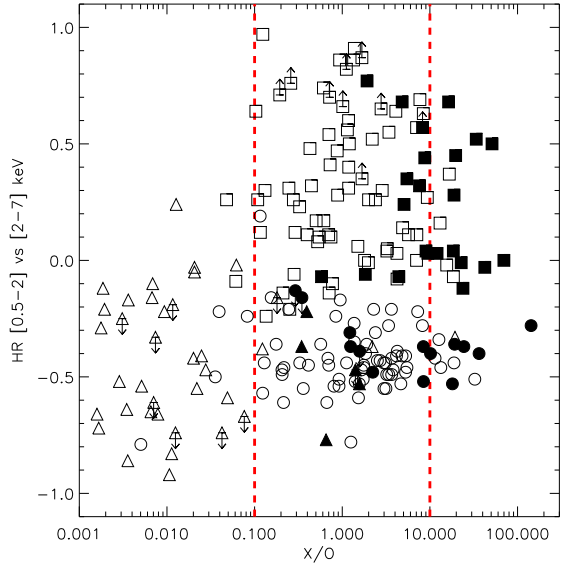}}
\end{center}
   \caption{The Hardness ratios (HR) versus X-ray-to-optical flux
     ratios (X/O). Open symbols indicate ptically bright (R$<25$)
     sources and filled symbols indicate OFS. Symbols are as in Figure
     \ref{figure:fig_8}. For clarity, we have omitted optically bright
     objects that are spectroscopically unidientified. Vertical lines
     are for X/O$=0.1$ and X/O$=10$ respectively.}
\label{figure:fig_hr_xo} 
\end{figure}

\section{X-ray-to-optical flux ratios} 
\label{section:xo}

X-ray-to-optical flux ratios (X/O) can yield important information on
the nature of X-ray sources (Maccacaro et al. \cite{maccacaro88}).  A
value of $0.1< {\rm X/O} <10$ is a clear sign of AGN activity since
normal galaxies and stars usually have lower X-ray-to-optical flux
ratios. In Figure \ref{figure:fig_hr_xo}, we show the hardness ratios
(HR) versus X/O for the X-ray sources in the CDF-S area. The majority
of X-ray unabsorbed and X-ray absorbed AGN/QSOs are inside the well
defined locus of active galactic nuclei ($0.1< {\rm X/O} <10$), while
``normal'' galaxies, for which the contribution to the X-ray flux is
mainly due to star-formation activity, have X/O$< 0.1$.  The OFS have
ratios characteristic of AGN and $\sim 49 \%$ of them show
intriguingly high values (X/O$ > 10$).\footnote{We note that $\sim 75
  \%$ of the X-ray sources in the CDF-S with X/O$> 10$ are OFS.} In
the spectroscopic follow-up of the CDF-S, three of the OFS with high
X/O have been identified (Szokoly et al.  \cite{szokoly04}; Vanzella
et al. \cite{vanzella04}): CDF-S/XID 45,54 and 265 have redshifts of
z$=2.291$,2.561 and 1.215, respectively. Two of them are classified as
X-ray absorbed QSOs and one as X-ray absorbed AGN (see Table
\ref{table:optfaintS}).  Recently, Mignoli et al.  (\cite{mignoli04})
have studied a sample of eleven hard X-ray selected sources with
X/O$>10$ using deep near-IR observations with ISAAC. All but one of
the sources have been detected in the K$_{\rm S}$ band with very red
colors (R$-$K$_{\rm S} > 5$).  They were able to provide a
morphological classification and the sample is dominated by elliptical
profiles (7/10). Using the morphological information and the R$-$K
colour the authors determined a minimum redshift for the sources in
the range z$_{\rm min}=0.80-1.45$. In the OFS inside the ``GOODS
area'' there are 20 objects with such high X/O values and we have
determined photometric redshifts for them. Three sources ( CDF-S/XID
133,523,614) have extremely low value of the odds parameter and we
will exclude them in the following analysis. Two different classes of
objects are present.  Twelve ( $\sim 71 \%$) have a best fitting SED
of an elliptical galaxy, the average column density is $\sim 1.0
\times 10^{23}$ cm$^{-2}$ and their predicted redshift range is
$0.9<{\rm z_{phot}}<2.7$ (with a mean redshift of $\sim 1.9$). The
remaining sources have a best fitting SED of either an irregular or
starburst galaxy, with a mean redshift of z$\sim 4$ and X-ray spectra
indicating a low value of absorption ($<{\rm N_{H}}> \sim 1 \times
10^{22}$ cm$^{-2}$).  One of these sources is the z$>7$ candidate
(CDF-S/XID 243) .\footnote{We are sampling an X-ray flux regime ($8
  \times 10^{-16} \lesssim {\rm F_X [2-10 ~ keV]} \lesssim 7 \times
  10^{-15}$ erg cm$^{-2}$ s$^{-1}$) much fainter than the one studied
  by Mignoli et al. (\cite{mignoli04}), $1 \times 10^{-14} \lesssim
  {\rm F_X [2-10 ~ keV]} \lesssim 1 \times 10^{-13}$ erg cm$^{-2}$
  s$^{-1}$ ( M.  Brusa PhD thesis).}\\ Summarizing, $\sim 71 \%$ of
the OFS with X/O$> 10$ are X-ray absorbed AGN and their photometry is
well reproduced by an unreddened early-type template at $0.9<{\rm
  z}<2.7$.  The other $29 \%$ of the sources do not show strong X-ray
absorption, they have bluer colours and among them there are some high
redshift X-ray unabsorbed/absorbed QSOs.  Finally, a significant
fraction ( $\sim 24 \%$) of the sources with high X/O are X-ray
absorbed QSOs.

\begin{figure}  
 \begin{center}
\resizebox{12cm}{!}{\includegraphics{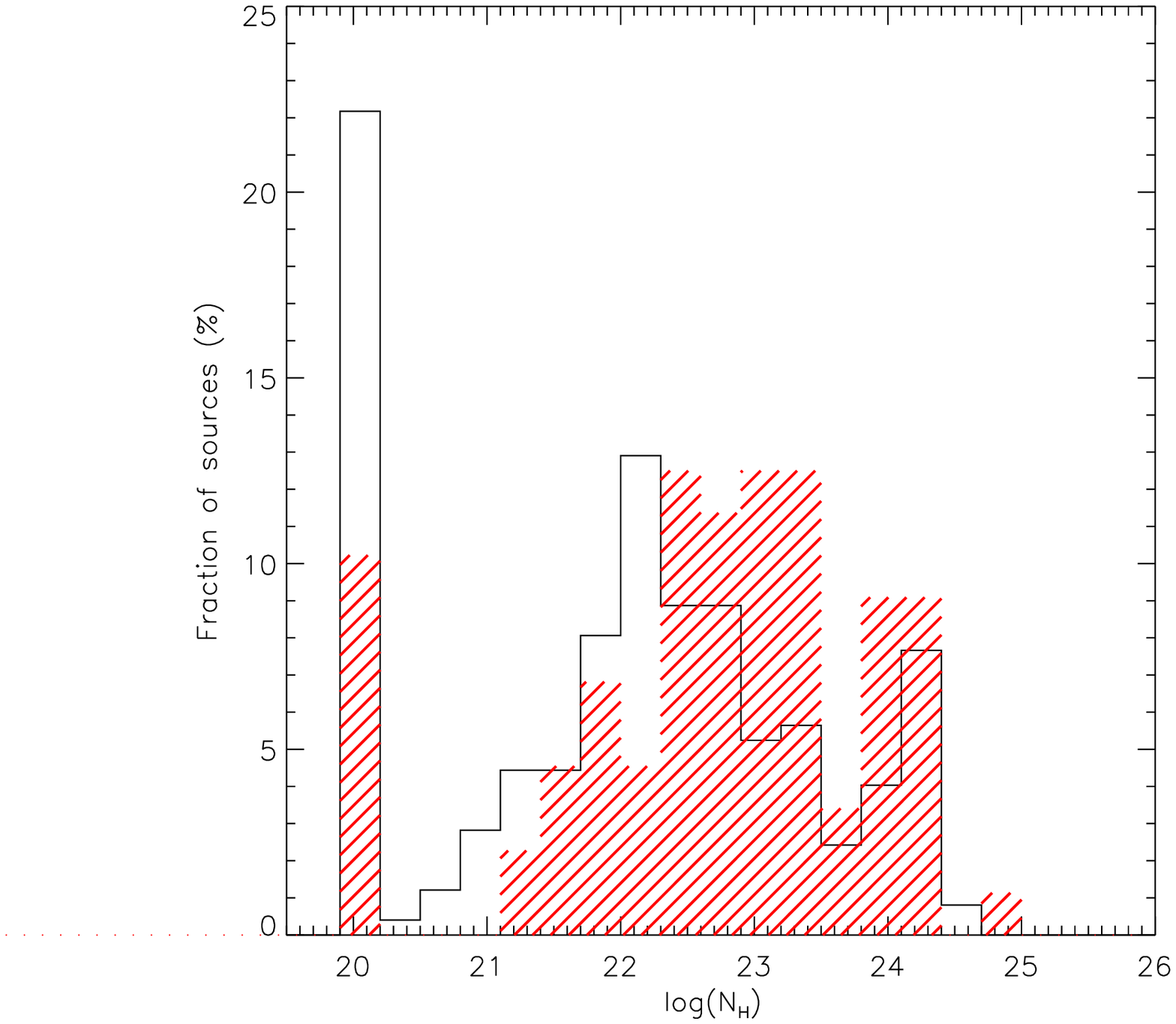}}
\end{center}
   \caption{Intrinsic N$_{\rm H}$ distributions for the OFS (hatched 
     histogram) and optically bright sources (open histogram).}
\label{figure:fig_nh_chandra} 
\end{figure}

\section{X-ray spectroscopy}
\label{section:xspec}

To further investigate the characteristics of the OFS we have
performed an X-ray spectral analysis of our sample. We use the X-ray
data accumulated in the 1 Msec {\it Chandra} exposure and XSPEC
(v11.1) for the fitting procedure. The spectral model adopted is a
power-law with an intrinsic absorber at the source redshift. An
additional photoelectric absorption component is fixed to the Galactic
column density in the CDF-S region of the sky ($\sim 8 \times 10^{19}$
cm$^{-2}$). For sources with less than 100 net counts in the [0.5-10]
keV band, we fix the photon index $\Gamma$ at 1.8 and derive the
column density N$_{\rm H}$ (see Tozzi et al.  \cite{tozzi05}).  To
estimate the intrinsic absorption that is affecting the X-ray source,
its redshift is needed: we use the spectroscopic redshift if known
otherwise we adopt the photometric redshift (Table
\ref{table:zphot_optf_1}). Thus, we are able to estimate the N$_{\rm
  H}$ value for 336\footnote{The remaining sources are either without
  a redshift estimate (four) or are stars (six).}  ($\sim 97 \%$) of
the 346 X-ray sources in the CDF-S.  We show in Figure
\ref{figure:fig_nh_chandra} the derived N$_{\rm H}$ distribution. The
open histogram is the distribution for optically bright sources while
the hatched histogram refers to OFS. A K-S test of the two
distributions gives a probability of $\sim 0.004 \%$ that they are
draw from the same population. We have already deduced from other
diagnostics (X/O, hardness ratios, optical/near-IR colours) that a
large fraction of these faint sources are absorbed.  Figure
\ref{figure:fig_nh_chandra} confirms and reinforces this picture
since, in this case, we are measuring directly the absorption from the
X-ray spectrum. Of the OFS $\sim 73 \%$ have a column density larger
than $10^{22}$ cm$^{-2}$; for comparison the fraction of bright
sources with such high N$_{\rm H}$ value is $\sim 55 \%$. X-ray
unabsorbed sources (the first bin in Figure
\ref{figure:fig_nh_chandra}) are approximatly three times more
numerous between the optically bright than the optically faint
sources.

\section{X-ray absorbed QSOs}
\label{section:qso2}

As mentioned in Section \ref{section:optf_spec}, a fraction of OFS
could be made of X-ray absorbed QSOs. We find that 11 ($\sim 23
\%$) of the OFS for which we have computed photometric redshifts have
an X-ray luminosity in the [0.5-10] keV band larger than $10^{44}$ erg
s$^{-1}$ and a N$_{\rm H} > 10^{22}$ cm$^{-2}$, and consequently X-ray
absorbed QSOs according to our definition in Section
\ref{section:catalogue}.

Several synthesis models of the XRB require a large population of
obscured QSOs. To compare our results with the Gilli, Salvati \&
Hasinger (\cite{gilli01}) and Ueda et al. (\cite{ueda03}) models, we
need to use the following definition of a X-ray absorbed QSO:
rest-frame L$_{\rm X}[2-10$ keV$] > 10^{44}$ erg s$^{-1}$ and N$_{\rm
  H} > 10^{22}$ cm$^{-2}$. We find that 44 sources of the 336 X-ray
sources (optically bright and faint) for which we were able to
determine N$_{\rm H}$, satisfy these criteria. These X-ray absorbed
QSOs contribute to the [2-10] keV XRB for a fraction of $\sim 15 \%$
(if we adopt the HEAO-1 measure of the total flux of the XRB in the
$[2-10]$ keV band)\footnote{We note that at the depth of the CDF-S
  survey the whole hard XRB has been resolved according to the HEAO-1
  measure (Rosati et al. \cite{rosati02}).}. Model B of Gilli, Salvati
\& Hasinger (\cite{gilli01}) predicts a $38 \%$ contribution from Type
II QSOs to the hard XRB, while the recent model by Ueda et al.
(\cite{ueda03}) requires a lower contribution from these sources
($\sim 16 \%$) in good agreement with what we have found.

\begin{figure*}
\resizebox{\hsize}{!}
{\includegraphics[]{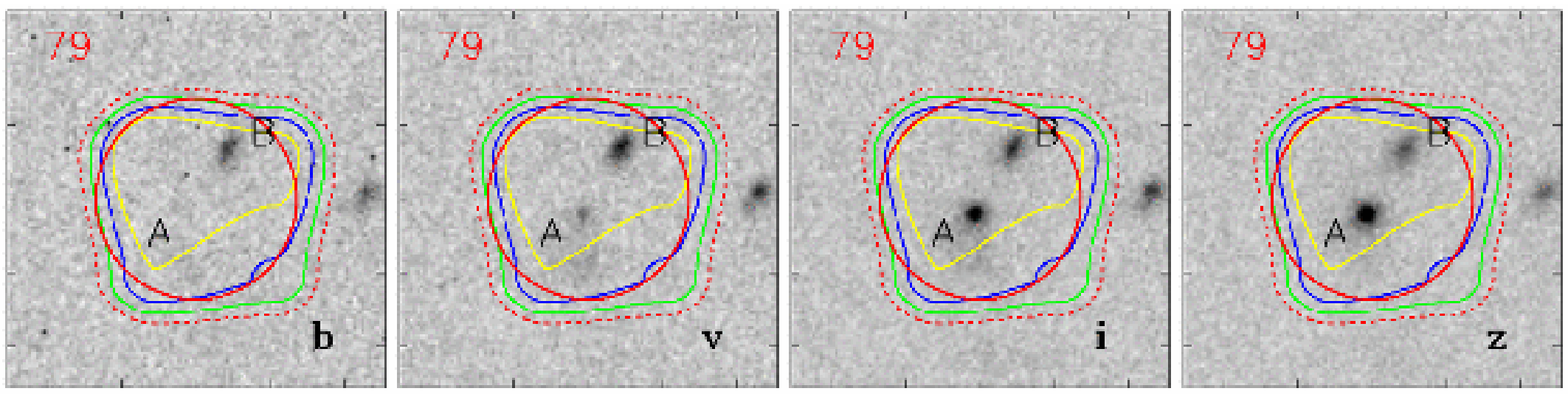}}\\
\resizebox{\hsize}{!}
{\includegraphics[]{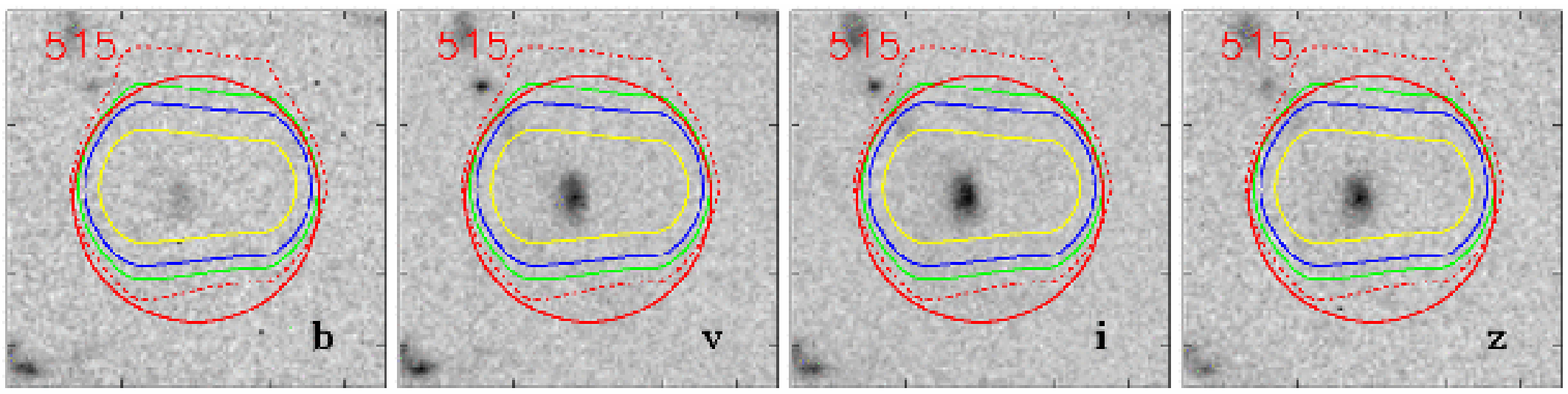}}\\
\resizebox{\hsize}{!}
{\includegraphics[]{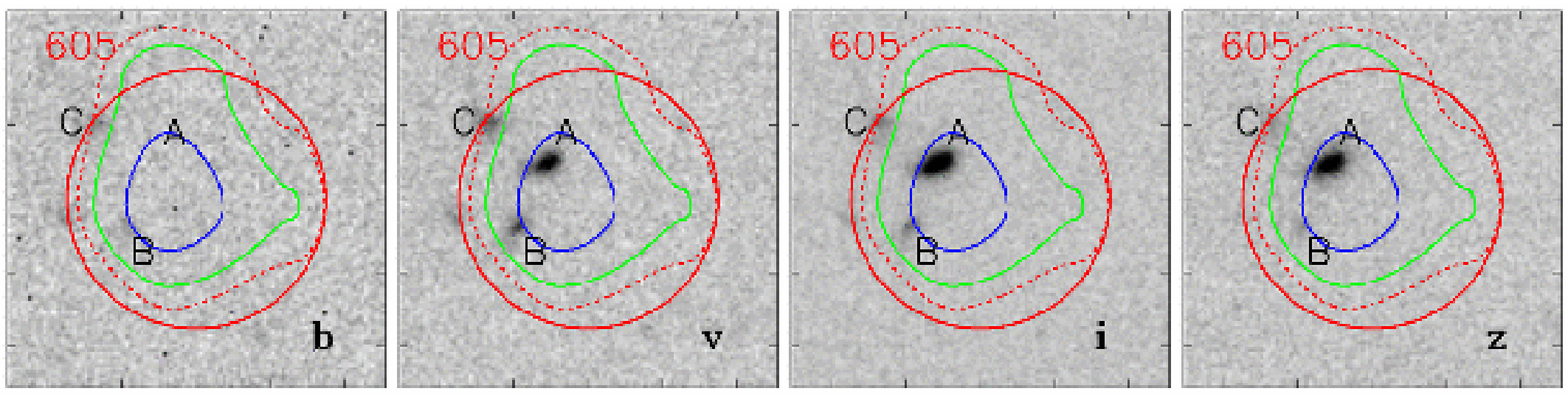}}\\ 
\caption{{\small{ Cutouts in the four UDF filters ( F435W, F606W, F775W, 
      F850LP) with X-ray contours of the three OFS inside this area.
      Images are 3\arcsec across. The $3 \sigma$ {\it Chandra}
      positional error is indicated by the solid circle.}}}
\label{figure:fig_UDF_cutouts}
\end{figure*}

 \begin{figure*}

   \parbox{12.3cm}{\resizebox{\hsize}{!}{\includegraphics[angle=0]{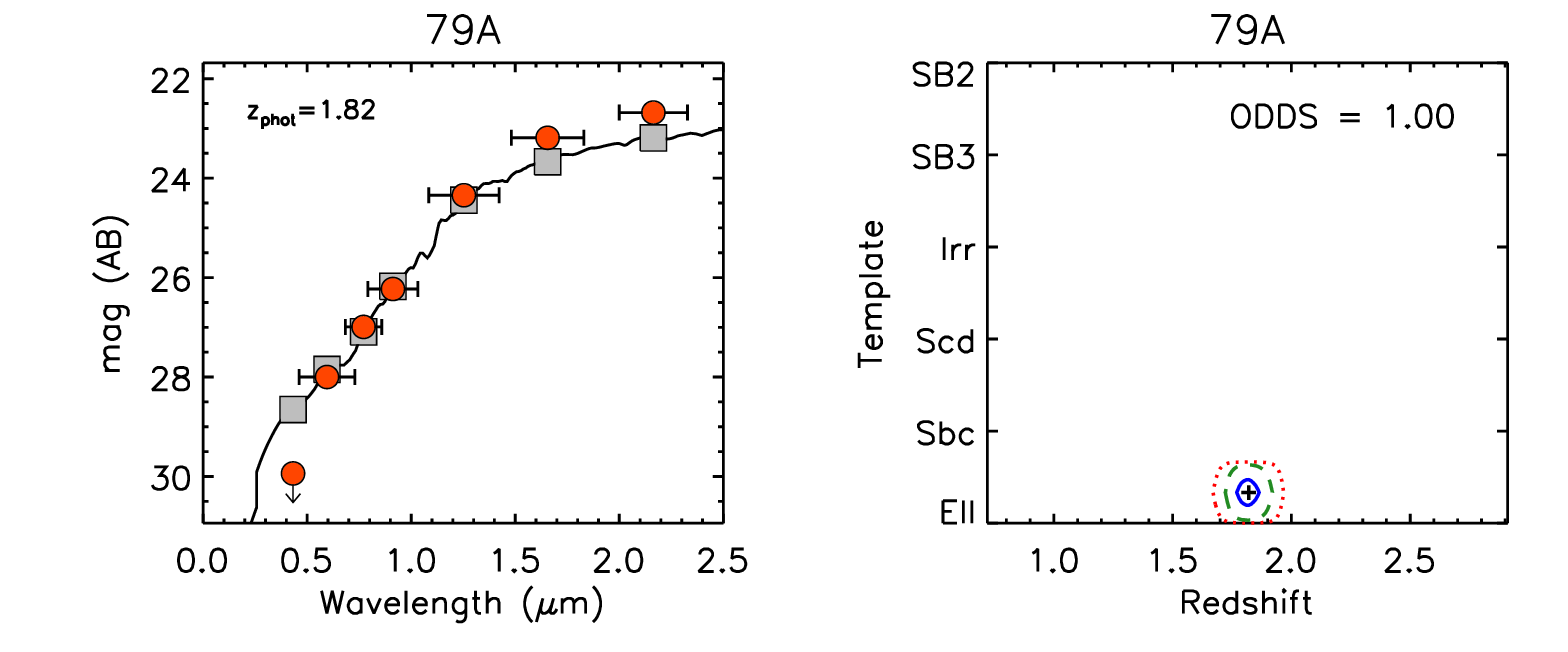}}}
   \\

   \parbox{12.3cm}{\resizebox{\hsize}{!}{\includegraphics[angle=0]{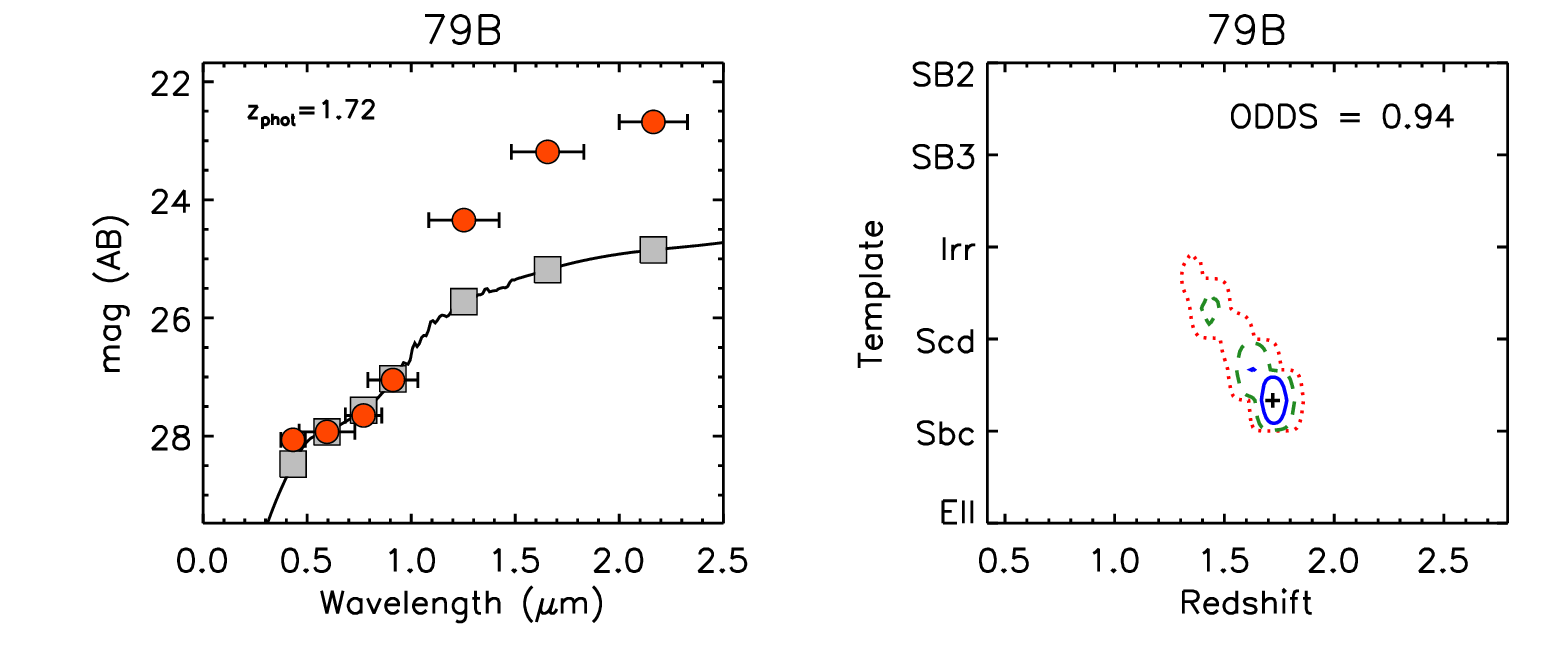}}}
   \\

   \parbox{12.3cm}{\resizebox{\hsize}{!}{\includegraphics[angle=0]{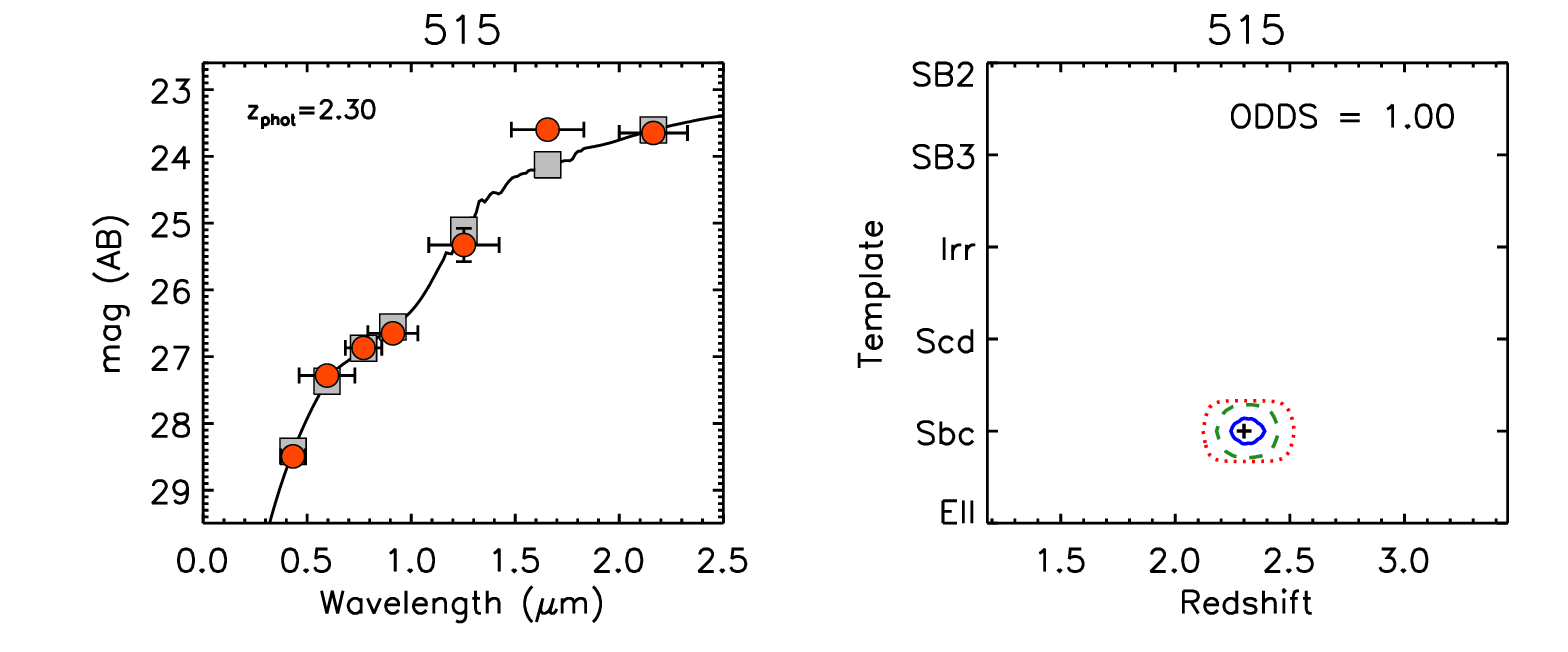}}}
   \\

   \parbox{12.3cm}{\resizebox{\hsize}{!}{\includegraphics[angle=0]{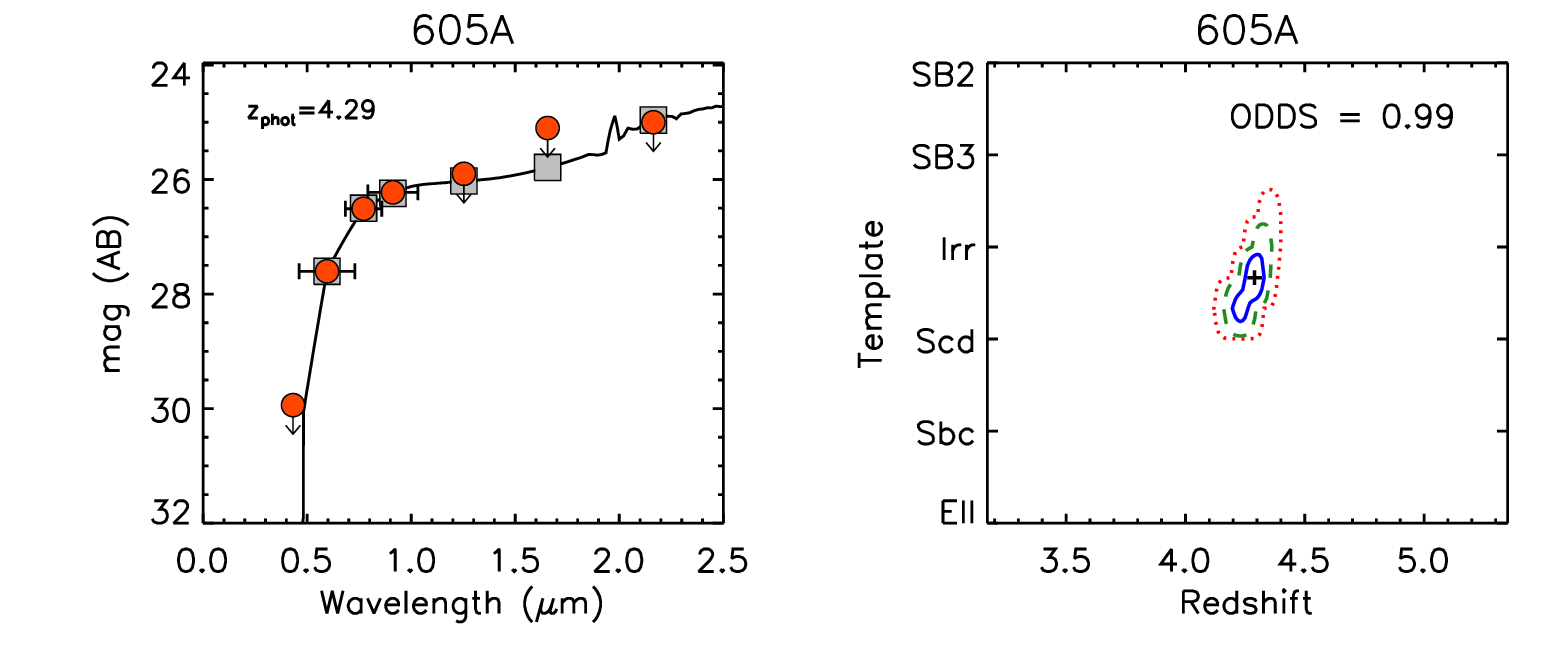}}}
   \\ \caption{Photometric redshifts for the counterparts of OFS in
     the UDF area. {\it Left side}: best fit template (solid line),
     the observed photometry (filled circles with error bars), and the
     best fit model photometry (filled squares). {\it Right side}: $1
     \sigma$ (dotted line), $2 \sigma$ (dashed line) and $3 \sigma$
     (continuum line) confidence contours for the photometric redshift
     determination in the SED template {\it vs} redshift plane. }
\label{figure:fig_cont_UDF} 
\end{figure*}

\section{Ultra Deep Field}
\label{section:udf}

\begin{figure*}
\parbox{12.3cm}{\resizebox{\hsize}{!}{\includegraphics[angle=0]{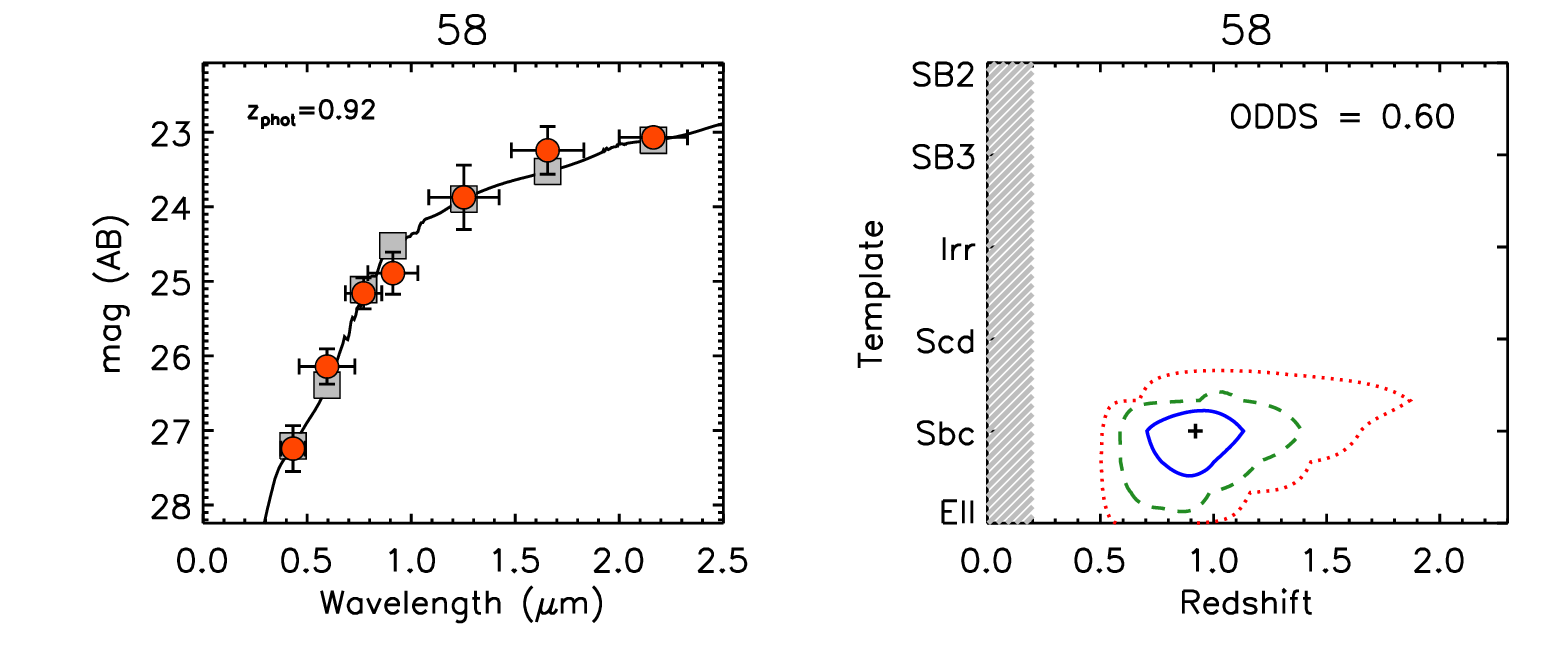}}}
\\
\parbox{12.3cm}{\resizebox{\hsize}{!}{\includegraphics[angle=0]{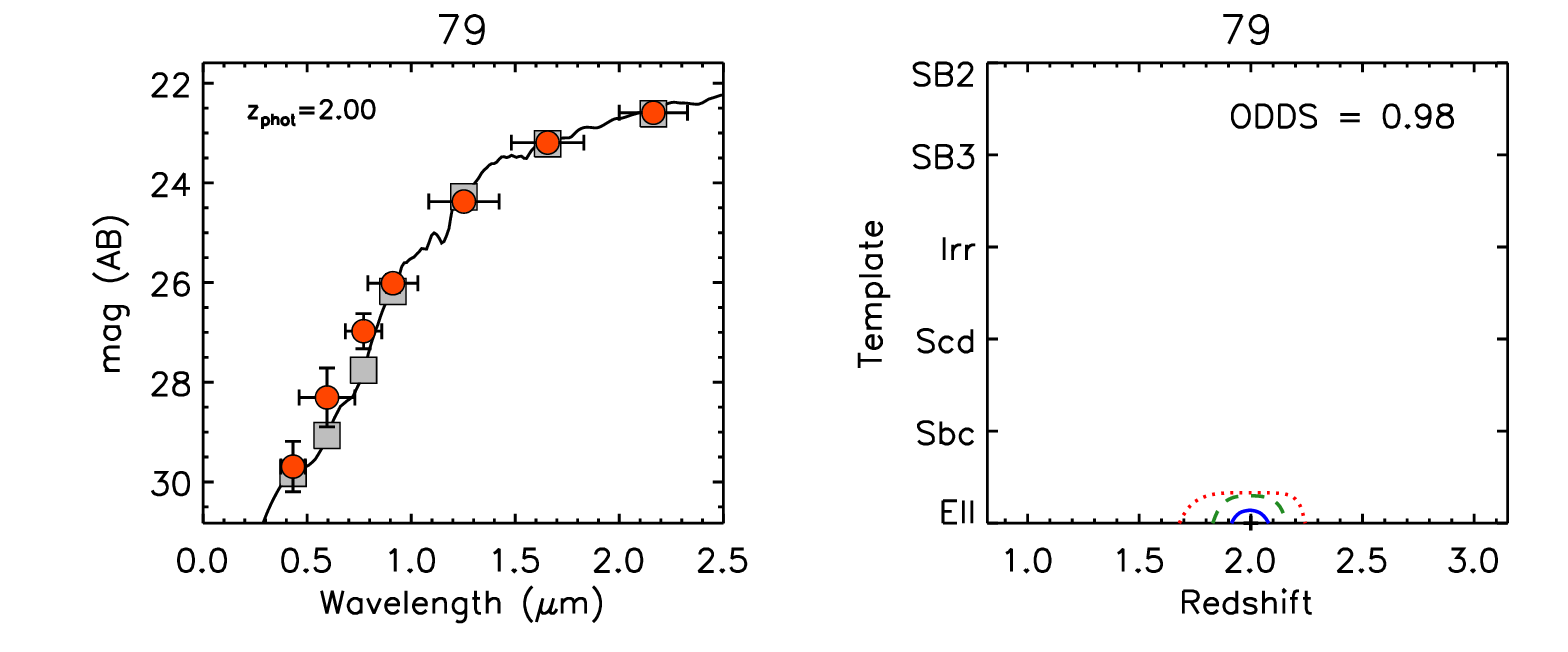}}}
\\
\parbox{12.3cm}{\resizebox{\hsize}{!}{\includegraphics[angle=0]{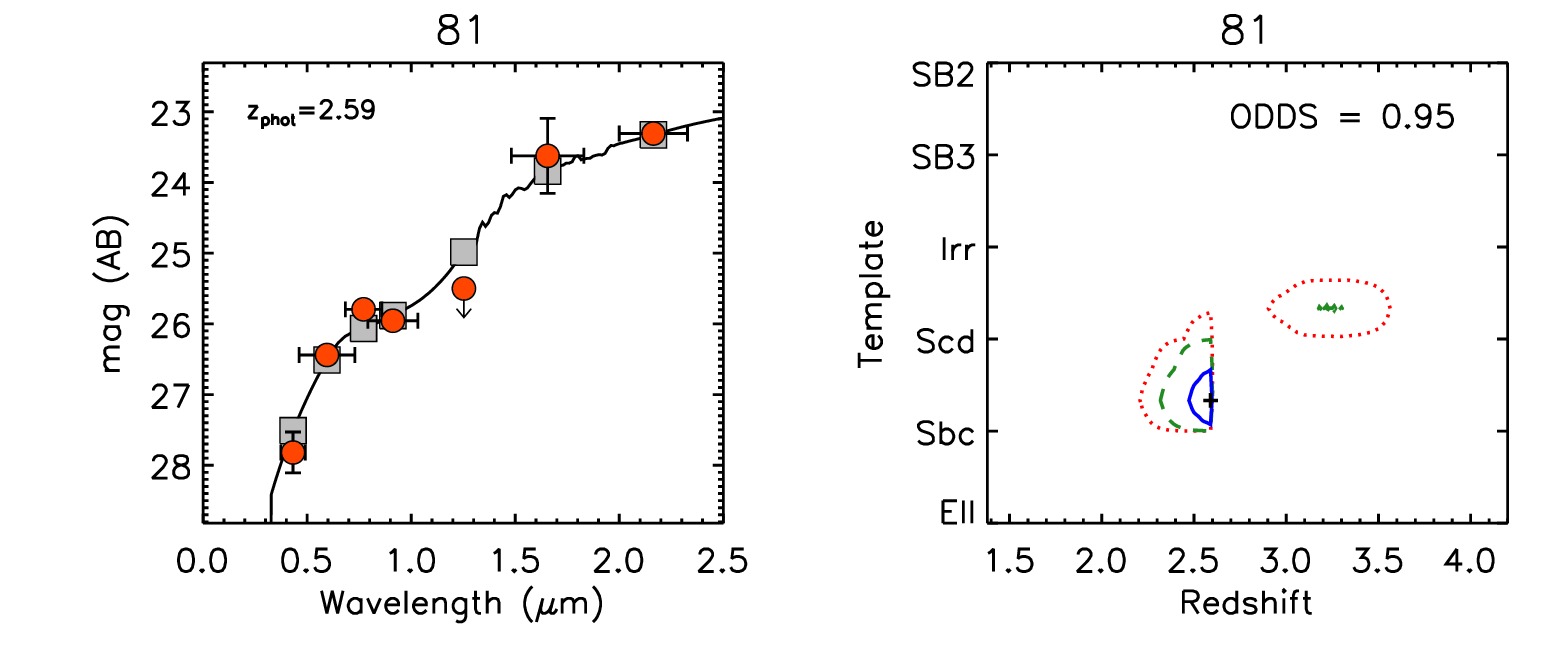}}}
\\
\parbox{12.3cm}{\resizebox{\hsize}{!}{\includegraphics[angle=0]{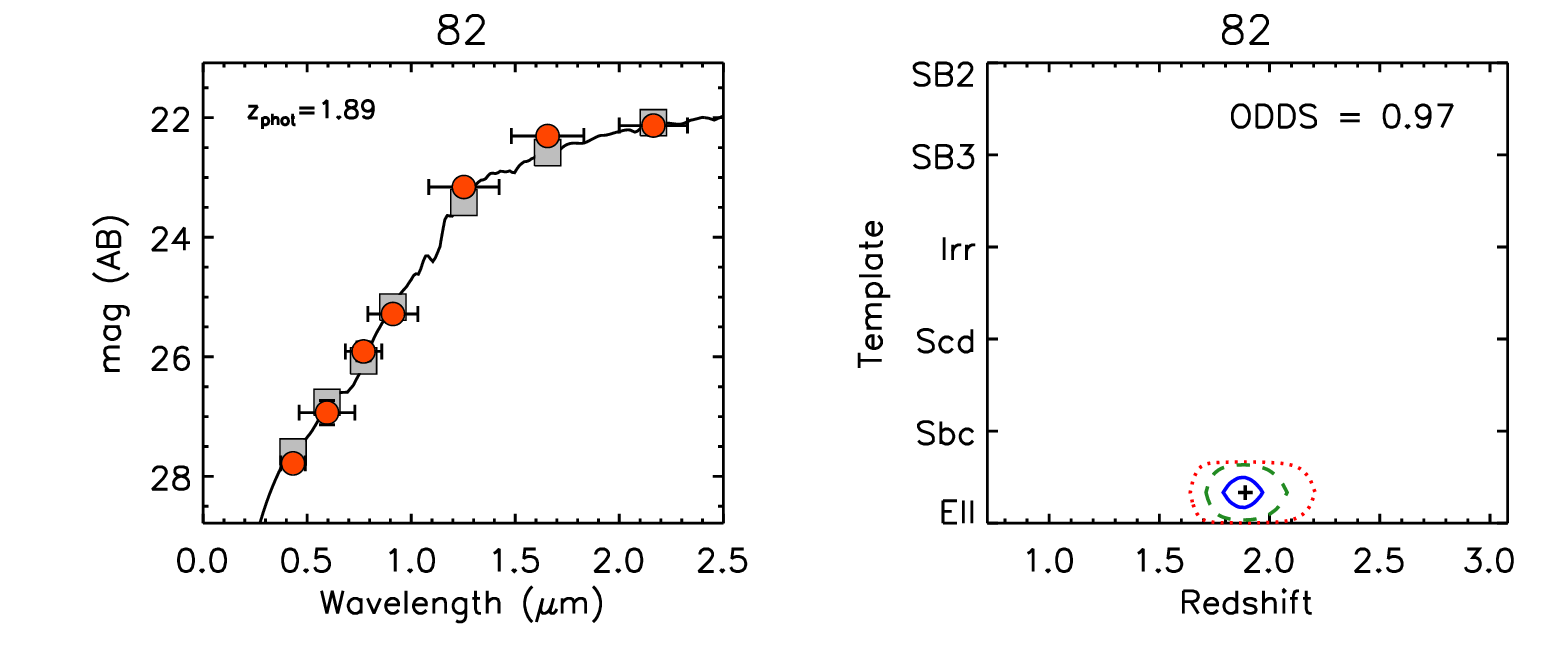}}}
\\
\caption{Photometric redshifts for OFS in the ``GOODS area'' with 
  odds$\ge 0.3$. {\it Left side}: best fit galaxy template (solid
  line) with overplotted the observed photometry (filled circles with
  error bars), and the best fit model photometry (filled squares).
  {\it Right side}: $1 \sigma$ (dotted line), $2 \sigma$ (dashed line)
  and $3 \sigma$ (continuum line) confidence contours for the
  photometric redshift determination in the SED template {\it vs}
  redshift plane. The cross indicates the best-fit solution. The
  shaded area refers to a low-probability region in the solution space
  (see text). The dot-dashed line for source 508 shows the redshift
  estimation based on the X-ray spectrum (see Sec. 4.3).  XIDs are
  from Giacconi et al.  (2002).  }
\label{figure:fig_cont_1}
\end{figure*}

The Hubble Ultra Deep Field (UDF) is a 400-orbit program to image one
single field inside the 'GOODS area' with the ACS camera on board of
HST (see Figure \ref{figure:fig_2}). Images have been taken in the
same four filters that were used for the GOODS survey: F435W ({\it
  b}), F606W ({\it v}), F775W ({\it i}) and F850LP ({\it z}). The data
have been recently released and go $1.0$ and $1.2$ magnitude deeper
than GOODS in the {\it b} and {\it z} band,
respectively.\footnote{http://www.stsci.edu/hst/udf/} \\Inside the UDF
area there are sixteen X-ray sources in the Giacconi et al.
(\cite{giacconi02}) catalogue and three of them are OFS: CDFS/XID
79,515,605.\footnote{We have also checked the Alexander et al.
  (\cite{alexander03}) catalogue, which includes all of the sixteen
  sources of Giacconi et al. (\cite{giacconi02}) plus one extra source
  with $\sim 12$ counts in the full {\it Chandra} energy band and with
  an optically bright counterpart.} We have extracted cutouts for
these faint sources in the four UDF bands and show them in Figure
\ref{figure:fig_UDF_cutouts}. The unprecedented depth and resolution
of these images allows us to observe the optical properties of OFS
with sub-arcsecond resolution. We indicate in the cutouts the $3$
$\sigma$ positional error circle obtained in Giacconi et al.
(\cite{giacconi02}), which takes into account the strong effect of
off-axis angle in the X-ray PSF and centroid. For CDFS/XID 79 two
different counterparts are clearly present. We name these sources 79A
and 79B ( see Figure \ref{figure:fig_UDF_cutouts}). They have
different colours: 79A is undetected in the {\it b} band and is
brighter than 79B in the {\it i} and {\it z} bands. In the cutouts of
CDFS/XID 515 there is only one optical counterpart inside the Chandra
error circle. Finally, for CDFS/XID 605 we find three optical
counterparts ( A, B, C) of which two are extremely faint in all four
bands ( B and C). We have selected from the publicly available
catalogue of the UDF ( h\_udf\_wfc\_V1\_cat.txt) source magnitudes in
the four bands for each of the optical counterparts. The magnitudes
available from this catalogue where computed using the {\it i} band
isophotes of each source, thereby producing isophotally matched
magnitudes that are suitable to compute colours. Combining these
colours with our near-IR photometry, we estimated photometric
redshifts as decribed in Section \ref{section:bpz} (Figure
\ref{figure:fig_cont_UDF}). We provide individual notes on each source
below.\\ Source CDFS/XID 79: the A counterpart is well fitted with an
early-type SED at redshift z$_{\rm phot} \sim 1.8$, similar to that
obtained using the GOODS photometry ( see Figure
\ref{figure:fig_cont_1}). However, we could not obtain a reasonable
fit to source 79B, probably because the near-IR photometry (which does
not resolve the two counterparts) is dominated by 79A.\\ Source
CDFS/XID 515: we obtain z$_{\rm phot} \sim 2.3$ and the confidence
contours are in good agreement with the value found with the GOODS
photometry ( see Figure \ref{figure:fig_cont_1}). \\Source CDFS/XID
605: in this case the depth of the UDF data is really a step forward
for the redshift determination. With the GOODS photometry we obtained
a ``double'' solution with a peak in the probability distribution at
low ( z$_{\rm phot} \sim 0.8$) and high ( z$_{\rm phot} \sim 4.7$)
redshift ( see Figure \ref{figure:fig_cont_1}).  With the UDF data we
have smaller errors on the photometry and can put a more stringent
upper limit in the {\it b} band where CDFS/XID 605 is undetected: this
source is now a strong candidate at high redshift with z$_{\rm phot}
\sim 4.29$, with a small uncertainty ($4.21-4.32$).

\section{Conclusions}

In this paper, we have taken advantage of the unique multi-wavelength
coverage of the GOODS survey in the Chandra Deep Field South to
constrain the nature and redshift distribution of optically faint
X-ray sources (R$\geq 25$). It is important to study their properties
since they are a significant fraction ($\sim 27 \%$) of the whole
X-ray sample.\\ Our study extends the earlier analysis by Alexandet et
al.  (\cite{alexander01}) by determining photometric redshifts for the
optically faint sources without spectroscopic redshifts. The
reliability of these photometric redshifts has been tested against the
spectroscopic redshifts available for the optically faint fources (
see Figure \ref{figure:fig_cont_spec_1}).  We find that a larger
fraction of the optically faint ( $76 \%$) sources are at redshift z
$> 1$ compared to the optically bright sample ( $49 \%$). This finding
reduces the disagreement between the observed redshift distribution in
{\it Chandra} deep fields (Barger et al. \cite{barger03}, Szokoly et
al. \cite{szokoly04}) and that predicted by XRB models based on the
ROSAT X-ray luminosity function, as the majority of the still
unidentified X-ray sources in these deep fields are optically faint.
However, the redshift distribution that we obtain including our best
photometric redshifts still peaks at z$\lesssim1$, while the current
XRB models predict z$\sim 1.3-1.5$.  One solution would involve
implementing a new X-ray luminosity function for AGN in the XRB
synthesis models, by combining Deep {\it Chandra} and XMM-{\it Newton}
fields with shallower surveys (Gilli et al. \cite{gilli03b}; Fiore et
al.  \cite{fiore03}). This new X-ray luminosity function will be able
to reproduce better the evolution with redshift of Seyfert-like
objects which make a large fraction of the observed peak at
z$\lesssim1$.

Several diagnostics indicate that the majority of the optically faint
sources are absorbed. Their hardness ratio distribution is harder
($98\%$ significance level) than that of the optically bright sample
indicating a large fraction of optically faint fources with a flat
X-ray spectrum which implies intrinsic absorption. Their
optical/near-IR photometry is dominated by the emission of the host
galaxies and their colours are in average redder that the optically
bright sources [$65\%$ are EROs, (R-K$\ge 5$) as compared to $11\%$ of
the optically bright sources]. We have performed an X-ray spectral
analysis and the distribution of N$_{\rm H}$ values shows that $\sim
73 \%$ of OFS have column densities larger than $10^{22}$ cm$^{-2}$
(for optically bright sources the fraction is of $55\%$).

We find that $\sim 23\%$ of the optically faint X-ray sources are
X-ray absorbed QSOs ( L$_{\rm X}[0.5-10 {\rm keV}]>10^{44}$ erg
s$^{-1}$ and N$_{\rm H}> 10^{22}$ cm$^{-2}$). Synthesis models of the
XRB include a significant contribution from X-ray absorbed QSOs
to the hard XRB (i.e.  $38 \%$ for model B of Gilli, Salvati \&
Hasinger \cite{gilli01}).  From the CDF-S survey we find a much lower
contribution from obscured QSOs, $\sim 15 \%$. This difference can not
be ascribed to the remaining fraction of the as yet unresolved XRB:
both the model and observational values are derived assuming the
HEAO-1 measure of the total flux of the XRB in the $[2-10]$ keV band
and the hard XRB has been resolved (using the HEAO-1 value) to the
depth of the CDF-S (Rosati et al.  \cite{rosati02}).\\ Our value is in
good agreement with the prediction ($\sim 16 \%$) based on the recent
model by Ueda et al.  (\cite{ueda03}).

Approximatly $50 \%$ of the OFS have high X-ray-to-optical ratios
(X/O$>10$). $71 \%$ of them are strongly X-ray absorbed ( $<$N$_{\rm
  H}> \approx 1.0 \times 10^{23}$ cm$^{-2}$) and their photometry is
well reproduced by the SED of an early-type galaxy with
$0.9<$z$_{phot}<2.7$. The remaining $29 \%$ is on average less
absorbed ( $<$N$_{\rm H}> \approx 1 \times 10^{22}$ cm$^{-2}$) and has
bluer colours reproduced by irregular or starburst galaxies with a
mean redshift of $\sim 4$. Among this second group we find a candidate
at redshift z$\gtrsim 7$. Finally, $\sim 24 \%$ of the sources with
high X-ray-to-optical ratios are X-ray absorbed QSOs.

\begin{acknowledgements}
  We are grateful to Narciso Benitez for his assistance with BPZ. We
  thank the referee, B.J. Wilkes, for a very detailed and useful report
  that improved the manuscript. DMA thanks the Royal Society for
  financial support. CN gratefully acknowledge support under NASA
  grants NAG-8-1527 and NAG-8-1133.
\end{acknowledgements}

\newpage
\begin{figure*}
\parbox{12.3cm}{\resizebox{\hsize}{!}{\includegraphics[angle=0]{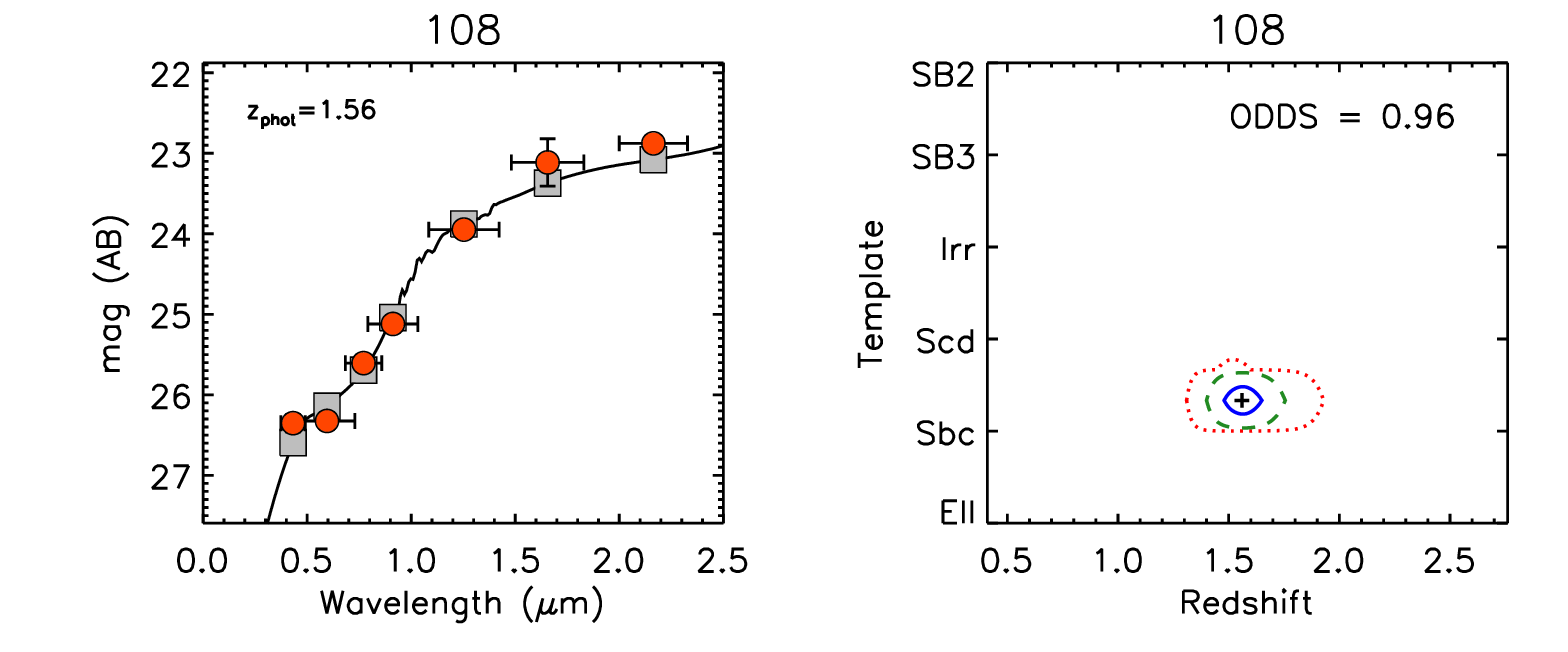}}}
\\
\parbox{12.3cm}{\resizebox{\hsize}{!}{\includegraphics[angle=0]{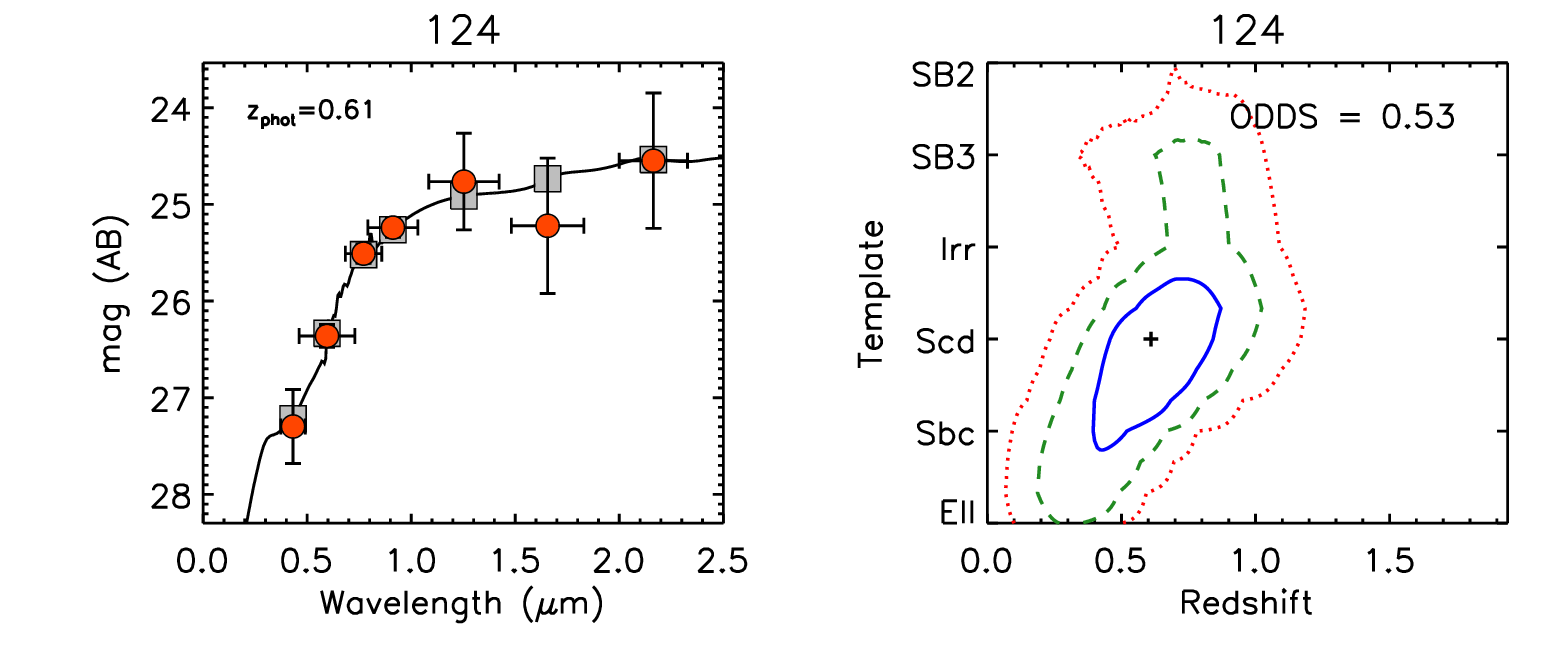}}}
\\
\parbox{12.3cm}{\resizebox{\hsize}{!}{\includegraphics[angle=0]{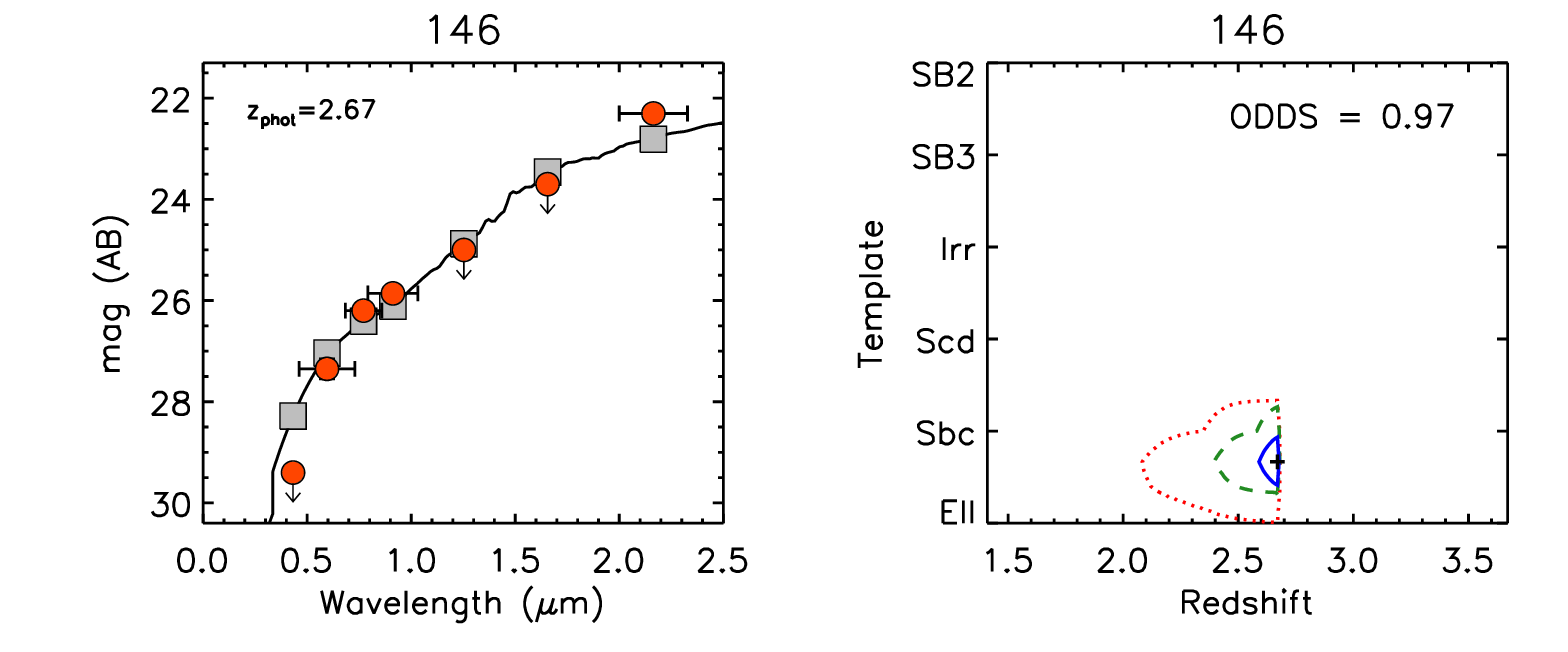}}}
\\
\parbox{12.3cm}{\resizebox{\hsize}{!}{\includegraphics[angle=0]{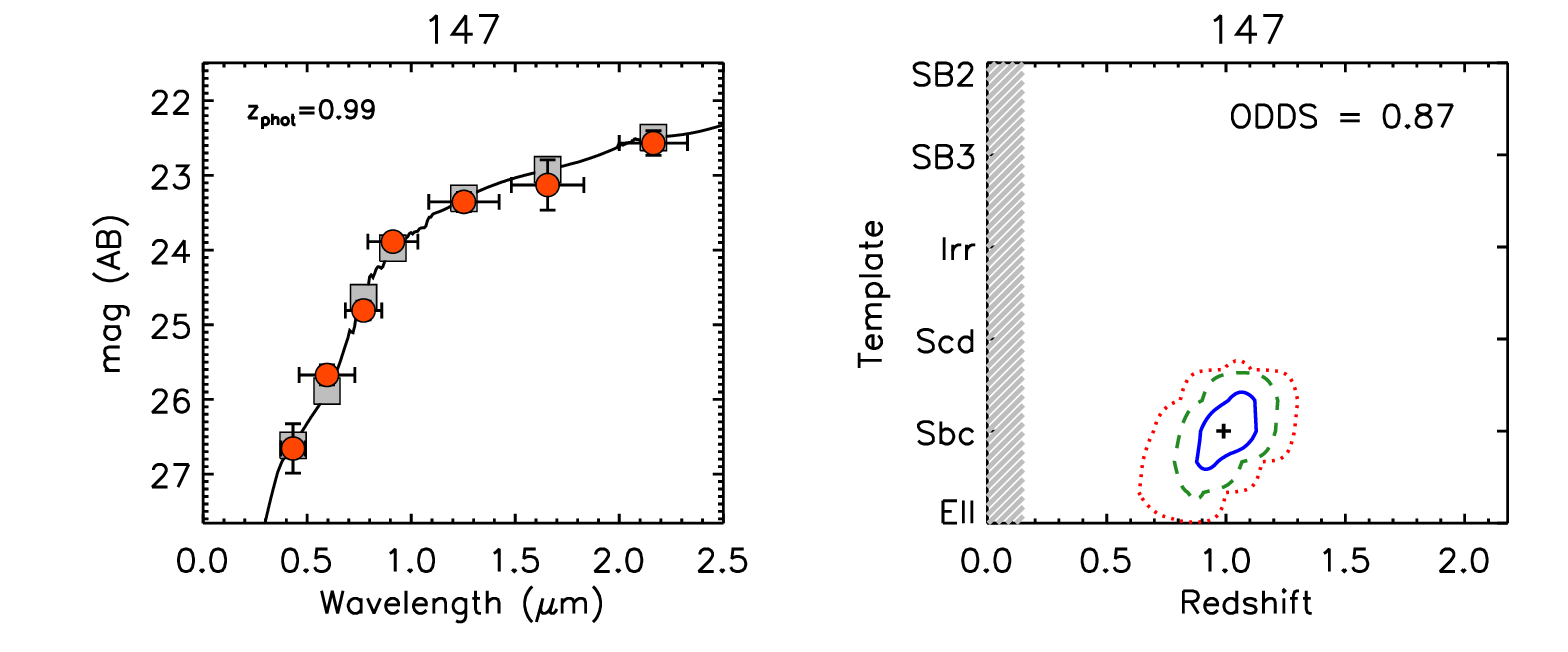}}}
\\
\label{figure:fig_cont_2}
\end{figure*}

\newpage
\begin{figure*}
\parbox{12.3cm}{\resizebox{\hsize}{!}{\includegraphics[angle=0]{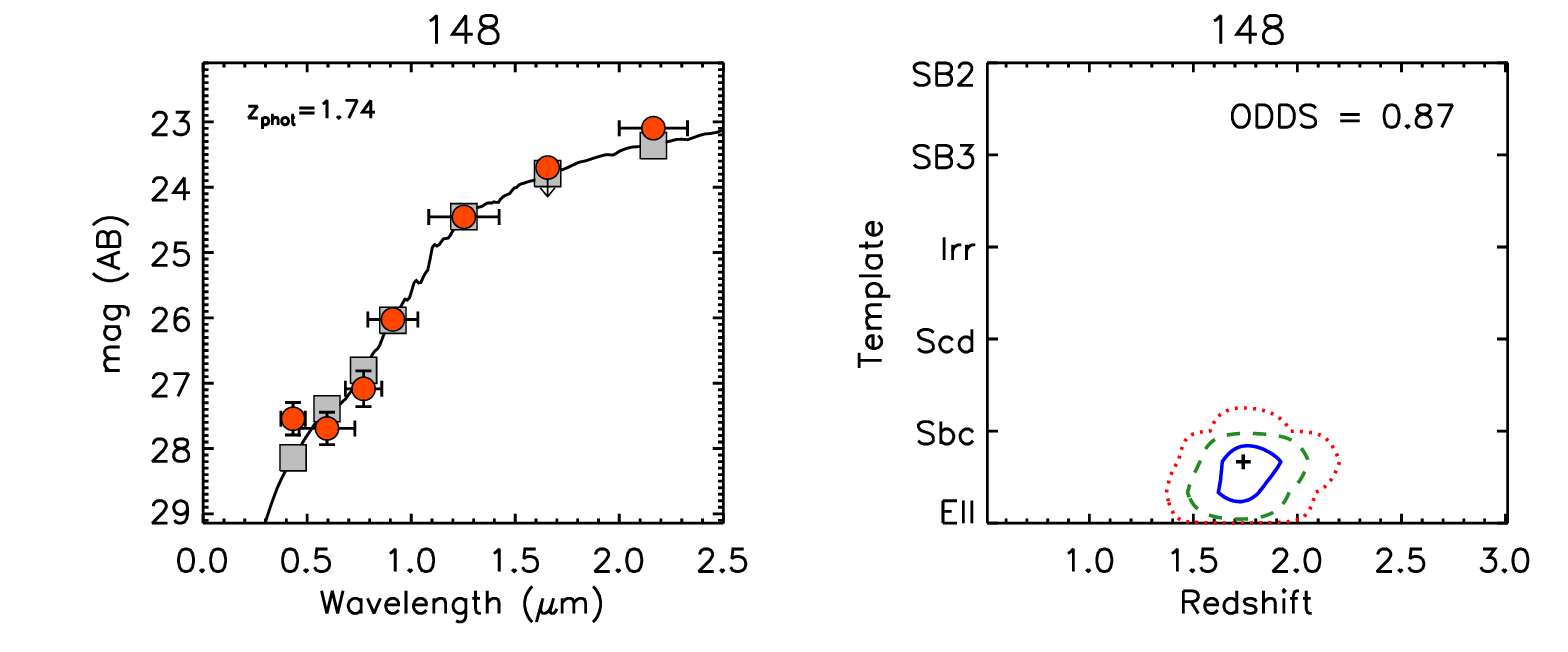}}}
\\
\parbox{12.3cm}{\resizebox{\hsize}{!}{\includegraphics[angle=0]{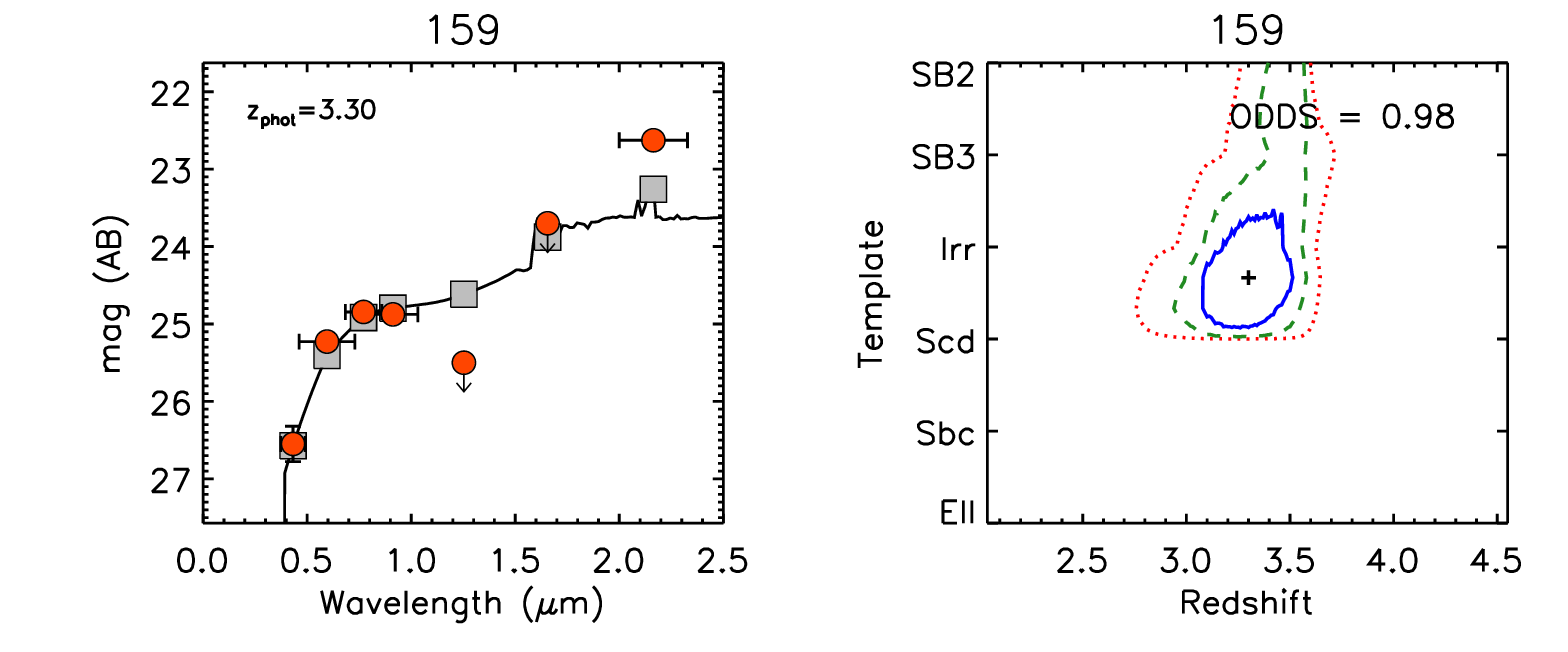}}}
\\
\parbox{12.3cm}{\resizebox{\hsize}{!}{\includegraphics[angle=0]{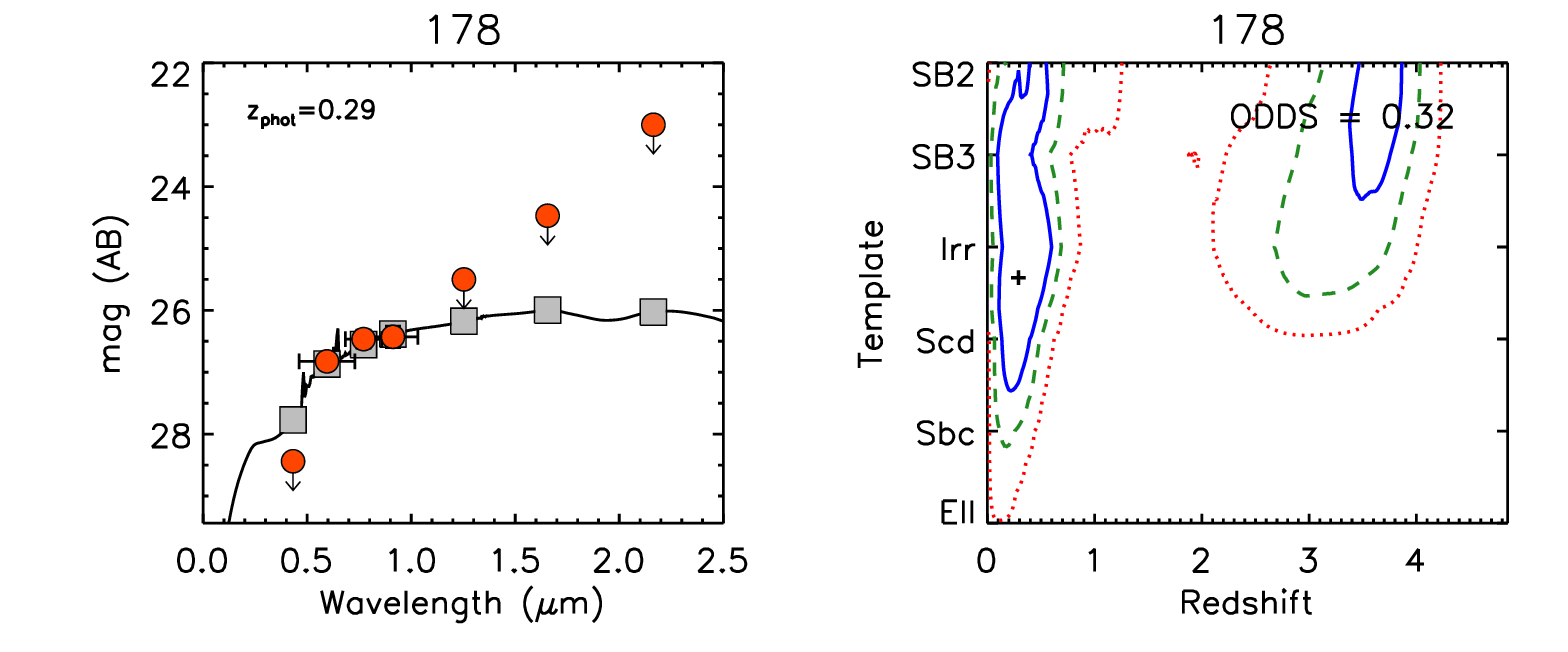}}}
\\
\parbox{12.3cm}{\resizebox{\hsize}{!}{\includegraphics[angle=0]{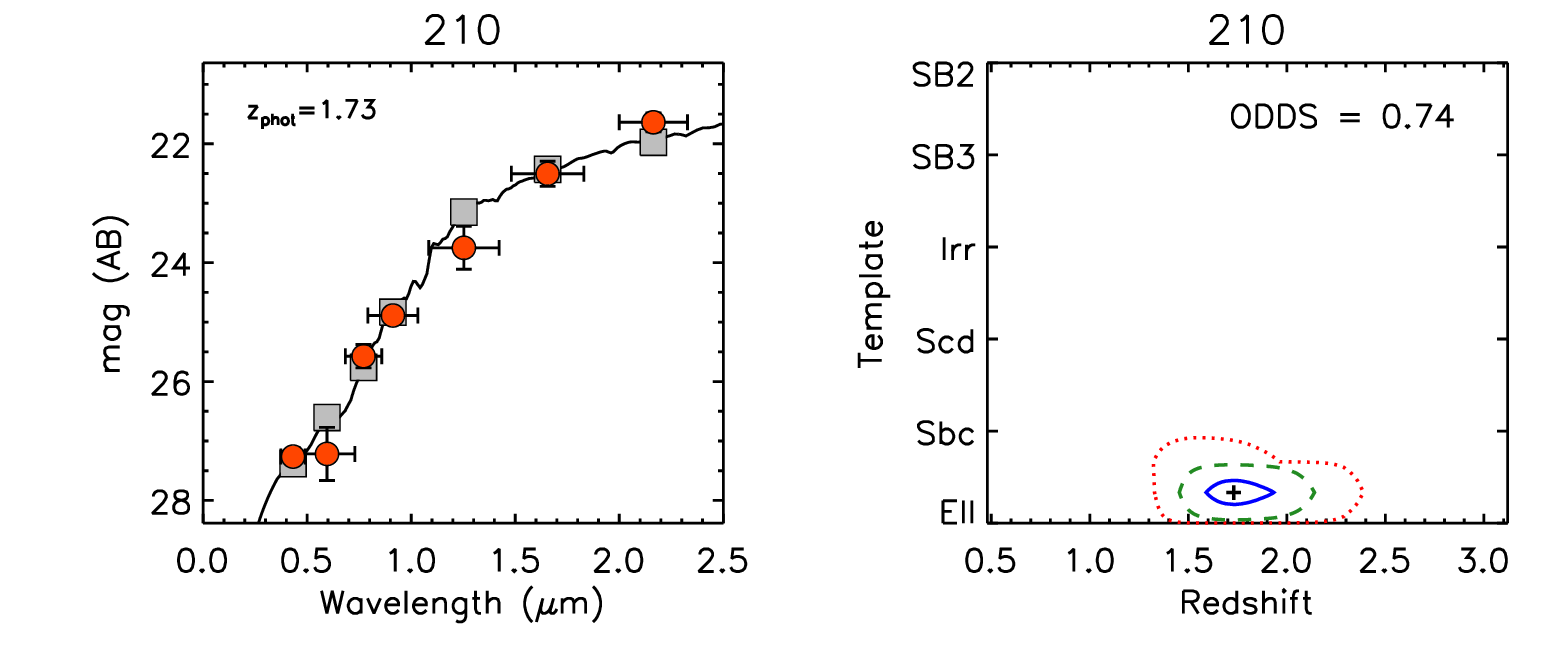}}}
\\
\label{figure:fig_cont_3}
\end{figure*}

\newpage
\begin{figure*}
\parbox{12.3cm}{\resizebox{\hsize}{!}{\includegraphics[angle=0]{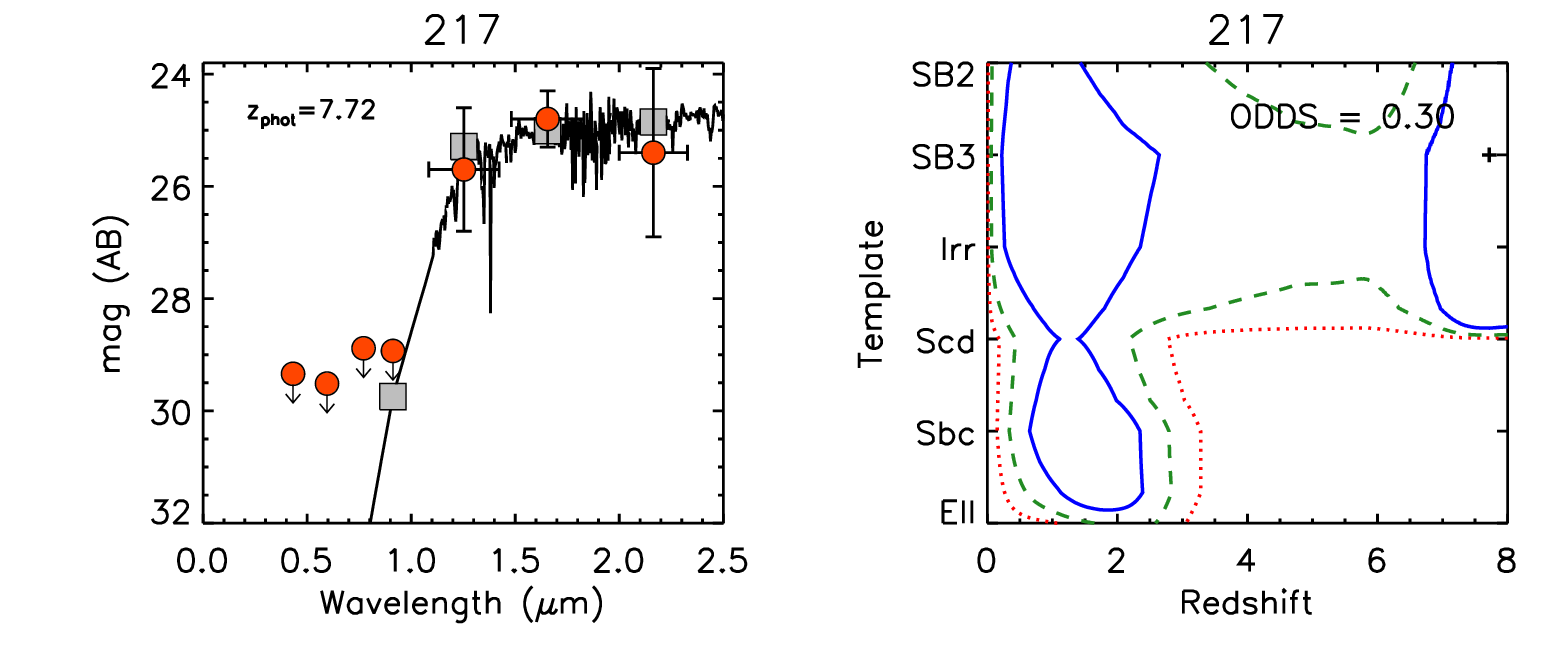}}}
\\
\parbox{12.3cm}{\resizebox{\hsize}{!}{\includegraphics[angle=0]{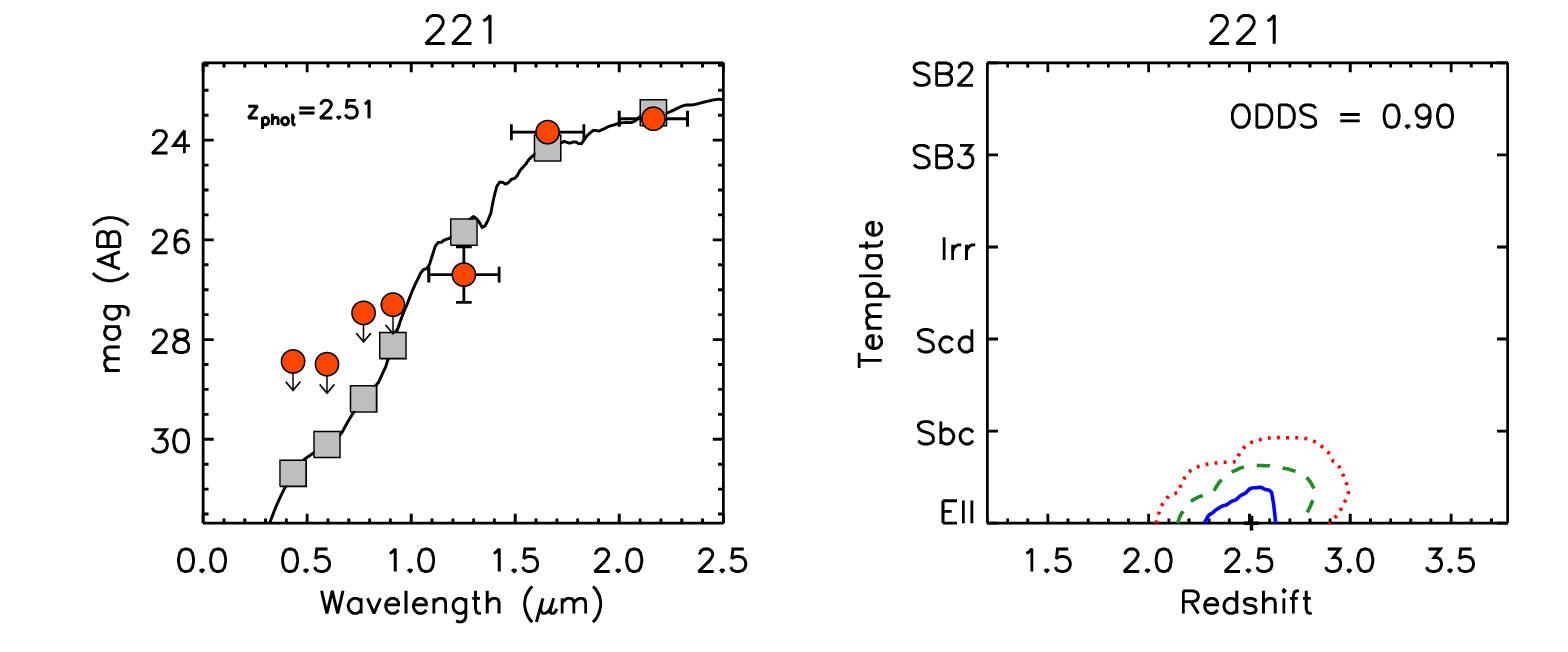}}}
\\
\parbox{12.3cm}{\resizebox{\hsize}{!}{\includegraphics[angle=0]{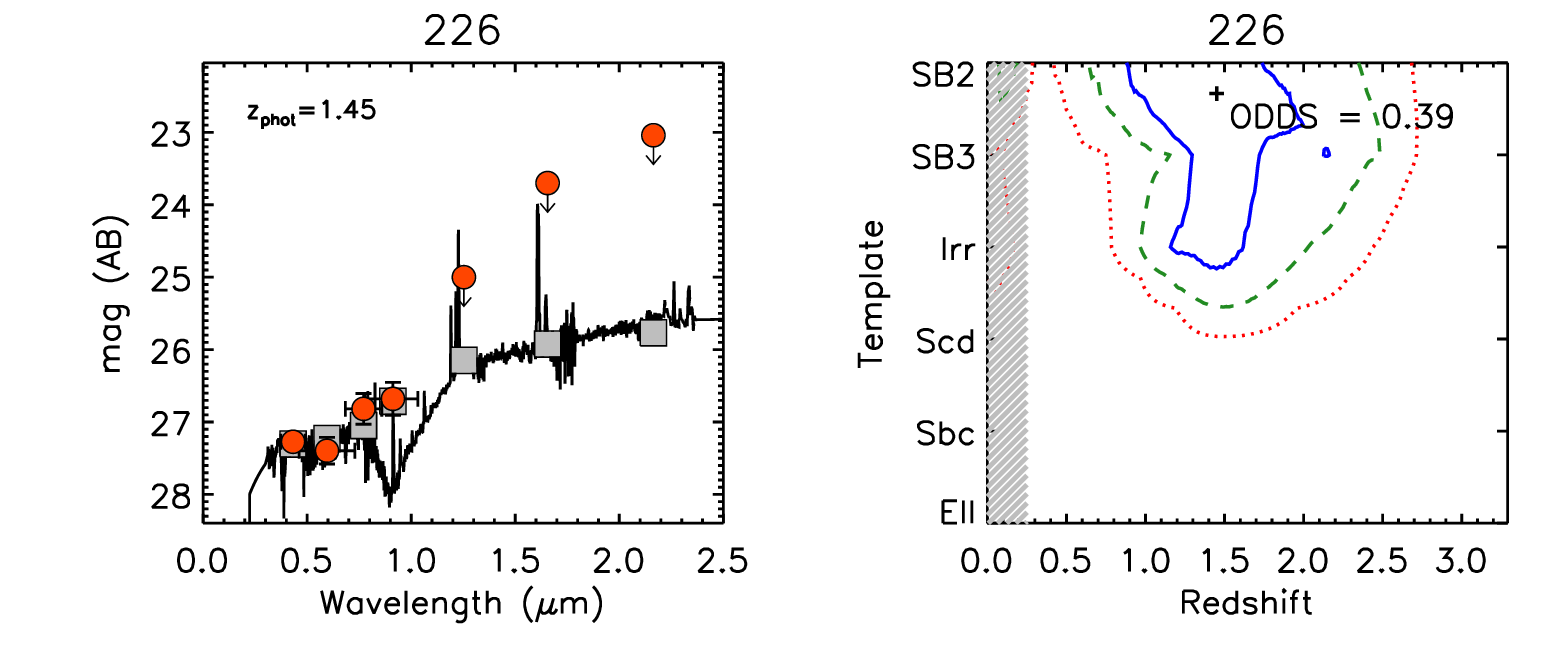}}}
\\
\parbox{12.3cm}{\resizebox{\hsize}{!}{\includegraphics[angle=0]{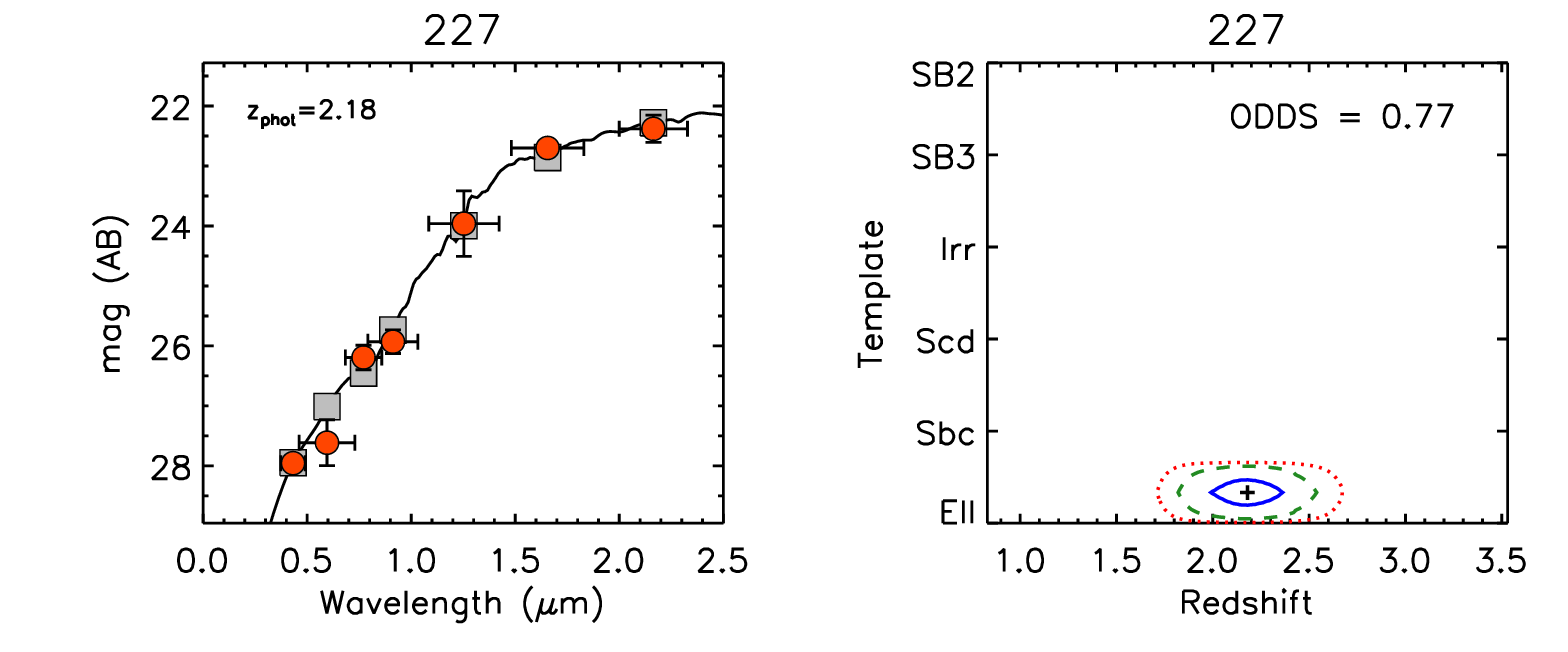}}}
\\
\label{figure:fig_cont_4}
\end{figure*}

\newpage
\begin{figure*}
\parbox{12.3cm}{\resizebox{\hsize}{!}{\includegraphics[angle=0]{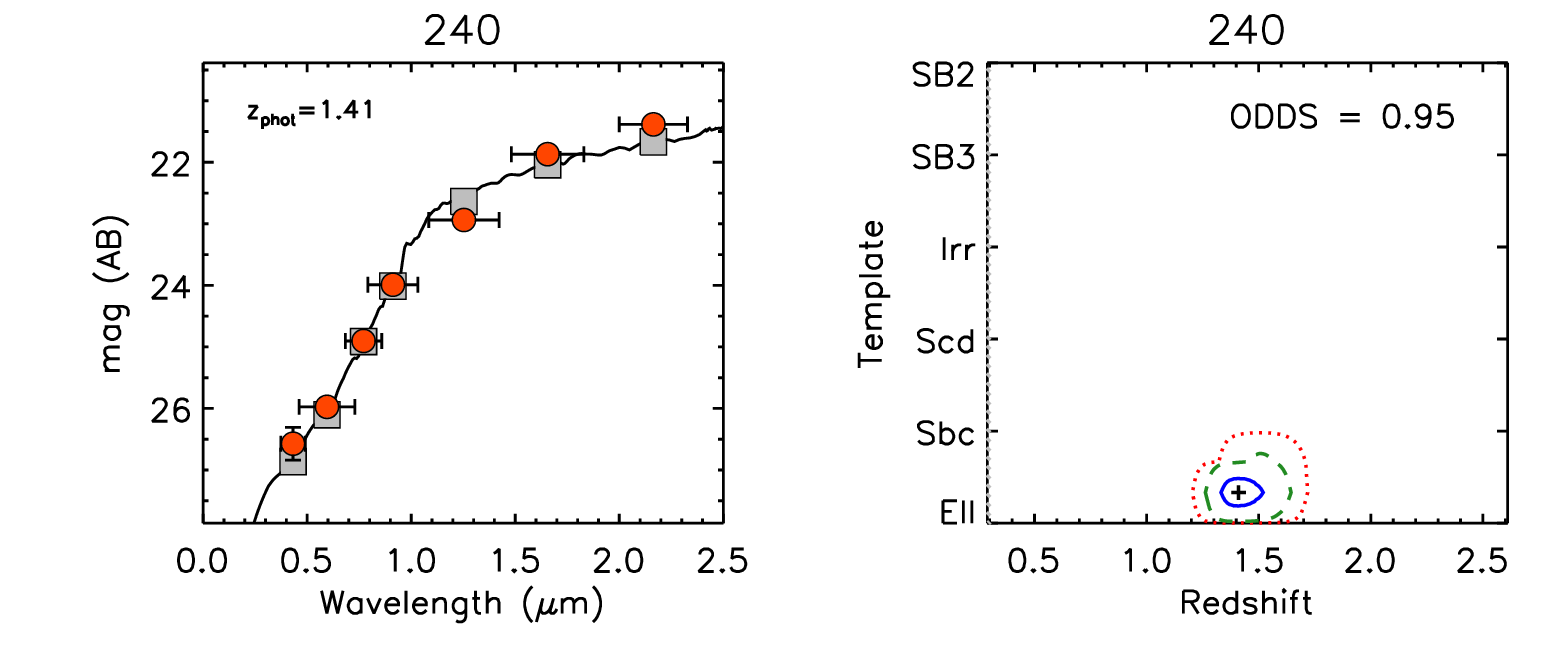}}}
\\
\parbox{12.3cm}{\resizebox{\hsize}{!}{\includegraphics[angle=0]{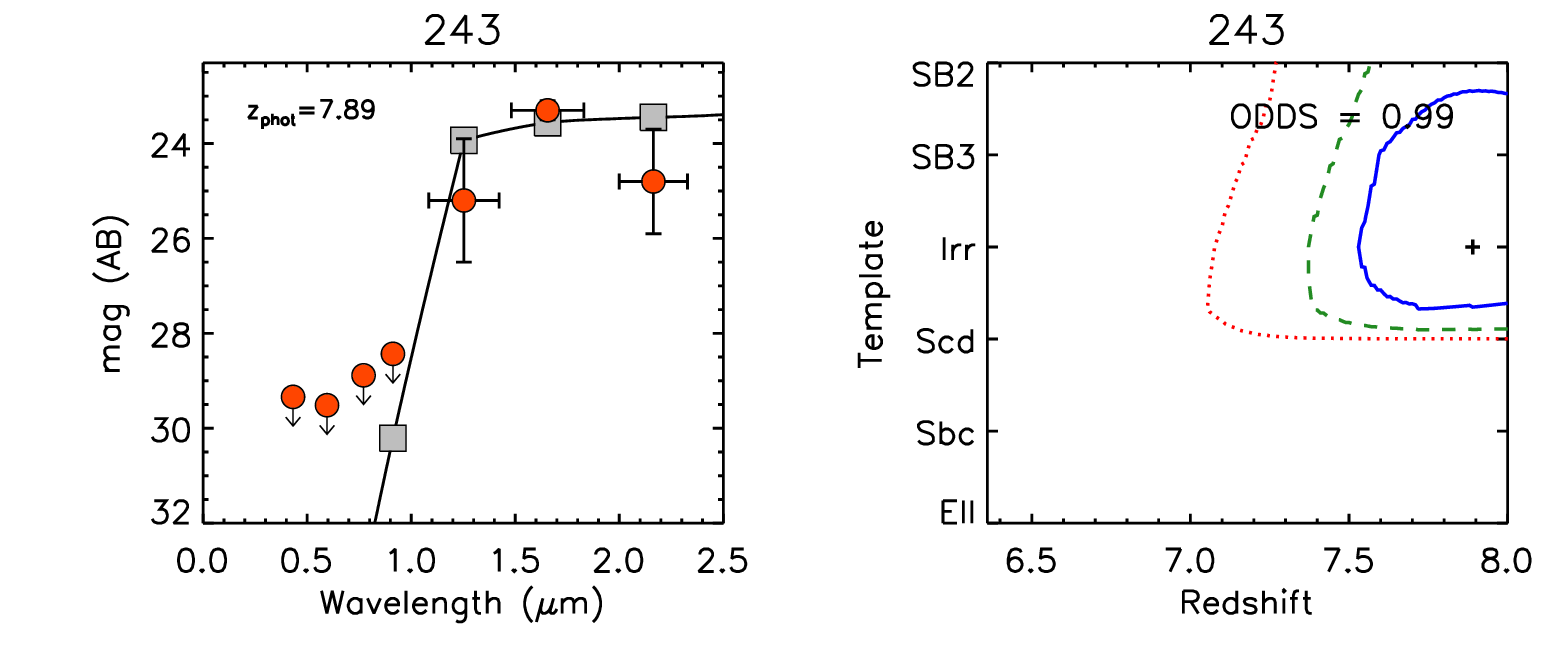}}}
\\
\parbox{12.3cm}{\resizebox{\hsize}{!}{\includegraphics[angle=0]{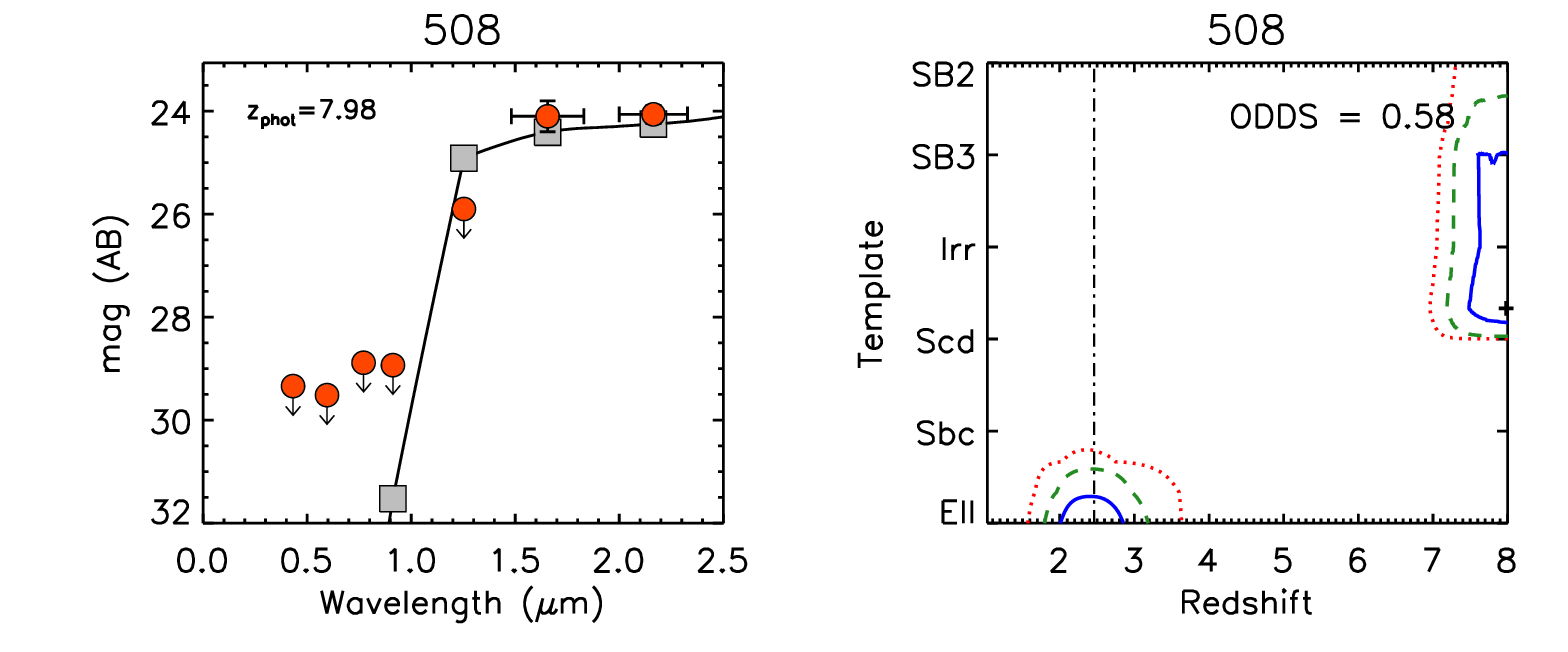}}}
\\
\parbox{12.3cm}{\resizebox{\hsize}{!}{\includegraphics[angle=0]{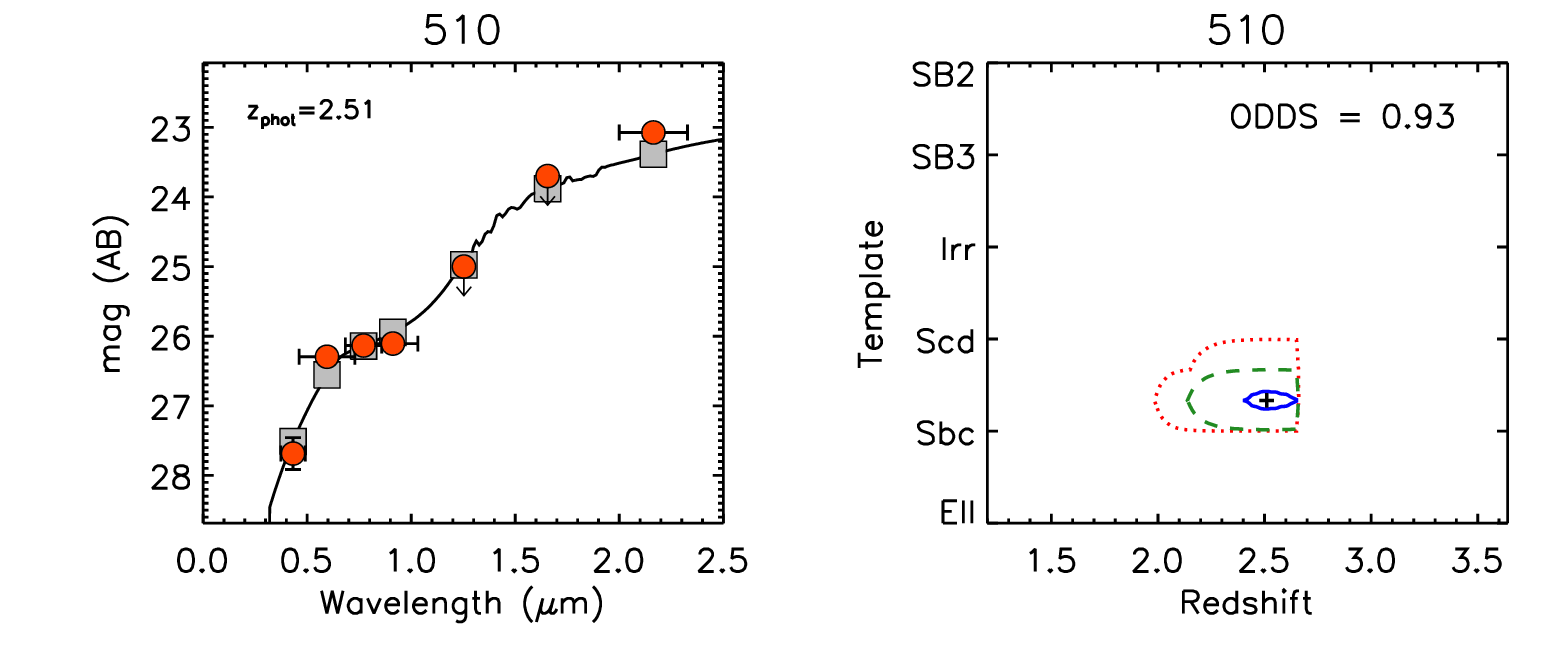}}}
\\
\label{figure:fig_cont_5}
\end{figure*}

\newpage
\begin{figure*}
\parbox{12.3cm}{\resizebox{\hsize}{!}{\includegraphics[angle=0]{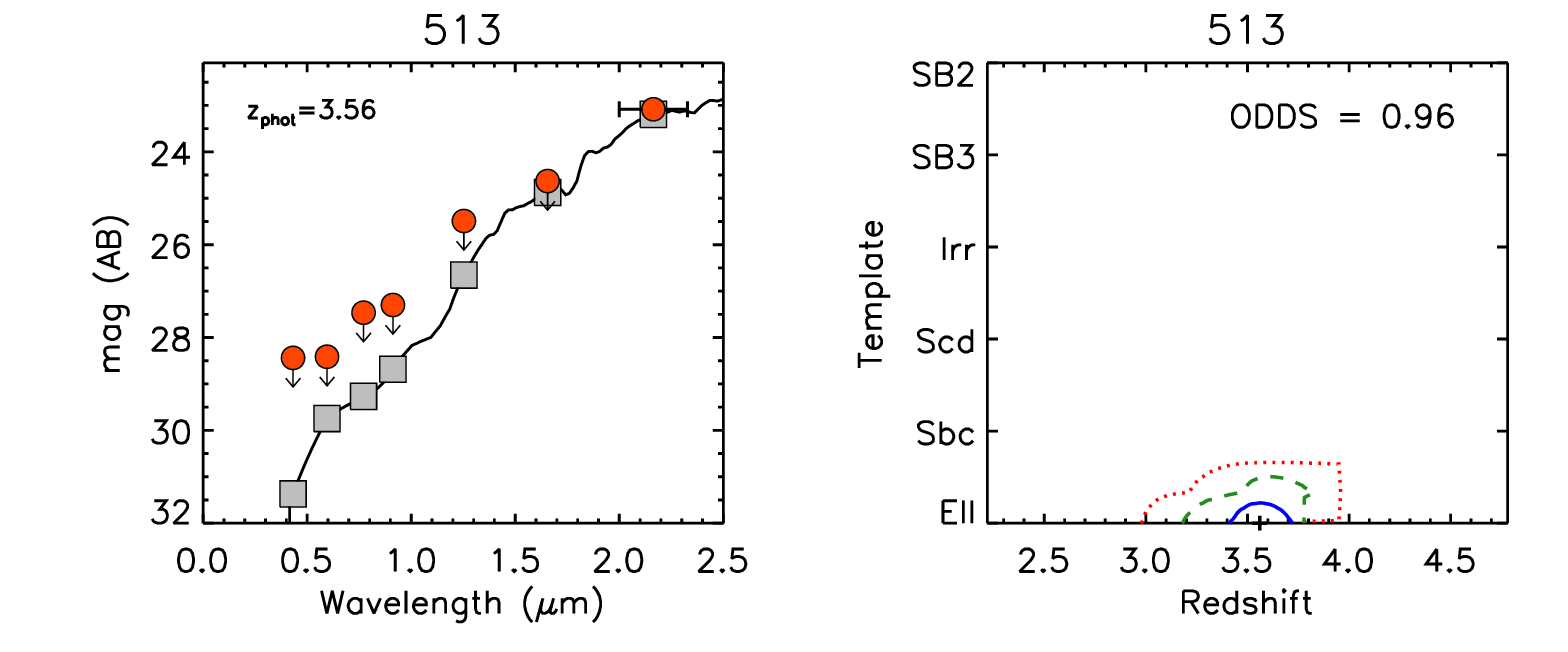}}}
\\
\parbox{12.3cm}{\resizebox{\hsize}{!}{\includegraphics[angle=0]{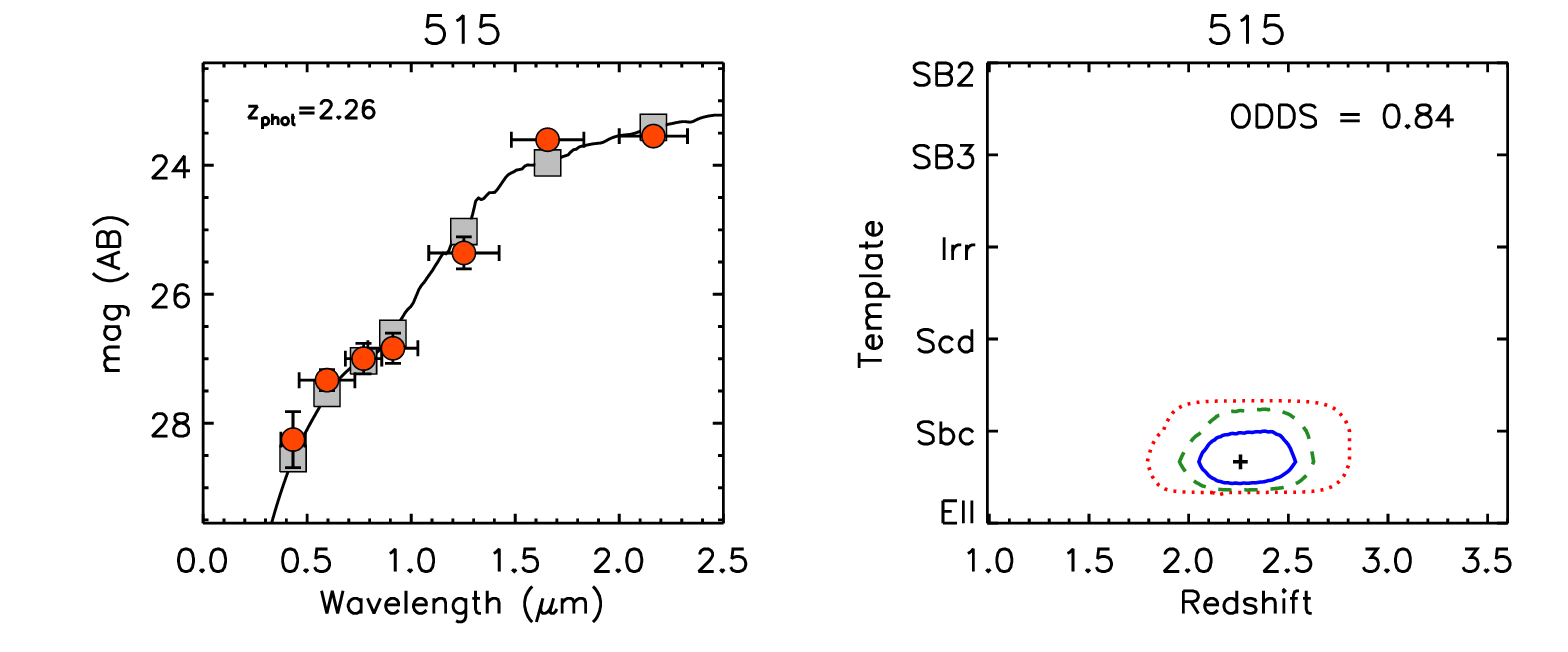}}}
\\
\parbox{12.3cm}{\resizebox{\hsize}{!}{\includegraphics[angle=0]{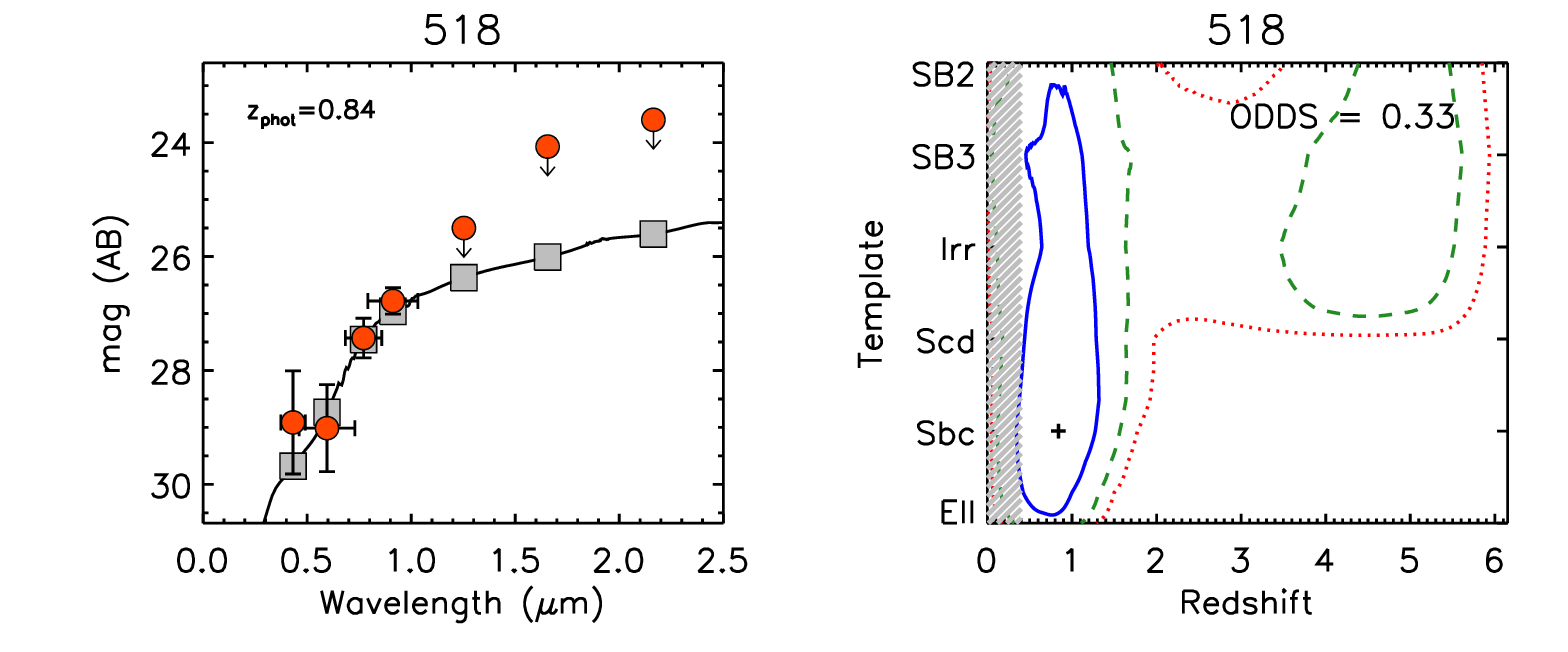}}}
\\
\parbox{12.3cm}{\resizebox{\hsize}{!}{\includegraphics[angle=0]{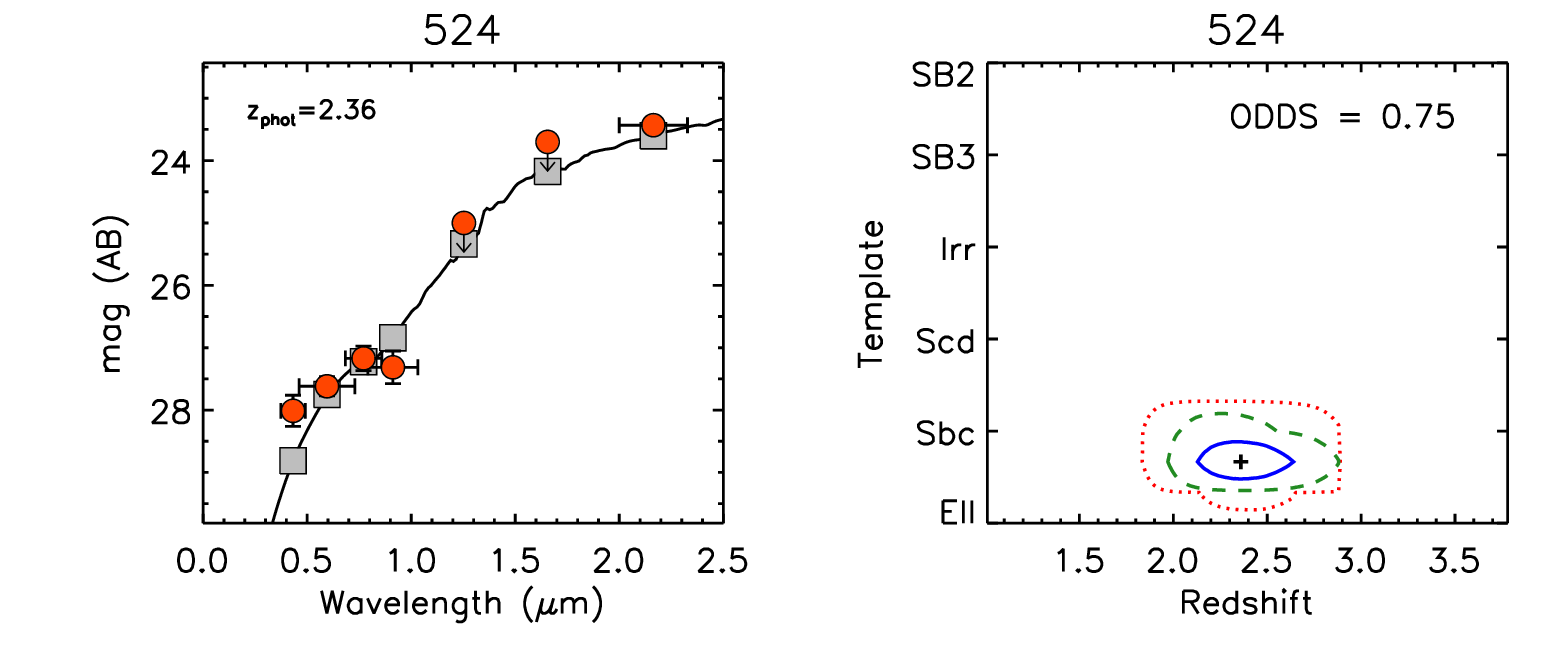}}}
\\
\label{figure:fig_cont_6}
\end{figure*}

\newpage
\begin{figure*}
\parbox{12.3cm}{\resizebox{\hsize}{!}{\includegraphics[angle=0]{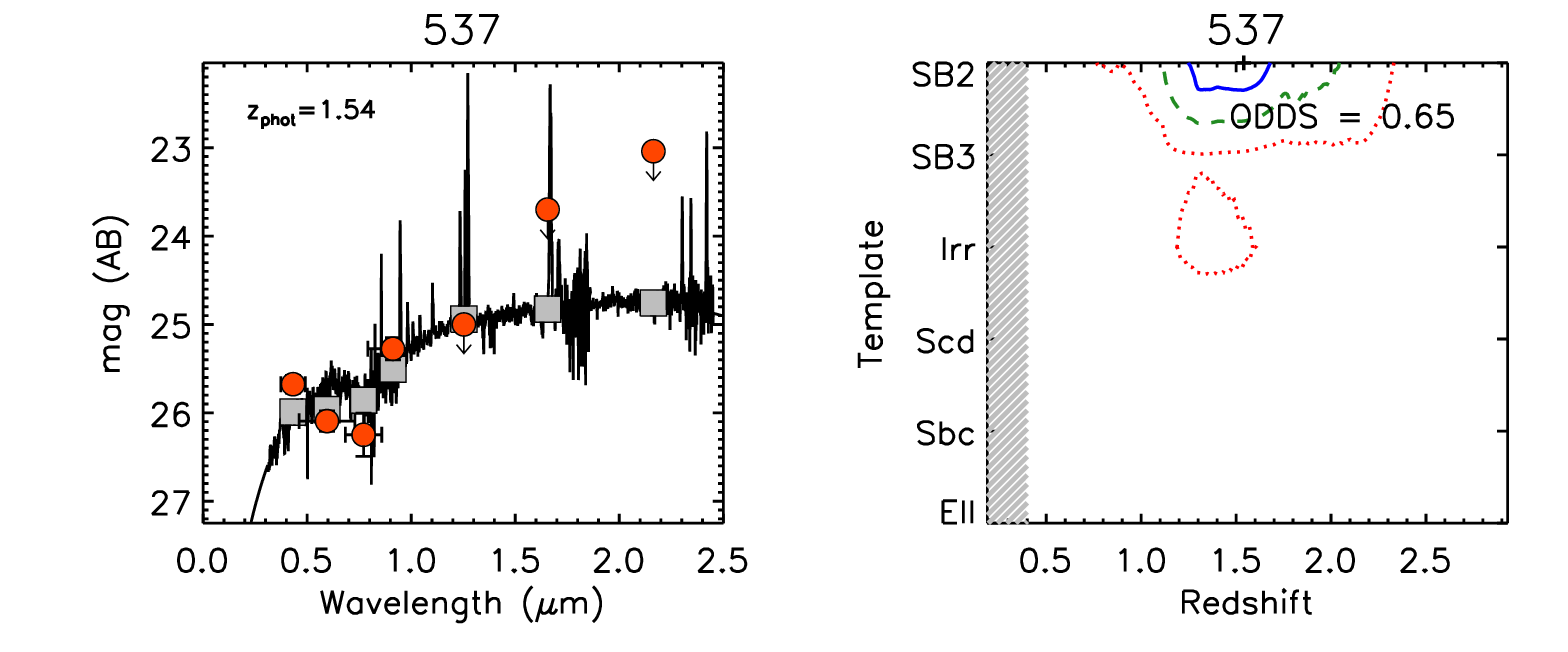}}}
\\
\parbox{12.3cm}{\resizebox{\hsize}{!}{\includegraphics[angle=0]{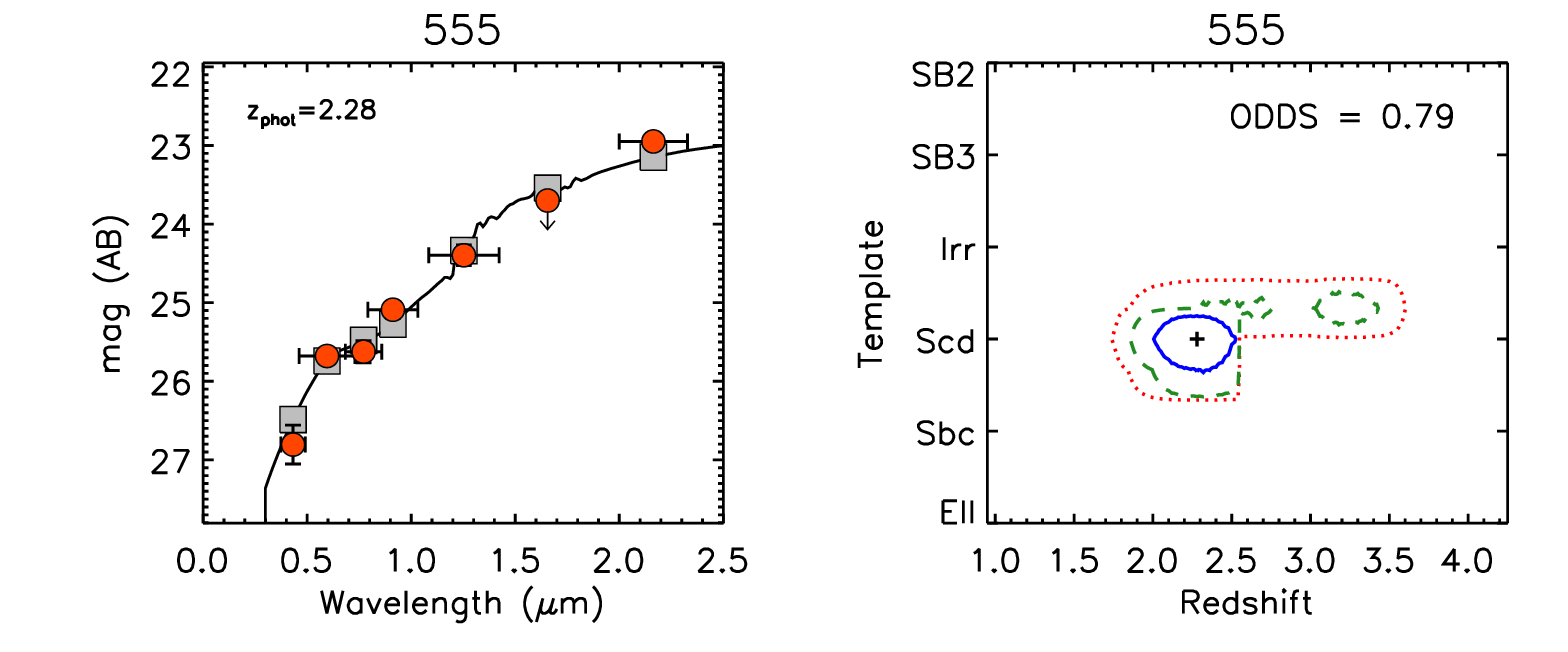}}}
\\
\parbox{12.3cm}{\resizebox{\hsize}{!}{\includegraphics[angle=0]{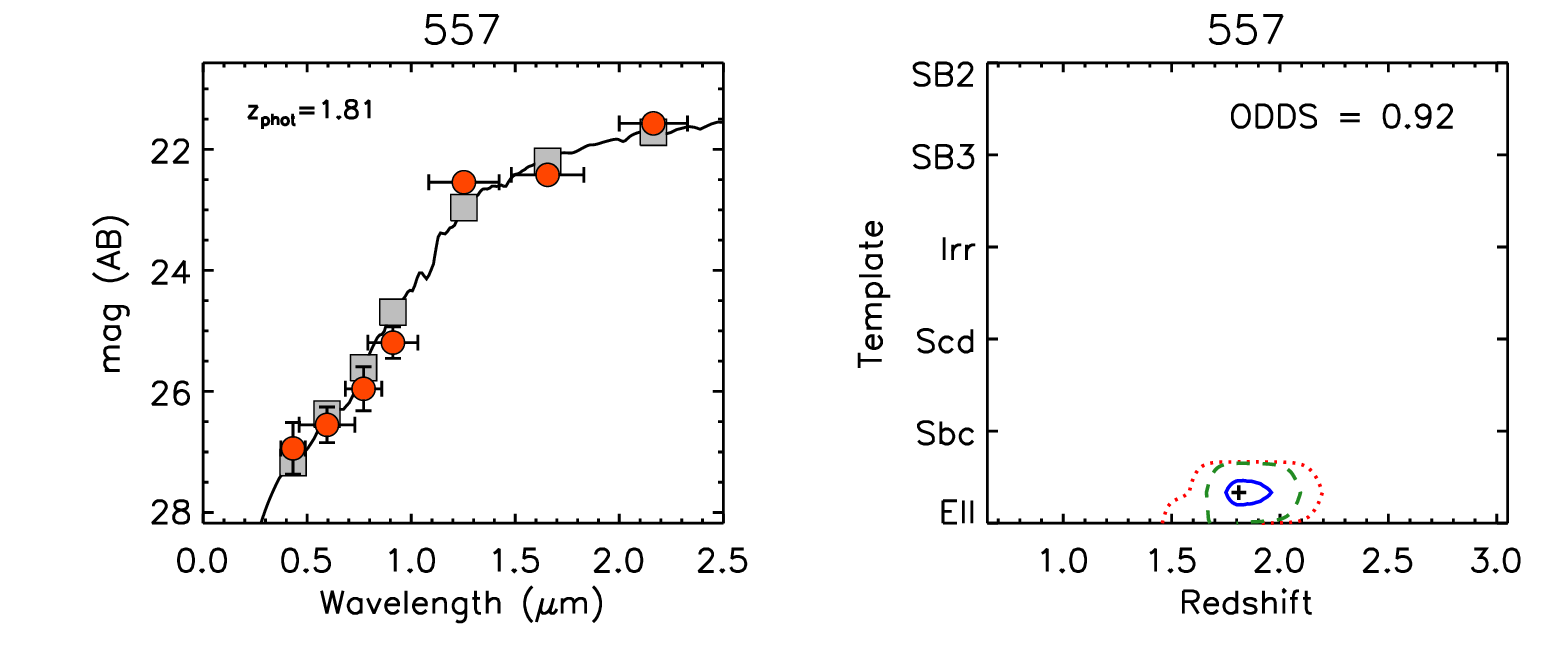}}}
\\
\parbox{12.3cm}{\resizebox{\hsize}{!}{\includegraphics[angle=0]{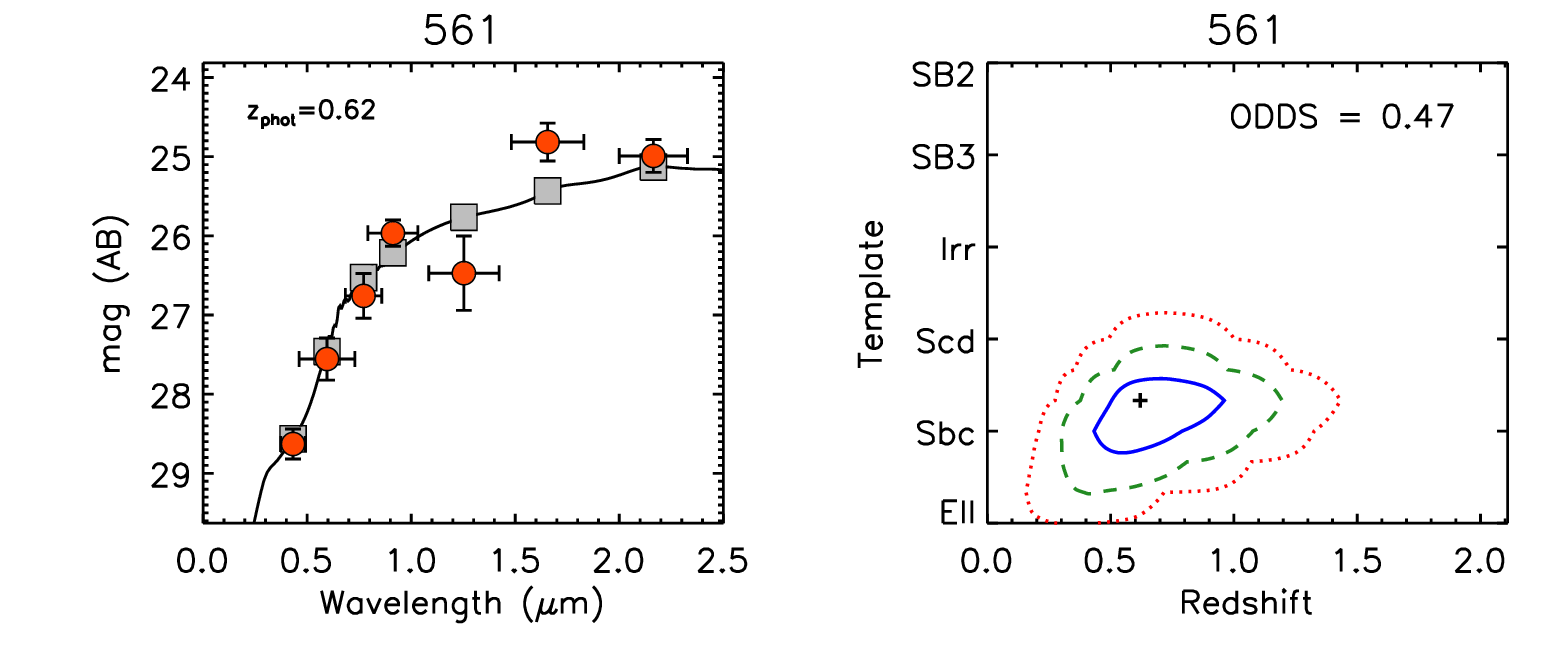}}}
\\
\label{figure:fig_cont_7}
\end{figure*}

\newpage
\begin{figure*}
\parbox{12.3cm}{\resizebox{\hsize}{!}{\includegraphics[angle=0]{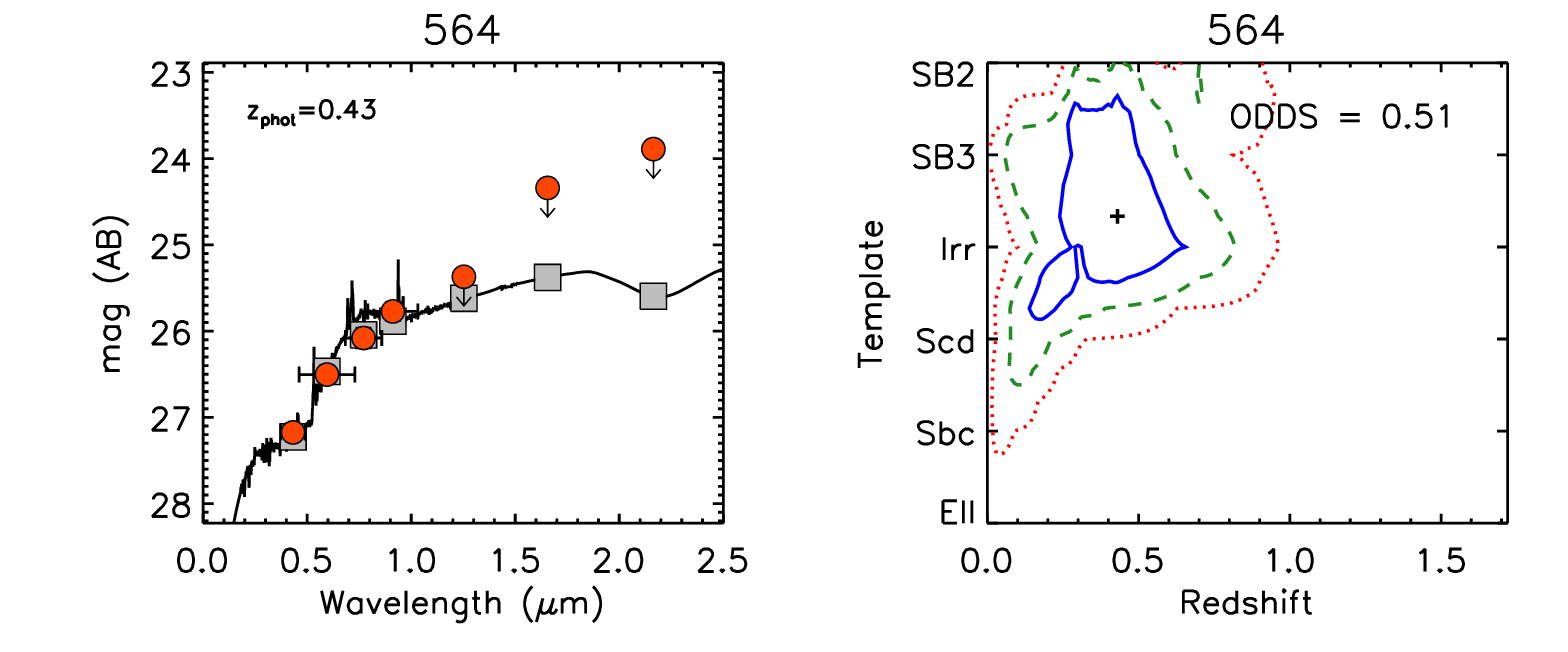}}}
\\
\parbox{12.3cm}{\resizebox{\hsize}{!}{\includegraphics[angle=0]{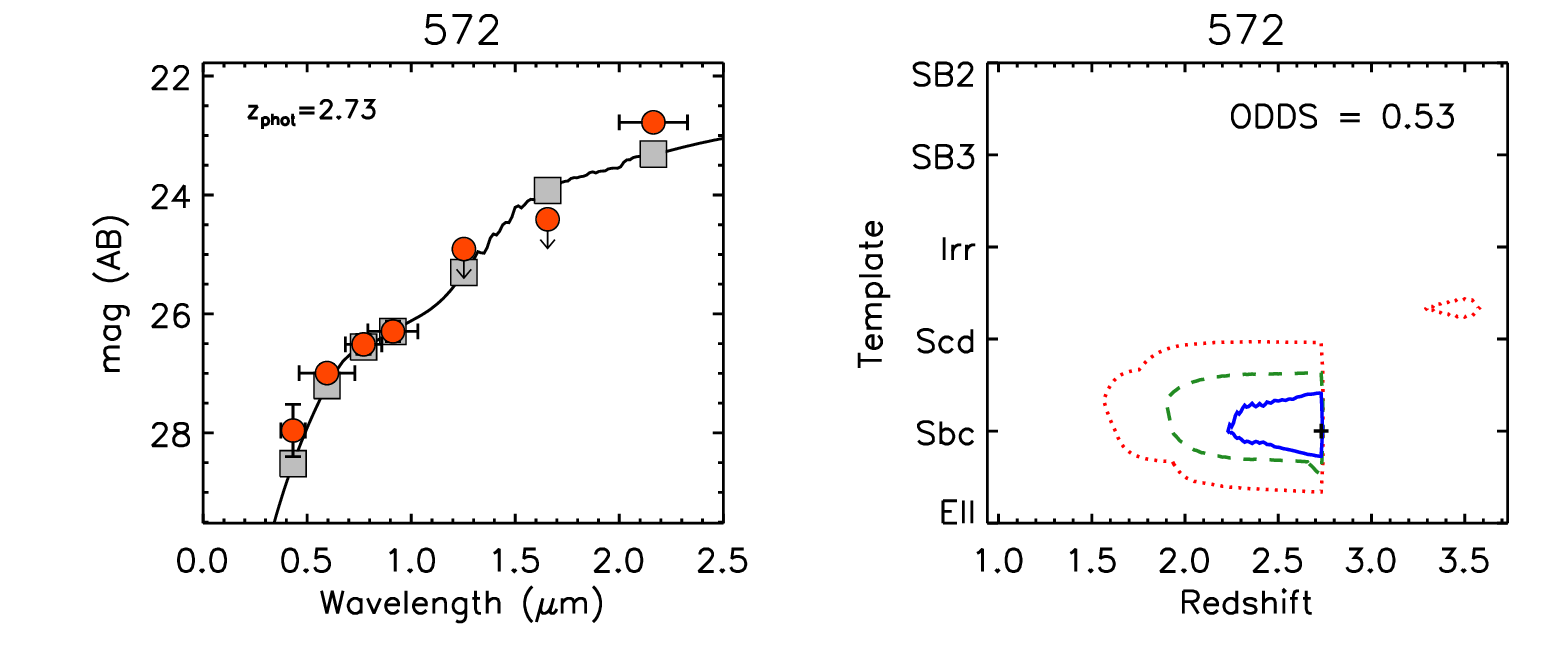}}}
\\
\parbox{12.3cm}{\resizebox{\hsize}{!}{\includegraphics[angle=0]{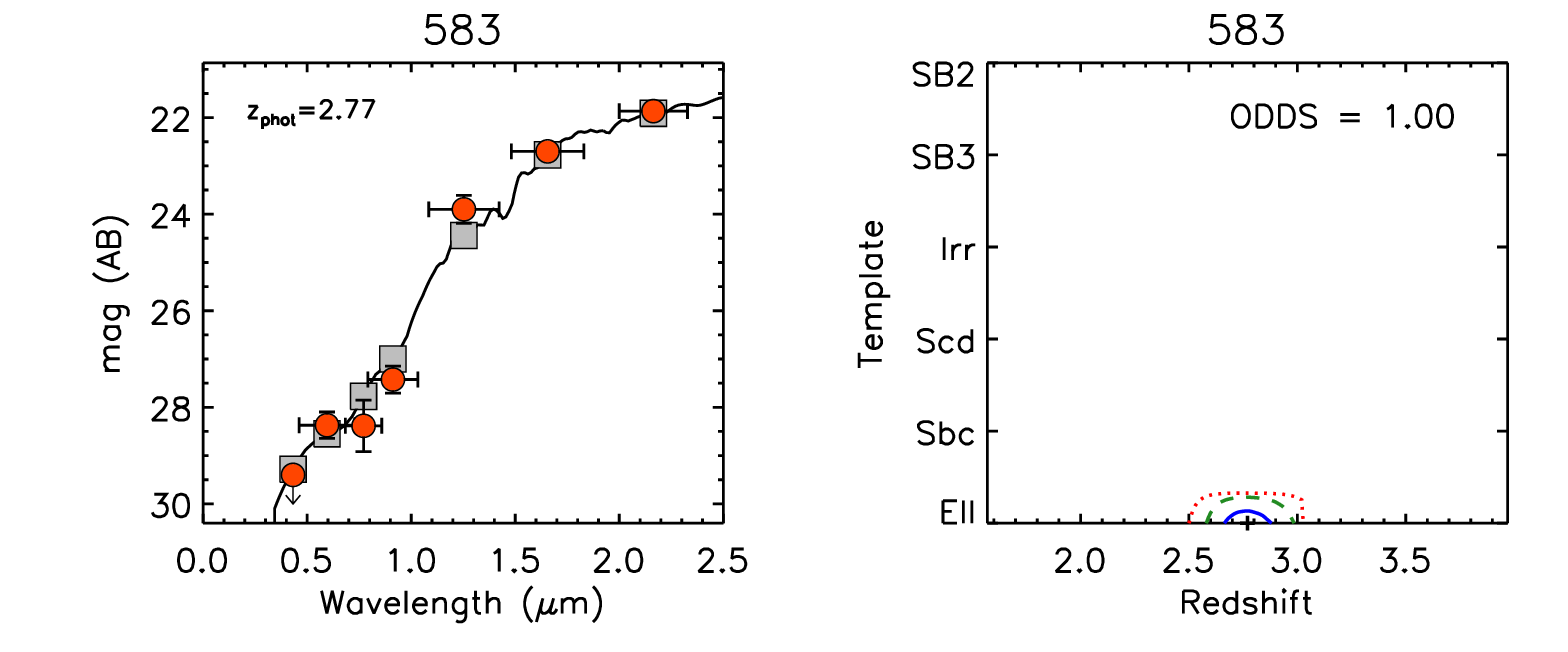}}}
\\
\parbox{12.3cm}{\resizebox{\hsize}{!}{\includegraphics[angle=0]{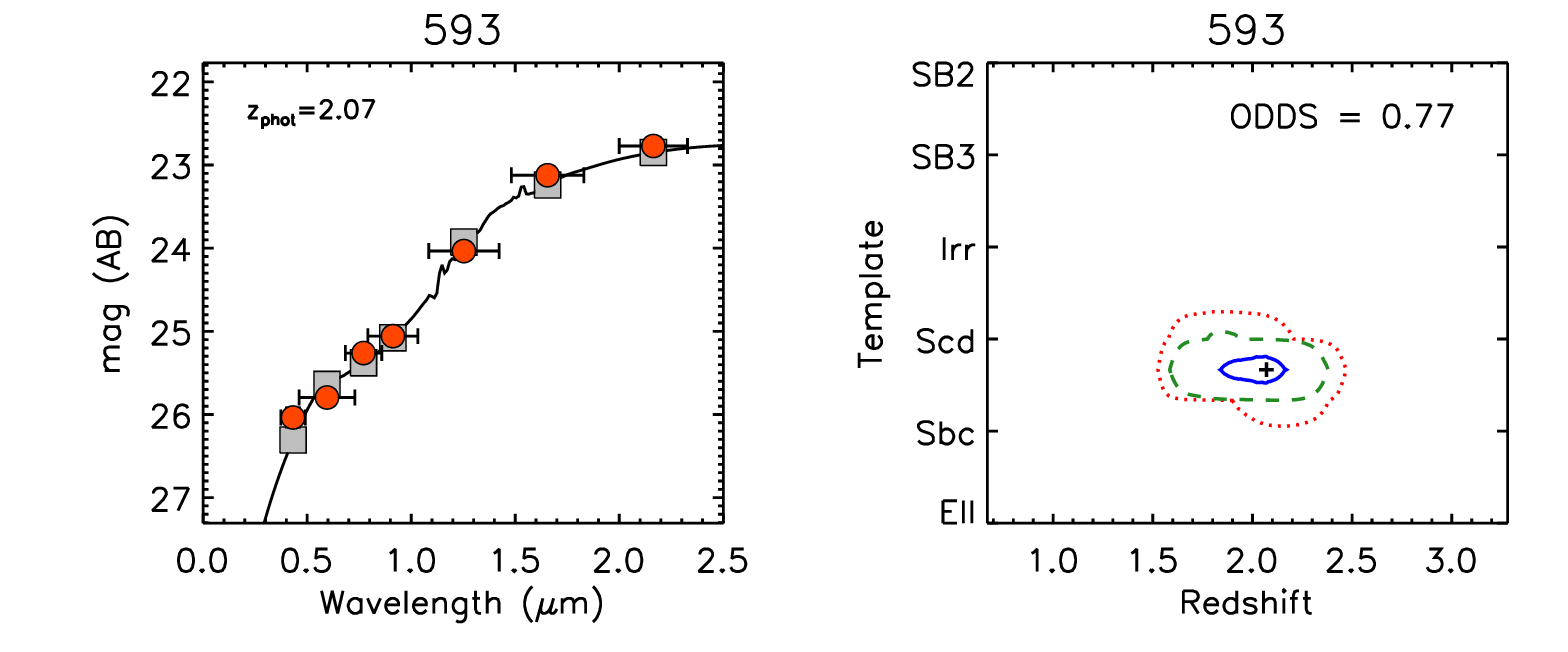}}}
\\
\label{figure:fig_cont_8}
\end{figure*}

\newpage
\begin{figure*}
\parbox{12.3cm}{\resizebox{\hsize}{!}{\includegraphics[angle=0]{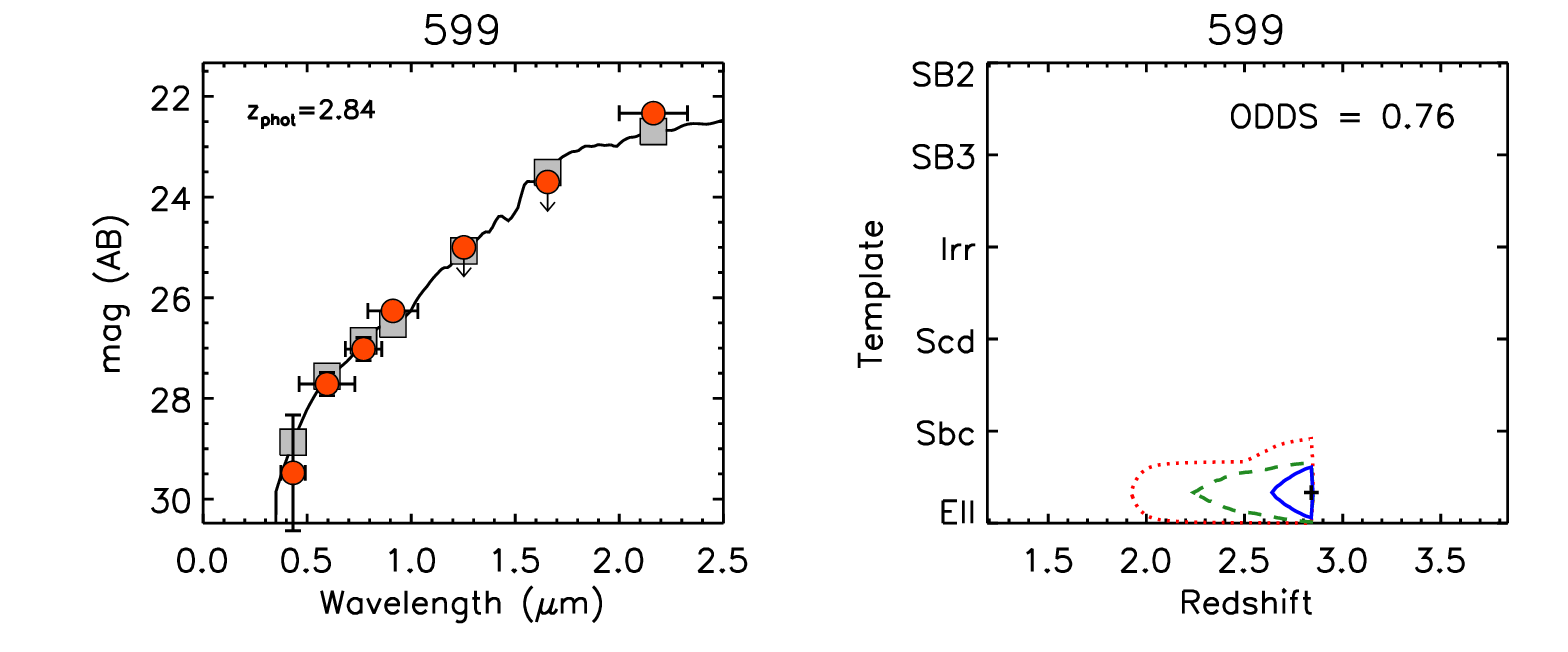}}}
\\
\parbox{12.3cm}{\resizebox{\hsize}{!}{\includegraphics[angle=0]{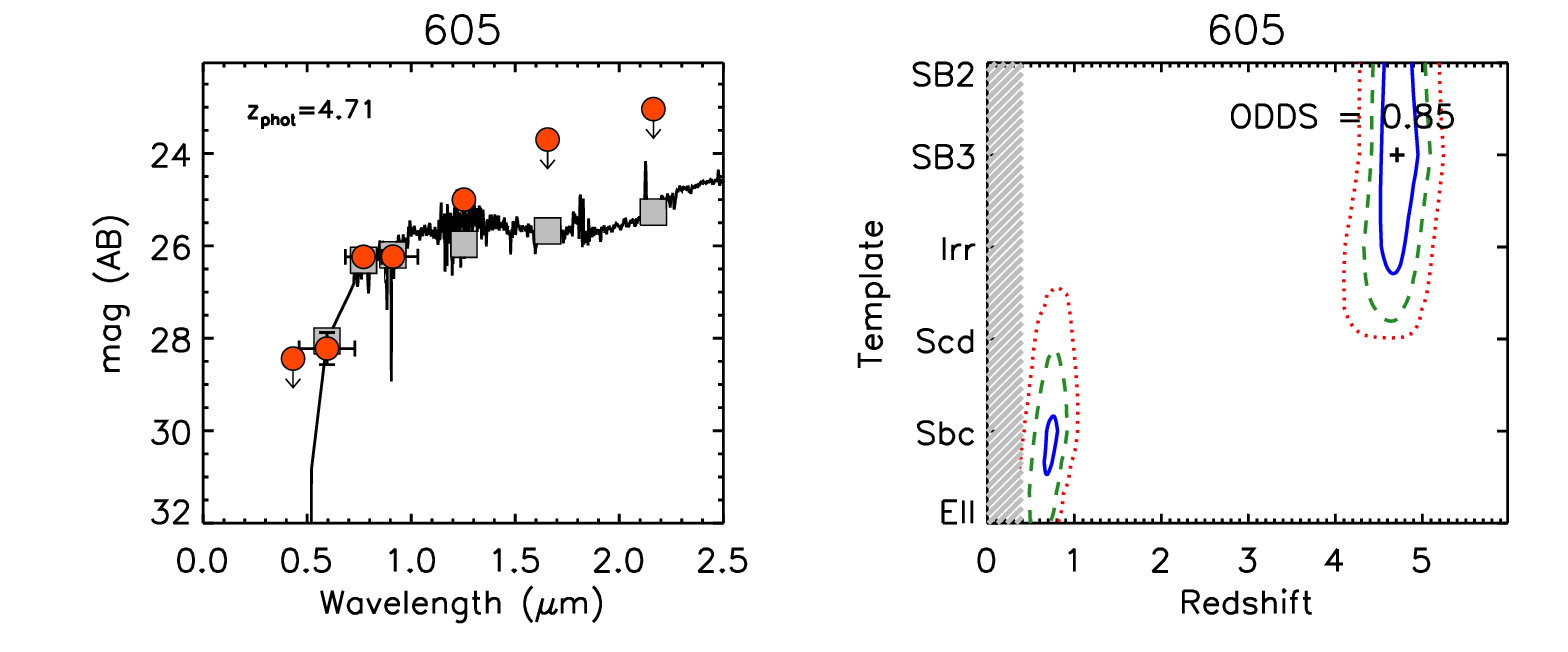}}}
\\
\parbox{12.3cm}{\resizebox{\hsize}{!}{\includegraphics[angle=0]{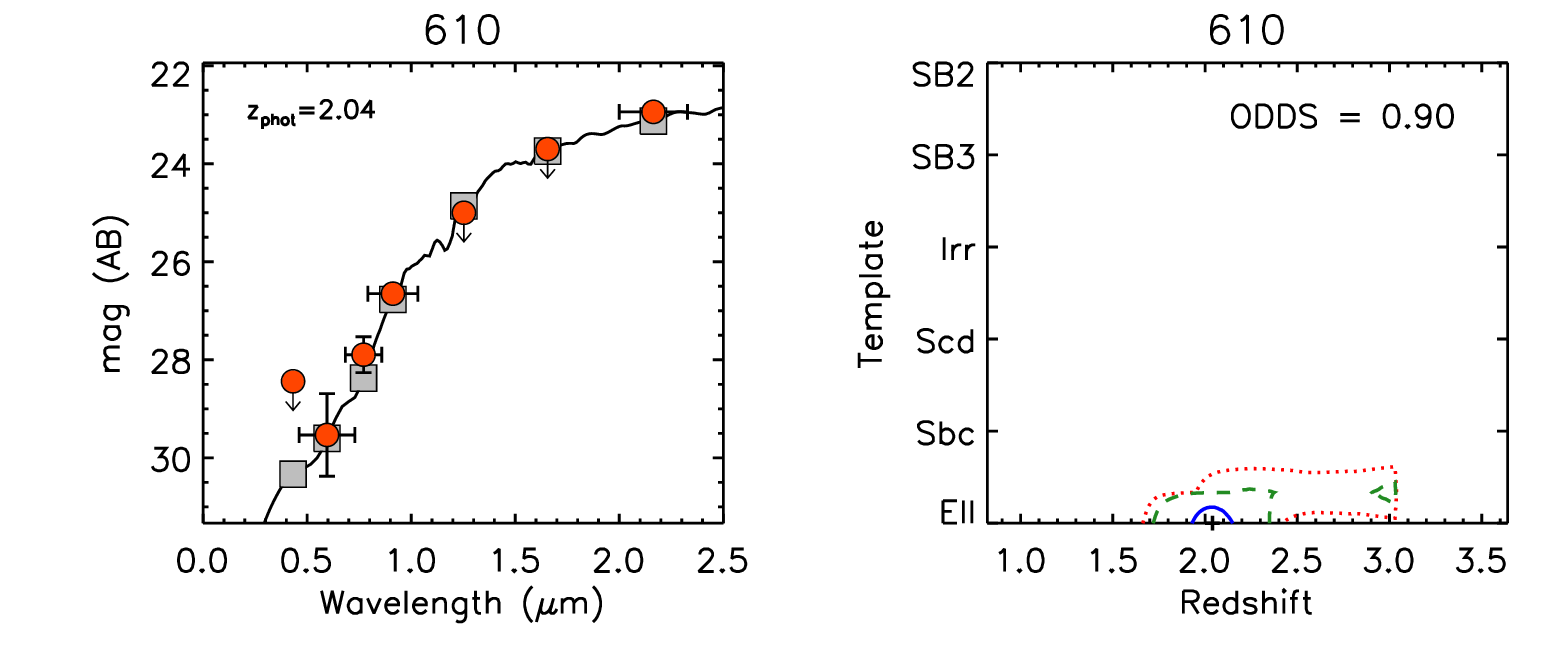}}}
\\
\parbox{12.3cm}{\resizebox{\hsize}{!}{\includegraphics[angle=0]{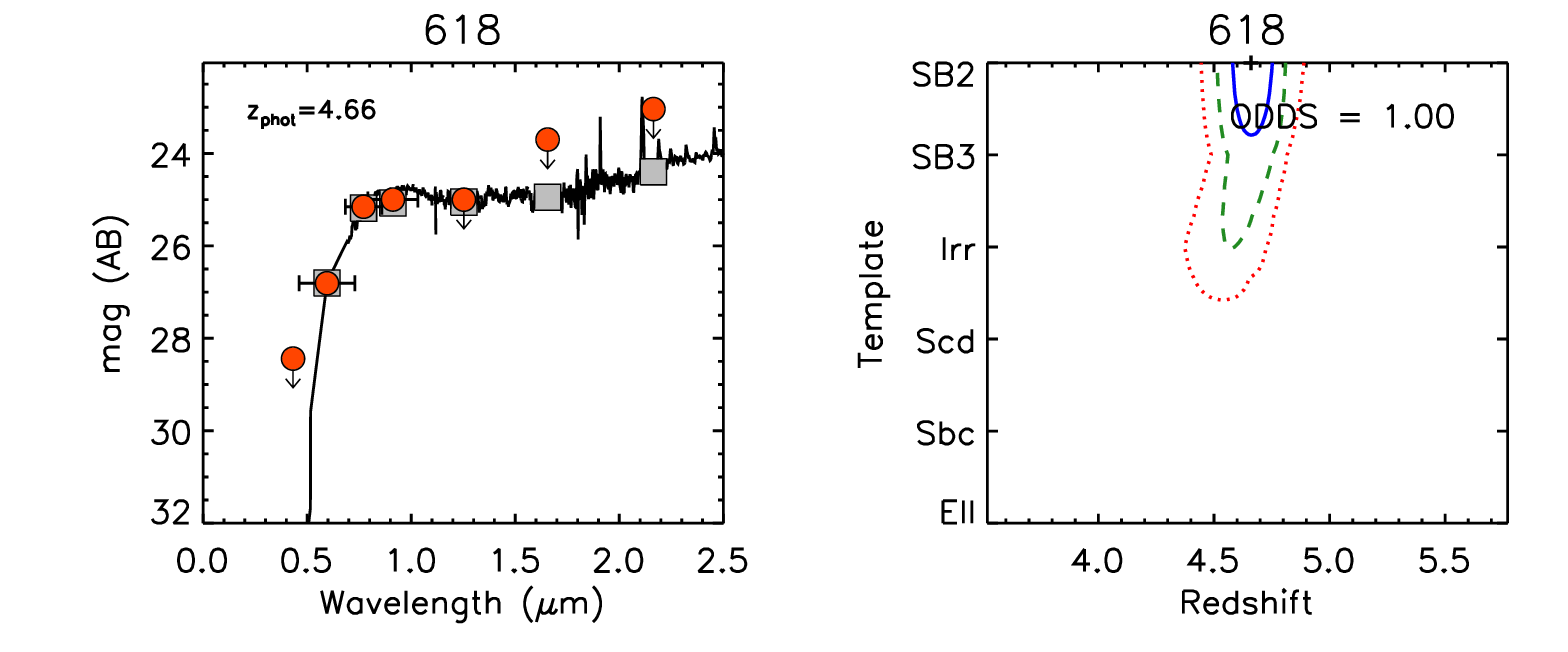}}}
\\
\label{figure:fig_cont_9}
\end{figure*}

\newpage
\begin{figure*}
\parbox{12.3cm}{\resizebox{\hsize}{!}{\includegraphics[angle=0]{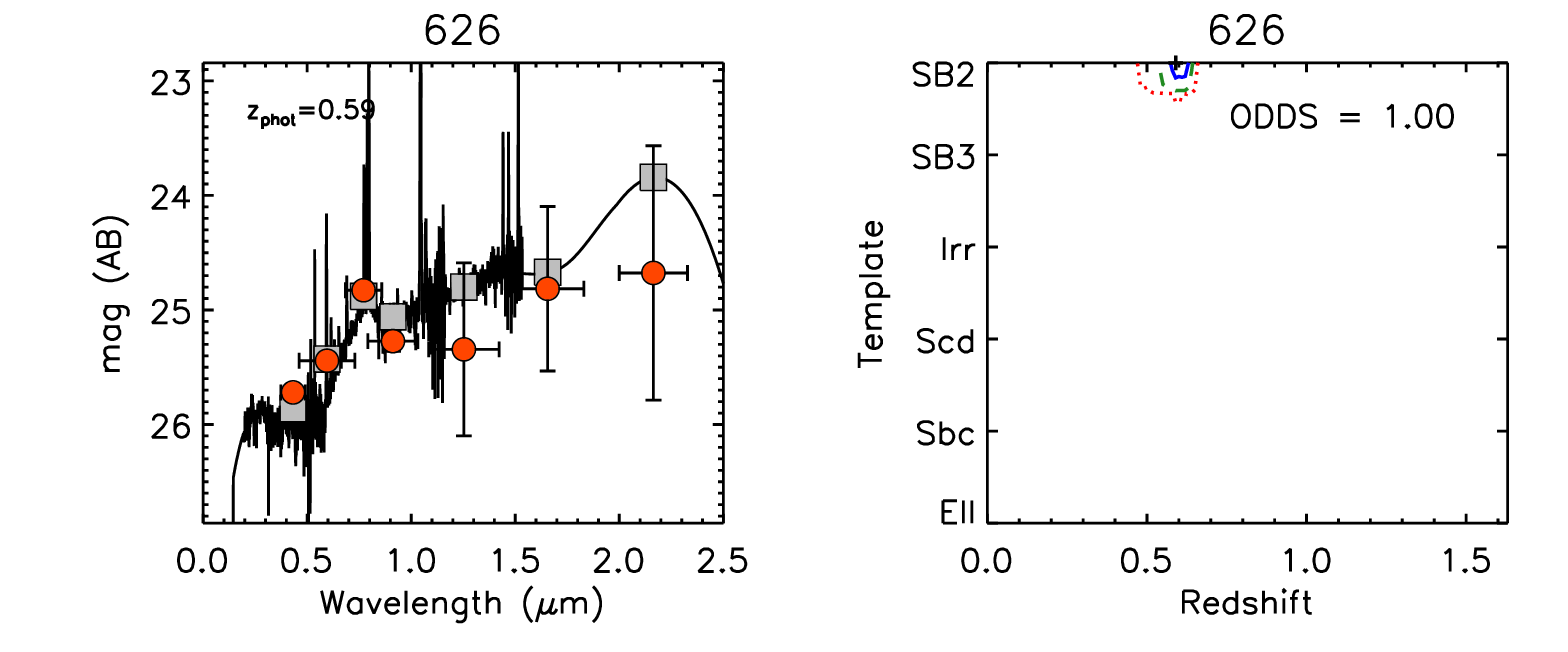}}}
\\
\parbox{12.3cm}{\resizebox{\hsize}{!}{\includegraphics[angle=0]{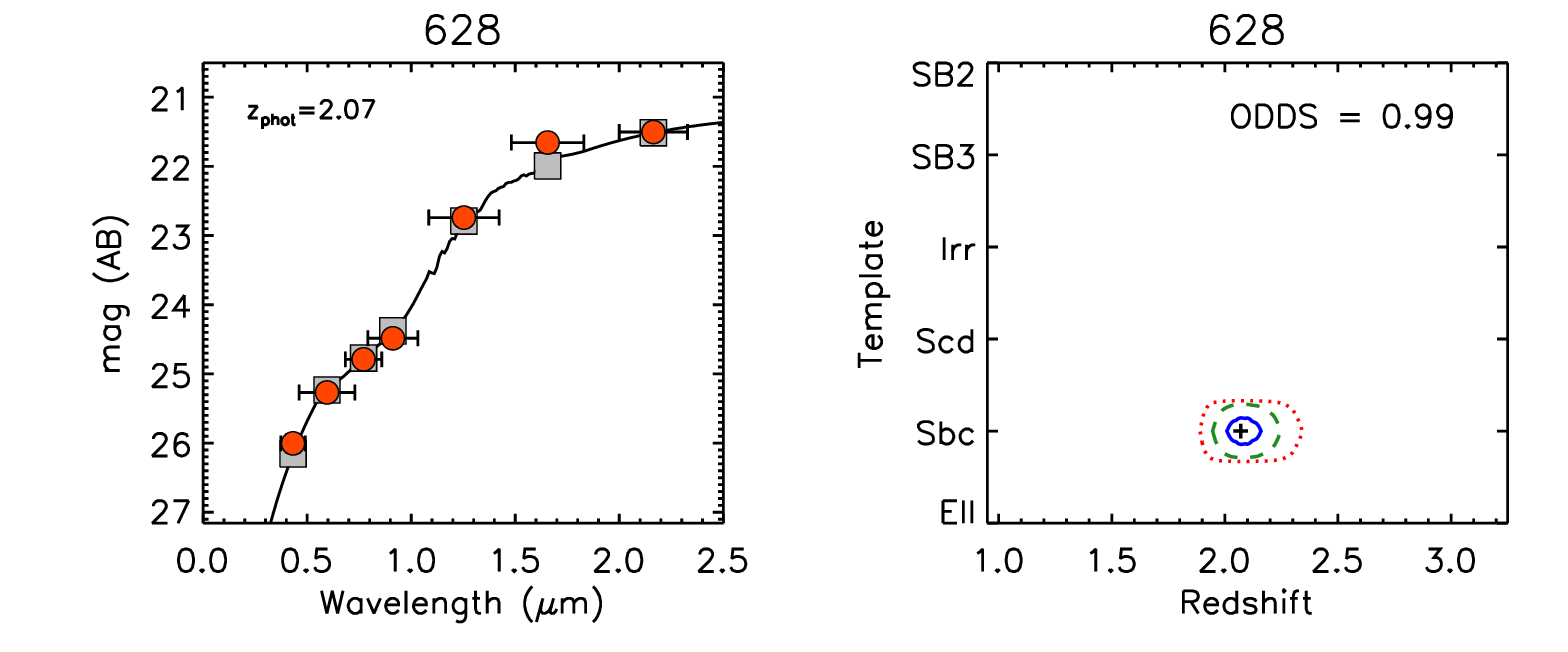}}}
\\
\parbox{12.3cm}{\resizebox{\hsize}{!}{\includegraphics[angle=0]{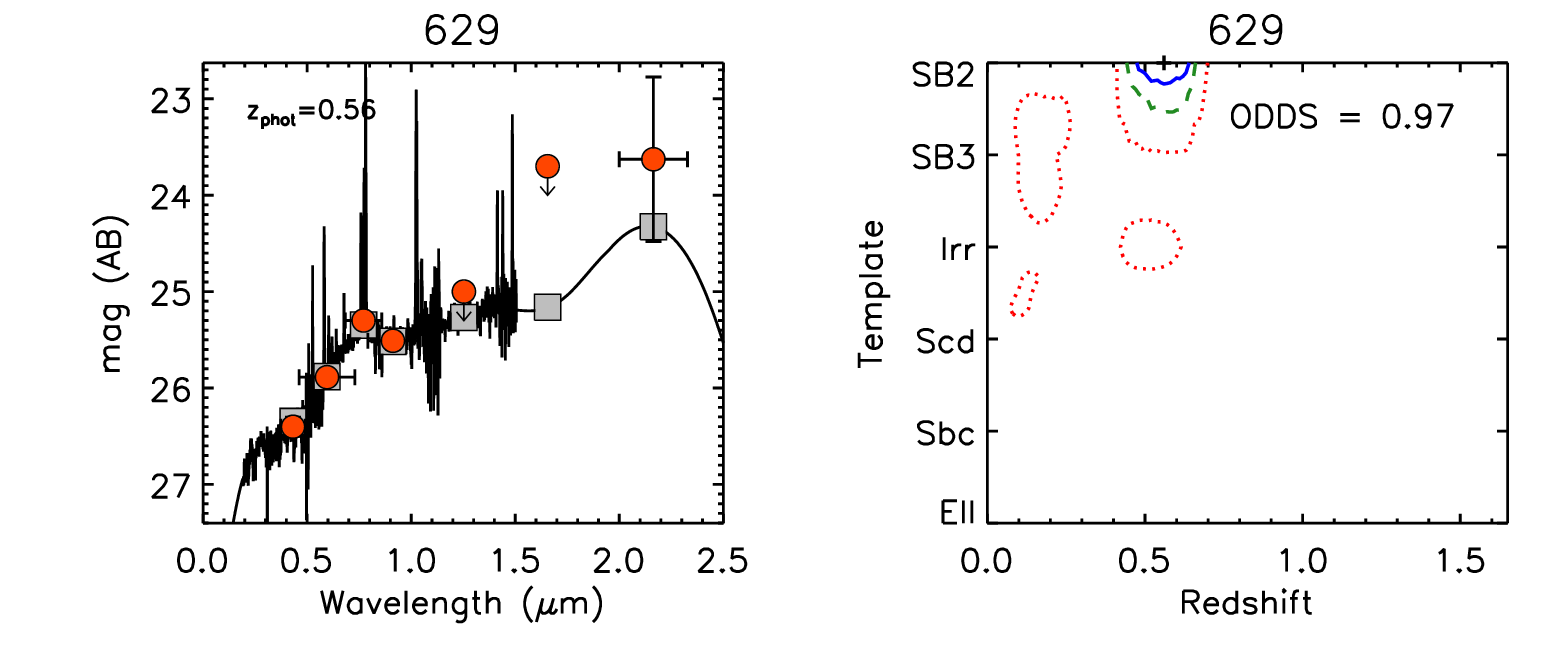}}}
\\
\label{figure:fig_cont_10}
\end{figure*}

\end{document}